\documentclass[12pt]{iopart}
\usepackage{graphicx}
\usepackage{amsfonts}
\usepackage[T1]{fontenc}
\newcommand{\Z}{\mathbb{Z}}

\begin{document}

\title[Nature of Light - Collin]{On the nature of light}

\author{Eddy Collin$^1$}

\address{$^1$Institut Néel - CNRS/UGA, 25 rue des Martyrs, 38042 Grenoble cedex 9, France}
\ead{eddy.collin@neel.cnrs.fr}
\vspace{10pt}
\begin{indented}
\item[] \today
\end{indented}

\begin{abstract}

Electromagnetism is at the heart of the Standard Model, especially through its unification with the weak nuclear force, leading to the {\it electroweak interaction}.
Indeed one of the great challenges of modern physics, in the framework of Quantum Field Theories, is to find a way to unify {\it all} known interactions. But despite all the successes of these theories, including the discovery of the Higgs boson, our basic description of light traveling in free space remains unsatisfactory. The four bosons that compose light are introduced in a rather trivial way, simply by quantizing the (scalar and vector) potential amplitudes. This leads to quite a few conceptual problems linked to the two virtual photons, the longitudinal and scalar ones. Moreover, {\it the spin} of the photon is rather poorly handled by the conventional Quantum Electro-Dynamics theory.
Therefore: what if the field's Lagrangian density, to some extent, {\it would not be} properly chosen? Here we look at these questions from a completely different point of view, {\it bypassing} the problems encountered in conventional theories. We choose a pragmatic approach that relies {\it only} on basic (Condensed Matter like) Quantum Mechanics, and a specific {\it gauge fixing} procedure for the potential field: we propose the concept of {\it gauge duality}, which leads to an original quantization scheme. Building on the Poincar\'e symmetries, all constants of motion are identified. Four bosons are introduced, responsible for a proper spin 1 pseudo-vector {\it and} parity and charge related operators. 
They emerge from scalar fields that can be viewed as {\it generalized fluxes} (in the sense of M. Devoret), with quantum conjugate {\it virtual charges} responsible for the "confinement" of light in space, within "virtual electrodes", somehow reproducing the {\it holographic principle} originally proposed for gravity.
{\it All observable properties} of light in free space then arise from a specific choice of eigenstates (a procedure replacing here the Ward identity of Quantum Field Theory). Real photons are thus the "helicity" bosons, while virtual ones correspond to a "parity charge".
Photon and anti-photon are (as expected) the same particle, linked through an internal gauge transformation.
\end{abstract}

%

\section{Introduction}
\label{intro}

{\it Context - } What {\it is} light?
 The question may seem to be a na\"ive one at first glance, but it is actually profoundly related to our modern understanding of the fundamental laws of Nature. The correct {\it description} of black-body radiation by Planck in 1901 started the construction of Quantum Theory \cite{planck}, when he postulated that the energy stored in an electromagnetic field can only be exchanged in discrete packets: {\it quanta}, named later {\it photons} \cite{photons}.
 What could have passed at the time for a mere mathematical trick is today well accepted; and {\it all} interactions in our Standard Model (SM) are mediated by so-called {\it gauge bosons}, that is packets of energy.
 
The properties of the free-field photons are known with remarkable accuracy \cite{refsphotons}: these are (stable) massless (traveling at speed of light $c$) bosons, with a {\it spin} $1$ internal degree of freedom which presents only two measurable projections (the particle's {\it helicity} $\pm \hbar$, with $\hbar$ Planck's constant, related to the two vacuum field polarizations), and having no electric charge (photons themselves do not interact with light).
This last point emerges naturally from the linearity of the constitutive electromagnetic equations, the well-known {\it Maxwell's equations}. 
It might be only a low energy approximation, since a (very weak) {\it vacuum polarizability} has been proposed (light-by-light scattering) \cite{lightbylight}. This shall be kept outside of our focus, which is solely on low energy physics. 
Since they represent the elementary bricks of electric $\vec{E}$ and magnetic fields $\vec{B}$, which couple to {\it all} charged particles, photons correspond to {\it the gauge bosons of the electromagnetic interaction}. \\

Let us be a bit more specific. Quantum Field Theories (QFT) aim at describing {\it all} elementary particles as excitations of elementary fields. 
Interactions are mediated by fields that possess a so-called {\it gauge symmetry}.
Those generate elementary particles which are {\it bosons}, namely integer spin particles which follow Bose-Einstein statistics. 
Their modeling is based on Gauge Theories (GT), the mathematical framework that deals with gauges and their properties. Gauge symmetries are nowadays promoted at the level of a {\it gauge principle}, namely the {\it reason} behind the differences between the 4 types of fundamental interactions; 
this is in strong contrast with the ancient belief that gauge symmetries represent only a mere redundancy of our mathematical descriptions \cite{gaugesym}.
We give below a brief account of the basis of this construction, in the Quantum Electro-Dynamics (QED) language, as can be found e.g. in Ref. \cite{cohen}. The aim is here to highlight the {\it failures} of the theory.
The starting point is the principle of least action, and a related {\it Lagrangian density} $\cal L$. The "good variables" that must appear in this Lagrangian are the potentials $\vec{A}$ (vector potential) and $V$ (scalar potential) from which $\vec{E}, \vec{B}$ derive, such that the 
Lagrange equations produce in the first place Maxwell's equations. The Lagrangian enables to define for $A_x,A_y,A_z$ and $A_t=V/c$ their conjugate momenta $\Pi_j$ ($j=x,y,z,t$). These momenta are then used to further define the {\it Hamiltonian density} $\cal H$. Follows the transcription into quantum-mechanical terms: the $A_j$, $ \Pi_j$ become {\it operators}, and one imposes the {\it canonical commutation relations} to them. The dynamics generated by the Lagrange equations then translates into the Heisenberg equations: the Hamiltonian turns out to be the {\it generator of time translations}  (see e.g. Ref. \cite{QcohenBook}). 

But this powerful (and presumably rock-solid) reasoning stems from a Lagrangian density {\it that must be given} in the first place. 
To do so, one builds on space-time symmetries (the Poincar\'e {\it external} symmetries), and on the electromagnetic gauge symmetry (the field's {\it internal} symmetry): the $\vec{A}, V$ fields are defined within (the 4-gradient of) a function $\Lambda(x,y,z,t)$, leaving the $\vec{E}, \vec{B}$ physical fields unchanged; this is also named {\it gauge invariance} in a more classical context. 
Let us now recall the "Standard Lagrangian Density" of (low energy) electrodynamics \cite{cohen,landau2}.  The Lagrangian density is given as ${\cal L} = \frac{1}{2} \epsilon_0 E^2 -  \frac{1}{2} \epsilon_0 \, c^2 B^2$ with $\vec{E},\vec{B}$ expressed in terms of the $A_j$s, 
which fulfills internal and external symmetry requirements ($\epsilon_0$ is the vacuum's permittivity).
Invoking the concepts of {\it topology}, any gauge field is understood as a fundamental geometrical property that generates the corresponding interaction. For electromagnetism \cite{barrett}, the transport of a charge $q$ state vector $|\Psi\!\!>$ immersed in a potential $\vec{A},V$ along a path $\gamma$ leads to the accumulation of a phase $\phi=\int_\gamma \vec{A}.d\vec{l}$. $\phi$ and $q$ are quantum conjugate variables, and the potential $\vec{A}$
is treated as a {\it connection}, the mathematical operation that enables transport while preserving the coherence of the theory.
The gauge invariance with respect to $\vec{A}$ implies an invariance with respect to $\phi$, an angle which in group theory 
results in the group $U(1)$ representation: this is the fundamental {\it gauge symmetry} underlying electromagnetism, which by means of Noether's theorem leads to the conservation of the charge $q$ \cite{barrett}.

However, the Standard Lagrangian Density seems to introduce {\it too many degrees of freedom}, as compared to what is needed in the Maxwell's description (using $\vec{E}$ and $\vec{B}$ fields).  
In particular, $\dot{V}$ does not appear in  $\cal L$, meaning that its conjugate momentum is identically zero, and  $V$ itself can be expressed as a function of the other field variables.
The redundancy (which is a direct consequence of the gauge symmetry) is solved by 
 {\it imposing} the so-called Coulomb gauge in which $V=A_z=0$. 
This (pragmatic) choice matches experimental expectations: only two degrees of freedom are experimentally accessible, 
which are described here by $A_x$ and $A_y$ (and their conjugate momenta). 
The corresponding Hamiltonian density writes  ${\cal H} = \frac{1}{2} \epsilon_0 E^2 +  \frac{1}{2} \epsilon_0 \, c^2 B^2$, which is nothing but {\it the total energy density} $\cal E$ of the electromagnetic field. 
Finally the photon spin operator, which is defined as $\vec{S} = \int \!\! \int \!\! \int \! d^3r \, \epsilon_0 \vec{E} \wedge \vec{A}$ (decomposing the total angular momentum $\vec{J}$ into spin and orbit $\vec{L}$) \cite{spin}, produces the expected helicity with $S_x=S_y=0$.
But all this comes at a cost: two "virtual" bosons have been lost, creating a clear asymmetry in the treatment of $x,y$ and $z,t$ components, which is incompatible with the original space-time symmetries (and the expected gauge symmetry, which we did break on pure mathematical grounds). 
Besides, the operator $\vec{S}$ {\it does not} verify the commutation rules imposed to a quantum spin \cite{QcohenBook}, while it must be a spin 1 property. \\

Although the model captures low energy experimental results, this is clearly unsatisfactory on an ontological level (and insufficient for high energy physics). 
The conventional approach to solve this issue is to {\it postulate} a "Manifestly Covariant Lagrangian Density" \cite{cohen,ward}, meaning that $A_i, \dot{A}_i$ ($i=x,y,z$) {\it and} $A_t, \dot{A}_t$ can all be treated on the same footing within $\cal L$.
But in order to still comply with Maxwell's equations, one requires the so-called Lorenz gauge: $\mbox{div} \vec{A} +\dot{V}/c^2=0$. This geometric relationship ensures as well that the potential fields $\vec{A}, V$ comply with Lorentz symmetry (as the original fields $\vec{E},\vec{B}$ do), without restraining the number of degrees of freedom. The various Lagrangian densities in use differ from each other by additive terms which are quadratic in the (derivatives of the) $A_j$s; these are also named {\it gauge choices}, and correspond to  {\it ad hoc} mathematical transformations that enable to simplify (or even perform) calculations \cite{gaugesym,ward}.  \\

\vspace*{-0.5cm}
This treatment also has a cost: the Hamiltonian density $\cal H$ does contain the four degrees of freedom, but the so-called scalar "virtual" photon (corresponding to $A_t, \Pi_t$) contributes as a {\it negative} energy density, which must be dealt with \cite{cohen}.
This problem is directly linked to a rather peculiar issue that appears here: the Hamiltonian density {\it is not} directly equal to the total energy density $\cal E$.
Besides, the commutation rules postulated for the scalar photon lead to {\it negative norms} for the quantum states, which also brings in an extra difficulty \cite{cohen}.
And after all, what shall we do with the two extra "virtual" photons, which cannot be accessed experimentally?
These issues are {\it mathematically rectified} by the gauge choices (in this respect the Standard Lagrangian can be viewed as the most abrupt possibility), and other mathematical techniques which have been developed over the years \cite{ gaugesym, cohen,ward}.
But note that all these approaches necessitate to {\it break the gauge symmetry}, which seems to be in violent contradiction with the gauge principle postulated in the first place.
For this reason specific theoretical tools have been developed to {\it restore} the gauge invariance, like the {\it Dressing Field Method} \cite{ gaugesym}.
As well, the spin $\vec{S}$ defined as $\int \!\! \int \!\! \int \! d^3r \, \epsilon_0 \vec{E} \wedge \vec{A}$ {\it is not} gauge invariant either.
One (historical) solution has been to split this expression, and keep only the part created by the transverse fields as being the spin, $\vec{S} = \int \!\! \int \!\! \int \! d^3r \, \epsilon_0 \vec{E}_{\perp} \wedge \vec{A}_{\perp}$, which is indeed gauge invariant  
\cite{spinPerp}. The helicity is recovered by appealing to the GT toolbox: the {\it Ward identity} ensures that
non-physical degrees of freedom disappear from measurable quantities (enforcing $S_x=S_y=0$) \cite{ward}.
However, already in Ref. \cite{spinPerp} the Authors realized that this operator {\it does not} fulfill the commutation rules of a quantum spin, which leaves us quite unsatisfied.
A recent approach proposes an even more drastic solution for $\vec{S}$, {\it postulating} an alternative spin operator as being $\int \!\! \int \!\! \int \! d^3r \, \vec{\Pi} \wedge \vec{A}$ (together with a related new orbital angular momentum $\vec{L}$) \cite{spinshit}. The point of this Reference is that the presented expressions possess the desired commutation properties. Nonetheless, these $\vec{S}$ and $\vec{L}$ {\it are not} gauge invariant either, and {\it are not} directly related to the work of Ref. \cite{spin}.
In direct alignment with the  dressing field method  of GT, these Authors restore then a satisfactory picture by {\it combining} the light field with the Dirac fields that describe charged particles; only such "inseparable" quantum entities could describe the measurable states of light. \\
 
This leaves us again not fully satisfied: Ockham's razor would certainly privilege a solution where light can be thought of {\it per se}, if such a solution exists.
What appears vividly from this (rather short) account on QFT, is that the gauge invariance is at the core of all our problems. The mathematical techniques of GT seem to be dictated by a {\it  necessity to correct problems}, and not by physical grounds, pushing us into mind-blowing ontological questions; see the excellent Ref. \cite{gaugesym} that also deals with the Philosophy of Physics aspect of it. 
Following C. Rovelli \cite{rovelli}, we note that the interaction Lagrangian density $\cal L_I$ of light with (charged) matter is directly dependent on the vector potential, which {\it is} a gauge-dependent quantity, while the Lagrangian density itself of the coupled system light$+$matter is gauge-invariant. 
So to speak, the light properties related to {\it internal degrees of freedom} (as the photon spin, or the photon charge) seem to be {\it intrinsically} gauge-dependent.
In this respect, the procedure of {\it gauge fixing} could carry an inherent physical meaning: it defines these internal properties, ensuring that they "can indeed be thought of per se", independently of matter. This would be a way to answer the question of C. Rovelli "where does $\vec{A}$ go, when no fermion is around?" \cite{rovelli}: $\vec{A}$ is firmly there, ready to interact. However, in order to be ontologically satisfactory, the gauge fixing procedure would need to leave {\it no spurious undefined redundancy } \cite{gaugesym}. \\

{\it Present work - } Let us face it: the most probable reason for all of these problems should certainly be that {\it we do not start from a fully satisfactory Lagrangian density} $\cal L$. Nature has to have a more "fancy" way of constructing conjugate variables than just referring to the potential amplitudes. And gauge symmetry must be deeply involved in it, or more precisely:  gauge invariance  seems to bring more troubles than fundamental understandings, and one must certainly consider   gauge fixing  as a more physical route \cite{gaugesym,rovelli}. The questions that shall then be answered are: what variables are the canonical quantum conjugate ones, and on what ground can we (fully) fix the gauge?
 
A final comment on the currently available models is in order: charge {\it and parity} (change under mirror symmetry) are introduced in a very indirect way. The charge must be zero (and therefore photons and anti-photons are the same particle), but the theoretical argument sounds like a tautology. As well, the light field parity is well defined when it interacts with matter \cite{QcohenBook,parity,parity2}; but what about the free-field? The definition of an intrinsic parity seems to be more of a convention than an actual property. This is again quite unsatisfactory: a well-defined photon field must certainly be described by {\it an observable} that accounts for the electric charge (which turns out to be 0), and another one for the {\it "parity charge"}, similarly to the "helicity charge" ($\pm \hbar$).

In the present manuscript we propose to tackle these issues in a completely different manner. We {\it shall not} 
start from a convenient Lagrangian density, nor redefine a proper spin operator; neither will we try to justify our postulates from a topological approach.
We will address all of the points raised in the above discussion from a lower level, in the language of  Condensed Matter Physics (CMP).
We will simply {\it take Maxwell's equations and the basic axioms of Quantum Mechanics (QM) as granted}
(the latter ones being the operatorial construction, with the Hamiltonian as generator of time translations). We assume that the Hamiltonian of the field {\it is} its total energy, and will construct the theory on the symmetries of {\it the Poincar\'e group} (space-time transformations that do not alter space-time intervals) \cite{ward}.
 As such, we shall address only the simplest types of traveling waves in free space; extending the approach to other cases is briefly discussed and shall be kept for future works.
Gauge fixing will be at the core of the modeling, building on results obtained in the framework of guided light transmission  \cite{waveguide}.

		\begin{figure}[h!]
		\centering
	\includegraphics[width=15cm]{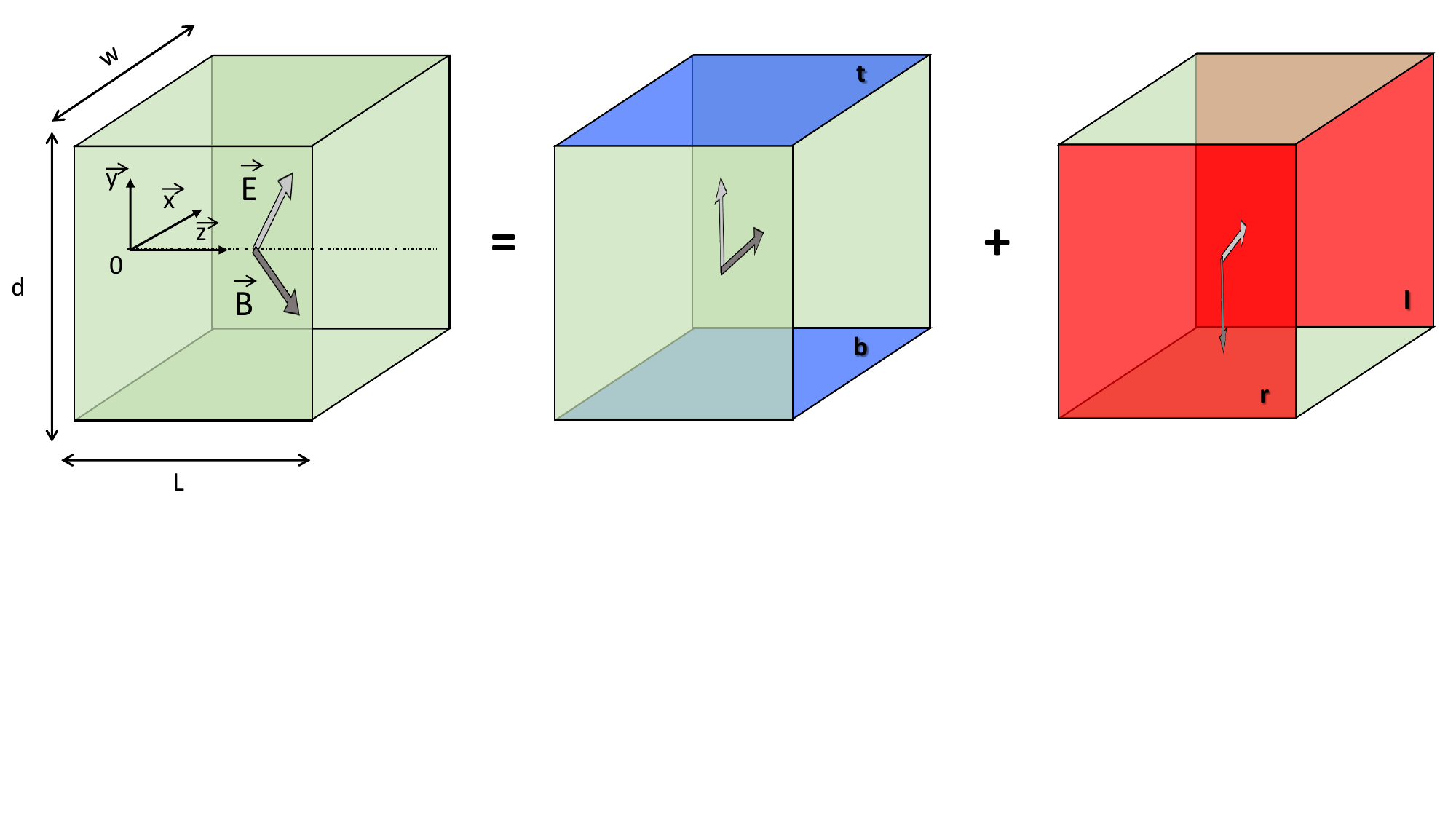}
    \vspace*{-3.7cm}			
			\caption{\small{ 
           An arbitrary TEM wave in free space (we schematize the homogeneous transverse fields, traveling here along $\vec{z}$) can be decomposed along two polarizations. Considering a small volume (green box), each of them is {\it equivalent} to the electromagnetic field confined within infinite parallel plate electrodes, which are here virtual (in blue and red). On each electrode, virtual charges and currents can be defined \cite{waveguide}. $t,b,l,r$ stand for top, bottom, left and right electrodes respectively. }}
			\label{fig_1}
		\end{figure}
		
\section{Starting point}
\label{beginning}

In this Section we recall the grounds on which we build. All field-related parameters are operators, but we omit the conventional {\it hat} notation for clarity.
Let us start by reminding Maxwell's equations in vacuum
\cite{cohen,pozar}:
\begin{eqnarray}
\mbox{div} \, \vec{E}(\vec{r},t)&=&0 , \label{max1} \\
\mbox{div} \, \vec{B}(\vec{r},t)&=&0 ,  \label{max2}\\
\vec{\mbox{rot}} \, \vec{E}(\vec{r},t)&=& -\frac{\partial \vec{B}(\vec{r},t)}{ \partial t} , \label{max3} \\
\vec{\mbox{rot}} \, \vec{B}(\vec{r},t)&=& +\frac{1}{c^2} \frac{\partial \vec{E}(\vec{r},t)}{ \partial t} . \label{max4}
\end{eqnarray}
We shall consider here only the free-field solution (in a flat Euclidean space), with {\it complete space-time symmetry}: the homogeneous (infinite) Transverse-Electric-Magnetic (TEM) wave, originating from time $- \infty$ and progressing towards $t \rightarrow +\infty$. For the writing, we further assume that light propagates in the $\vec{z}$ direction. Mathematical expressions are given in \ref{gauge}.

$\vec{E},\vec{B}$ fields are perpendicular, and can be decomposed along two polarizations as shown in Fig. \ref{fig_1} \cite{landau2,pozar}. Imagine that we cut a small box in free space, as depicted in the Figure (of size $w \times d \times L$). 
Following the reasoning of Ref. \cite{waveguide}, we first point out that for each polarization taken independently, the field is {\it strictly equivalent} to the one created within two parallel metallic plates that would extend to infinity (beyond the size of the original box, which defines only a small portion of them). We call them {\it virtual electrodes}, and they appear in blue and red in Fig. \ref{fig_1}. These electrodes come along with {\it virtual surface charges and currents} $\sigma_s$ and $\vec{j}_s$ (with $s=t,b,l,r$ for top, bottom, left and right), obtained from the boundary conditions:
\begin{eqnarray}
\vec{n} \times \vec{E}& = &0, \label{bound1} \\
\vec{n} \cdot \vec{E} &=& +\frac{\sigma_s}{\epsilon_0} , \label{bound2}  \\
\vec{n} \cdot \vec{B}& =& 0 ,  \label{bound3}  \\
\vec{n} \times \vec{B}&=& + \frac{1}{\epsilon_0 \, c^2}  \, \vec{j}_s , \label{bound4}
\end{eqnarray}
with $\vec{n}$ the vector normal to the electrode's surface, pointing towards the inside of the imaginary box. 
Combining these with Eq. (\ref{max4}), one obtains:
\begin{equation}
\mbox{div}_{\vec{n}} \,\, \vec{j}_s + \frac{\partial \sigma_s}{\partial t}=0 , \label{conserve} 
\end{equation}
which is nothing but {\it charge conservation} ($\mbox{div}_{\vec{n}}$ being the 2D divergence operator calculated in the plane perpendicular to $\vec{n}$).
We argue that these charges and currents are not just a curiosity: they can be seen as {\it the reason} for light confinement within the small volume, exactly like in the case of guided light when the electrodes are real \cite{waveguide}. 
In this view, the boundary conditions Eqs. (\ref{bound1}-\ref{bound4}) are an essential ingredient of the theory.
Note that even though the blue and red electrodes of Fig. \ref{fig_1} do not physically exist, and could not possibly exist simultaneously, they do have a perfectly well defined practical meaning: if one places a non-invasive charge (or current) detector within one of these virtual planes, one {\it would detect} charge and current dynamics. Precisely because the boundary conditions are those of a {\it real metallic plate}, which is what the sensor is made of. All light properties will stem from these boundary surfaces, as will be shown below. \\
		
The problem at hand is invariant through all continuous symmetries of the so-called Poincar\'e group: the space-time symmetries that leave intervals between events unchanged \cite{ward}.
These are translations in space, shift in time, rotations in space, and the {\it Lorentz boosts} between equivalent reference frames.
The Poincar\'e group is 10-dimensional. 
One should also add space-time {\it discrete} symmetries, with two of which that play a specific role (\ref{PandT}): parity $\cal P$ (inversion $z \rightarrow -z$), and time reversal $\cal T$ ($t \rightarrow -t$).
We build on the QM formalism, with Heisenberg's equation describing the time evolution of any operator $G$ with no explicit time-dependence \cite{QcohenBook}:
\begin{equation}
\frac{d G}{d t} = + \frac{\mbox{i}}{\hbar} \left[  H , G \right] , 
\end{equation}
in which $H$ is the Hamiltonian of the electromagnetic field confined in our small box; we {\it define it} as being the total energy of this field.
Applying Noether's theorem leads to {\it 10 constants of motion}, with the obvious ones \cite{cohen,QcohenBook, ward}: 
\begin{itemize}
\item Total energy itself, $H= \int\!\!\int\!\!\int \! d^3r  \left( \frac{1}{2} \epsilon_0 E^2 +  \frac{1}{2} \epsilon_0 \, c^2 B^2\right)$, for the time-shift invariance,
\item Total momentum $\vec{P}=\int\!\!\int\!\!\int \!  d^3r \left(   \epsilon_0 \vec{E} \wedge \vec{B} \right)$, for the space-translation invariance,
\item Total angular momentum $\vec{J}=\int\!\!\int\!\!\int \! d^3r \,\, \vec{r} \wedge \left(   \epsilon_0 \vec{E} \wedge \vec{B} \right)$, for the space-rotation invariance. 
\end{itemize}
These correspond to 7 conserved quantities, 
which we shall name {\it external properties}.
Thus 3 are still missing (related somehow to the Lorentz boosts), which {\it must be internal attributes}. 
How can they emerge from the above expressions? \\

This is were the concepts of {\it potential fields} and {\it gauge} enter into play. The $\vec{E},\vec{B}$ (real) fields can be defined from $\vec{A}, V$ (potentials) as \cite{cohen,pozar}:
\begin{eqnarray}
\vec{E}(\vec{r},t) & = &  - \vec{\mbox{grad}} \,V(\vec{r},t)-\frac{\partial \vec{A}(\vec{r},t)}{\partial t}, \\
\vec{B}(\vec{r},t) & = & + \vec{\mbox{rot}} \,\vec{A}(\vec{r},t) , 
\end{eqnarray}
which remain unchanged if one preforms the replacement:
\begin{eqnarray}
\vec{A}(\vec{r},t) & \rightarrow & \vec{A}(\vec{r},t) + \vec{\mbox{grad}} \, \Lambda(\vec{r},t), \\
V(\vec{r},t) & \rightarrow & V(\vec{r},t) -\frac{\partial \, \Lambda(\vec{r},t)}{\partial t}, 
\end{eqnarray}
where $\Lambda(\vec{r},t)$ is an arbitrary function called {\it gauge}. Requiring that $\vec{A}, V$ are compliant with special relativity, one must impose a geometric relationship called the {\it Lorenz gauge} between potential field components \cite{cohen}:
\begin{equation}
\mbox{div} \vec{A} (\vec{r},t)+\frac{1}{c^2} \frac{\partial V(\vec{r},t)}{\partial t}=0 . \label{lorz}
\end{equation}
Not all gauge functions are relevant: the symmetries of the problem limit rather strongly the possible expressions for $\Lambda$, as discussed in \ref{gauge} and Ref. \cite{waveguide}. 
As well, a gauge related discrete symmetry must exist that can "transform" photons into anti-photons, affecting {\it only} internal properties (see \ref{phidual}): this is the conjugation $\cal C$ symmetry. \\

The introduction of the vector potential $\vec{A}$ enables to decompose the total angular momentum, following Ref. \cite{spin}, into:
\begin{eqnarray}
\vec{J} & = & \int \!\!\! \int \!\!\! \int \! d^3r \, \left( \epsilon_0 \vec{E} \wedge \vec{A} \right) \nonumber \\
        & + & \int \!\!\! \int \!\!\! \int \! d^3r \, \left( \sum_{i}^{x,y,z} \epsilon_0 E_i \,.\, \vec{r} \wedge \vec{\mbox{grad}} A_i \right)  \nonumber  \\
        & - & \int \!\!\! \int \! d^2r \, \left( \vec{r} \wedge \vec{A} \, . \, \epsilon_0 \vec{E} \times \vec{n}  \right) , \label{humblet}
\end{eqnarray}
see details in \ref{spindecompe}.
Since the first term  does not contain $\vec{r}$ explicitly, it has been attributed historically to the spin component  \cite{spin}. But there is actually {\it no reason why } it should be exactly equal to $\vec{S}$. Similarly, the second term has been assigned to the orbital component $\vec{L}$; again, there is no a priori argument to guarantee a strict equality between the two quantities. We therefore propose:
\begin{eqnarray}
\vec{K}_S + \vec{S} & = & \int \!\!\! \int \!\!\! \int \! d^3r \, \left( \epsilon_0 \vec{E} \wedge \vec{A} \right) , \label{Seq} \\
\vec{K}_L + \vec{L} & = &\int \!\!\! \int \!\!\! \int \! d^3r \, \left( \sum_{i}^{x,y,z} \epsilon_0 E_i \,.\, \vec{r} \wedge \vec{\mbox{grad}} A_i \right) , \label{Leq}\\
\vec{K}_S + \vec{K}_L & = &  \vec{K}  .
\end{eqnarray}
The spin $\vec{S}$ is an internal degree of freedom: so should the pseudo-vector $\vec{K}$ also be \cite{pseudo}. We argue that it must 
itself contain a "parity-related" component $\vec{\Pi}$ (not to be confused with the parity operator, or the conjugate momenta discussed in Introduction), and an "electric charge-related" one $\vec{C}$, such that:
\begin{equation}
\vec{K}  = \vec{\Pi} + \vec{C} .
\end{equation}
Both $\vec{\Pi}$ and $\vec{C}$ (and obviously $\vec{S}$) are affected by the discrete symmetries $\cal C, P$ and $\cal T$, in accordance with their roles as internal properties.
The 3 last constants of motion must thus arise from these expressions: the "helicity charge" from $\vec{S}$, the "parity charge" from  $\vec{\Pi}$ and finally the "electric charge" from $\vec{C}$. \\

The last term of Eq. (\ref{humblet}) is a surface integral performed over the boundaries of our small imaginary box. We pose:
\begin{eqnarray}
\vec{m}_s & = & \left( \vec{r} \wedge \vec{A} \, . \, \epsilon_0 \vec{E} \times \vec{n}  \right) , \\
\vec{M} & = & \int \!\!\! \int \! d^2r \,\, \vec{m}_s .
\end{eqnarray}
We argue that there is a profound misconception about this term. 
The {\it conventional assumption} is that the fields must fall-out quickly enough at larges distances, such that the surface integral tends to zero, ensuring that $\vec{M}$ can be dropped \cite{spin}. 
This does not apply when one considers an ideal TEM wave as we do here. 
On the contrary, when defining a small volume in space, we pointed out at the beginning of this Section that the surfaces of this box can be seen as virtual electrodes hosting virtual surface charges $\sigma_s$ and currents $\vec{j}_s$, essential to our understanding. Following the same reasoning, the angular momentum surface density $\vec{m}_s$, which is also defined on the virtual electrodes only ($s=t,b,l,r$), can be seen as {\it the generator of the internal degrees of freedom}.   Eq. (\ref{humblet}) can then be inverted, in order to write:
\begin{equation}
\vec{M} = \vec{\Pi} + \vec{C} + \vec{S} + \vec{L} - \vec{J} .
\end{equation}
The whole approach consists then in {\it deriving} the properties of the electromagnetic field in the box from $\sigma_s$, $\vec{j}_s$ and $\vec{m}_s$: it essentially means that we transform the 3D (and difficult) QFT problem into a 2D (presumably simpler) one \cite{waveguide}. 
This is actually nothing but the expression of the {\it holographic principle} in the framework of electromagnetism, which was originally proposed for quantum gravity \cite{susskind}.
As such, the notion of {\it light confinement} applied to the virtual electrodes shall be taken strictly: what happens inside the small volume is dictated only by the surfaces, and it is in this sense cut from the outside world (as for real electrodes).

\section{New paradigm}
\label{paradigm}

Following Ref. \cite{waveguide}, the surface charges $\sigma_s$ (which are themselves directly related to the currents $\vec{j}_s$ though charge conservation) can be shown to stem form quantities $Q (z,t)$ and $Q\,' (z,t)$, the {\it virtual charges amplitudes} confined on the corresponding electrodes. We use the same convention as those of the previously cited Reference, and non-primed  quantities are related to $t,b$ (top and bottom) electrodes, while primed ones denote parameters corresponding to $l,r$ (left and right). 
The quantum conjugate variables of these charges are respectively $\varphi (z,t)$ and $\varphi\,' (z,t)$, the {\it generalized fluxes} in the sense of M. Devoret \cite{devoret}.
Derivations can be found in \ref{genflux}.

A key result is the fact that $\varphi $ and $\varphi\,' $ can be directly linked to the potentials through 
equations of the type \cite{devoret}: $\partial \varphi /\partial t = \Delta V$  (and equivalent ones involving $\Delta A$ and primes), see \ref{propG}.
The quantities $\Delta V, \Delta V'$ and $\Delta A,\Delta A'$  are the virtual electrodes {\it potential differences}, as introduced in \ref{gauge}.
But this actually implies a {\it specific gauge choice}, that we name the "$\varphi$-gauge" (or Devoret gauge) \cite{waveguide}.
Besides, Eqs. (\ref{Seq},\ref{Leq}) {\it are not} gauge invariant either. And indeed, another specific gauge choice (that we name the "$s$-gauge") leads to $\vec{S}=S_z \, \vec{z}$: 
one recovers the expected spin helicity, which is not a surprise since this gauge is essentially the Coulomb gauge of the standard QED model.  
As for the potential differences, we then define 
{\it angular momentum surface density differences} $\Delta \vec{m}_s, \Delta \vec{m}_s'$, see \ref{gauge}.
The properties of these two gauges, which components are given in \ref{gauge} as well, are discussed thoroughly in \ref{propG}. \\

How much {\it ontological meaning} are we ready to put in Devoret's relationships which link fluxes $\varphi$ to potential differences $\Delta V$? 
Knowing that if we take it as meaningful, we {\it must reject} gauge invariance. Besides, we must then consider the "$s$-gauge" leading to helicity as being as relevant, providing adapted expressions for the $\Delta \vec{m}_s$... But then, how could we chose one gauge or the other for a physical description? 
Only one possibility seems to emerge: a strong principle of {\it gauge fixing} must be invoked \cite{gaugesym,rovelli}, which treats the two above mentioned ones on the same footing. \\


These considerations bring us to our new postulate, on which the following modeling is constructed:
\begin{center}
\fbox{
    \parbox{0.7\textwidth}{
    {\it Gauge duality}: the electromagnetic field must be generated from the {\it superposition} of the two relevant gauges, the one that {\it defines} $\varphi ,\varphi\,'$ with $\Delta V,\Delta V' \neq 0$ called "$\varphi$-gauge", and the one that {\it defines} the helicity through $\Delta \vec{m}_s, \Delta \vec{m}_s'$ with $\Delta V , \Delta V'= 0$, called "$s$-gauge".
    }
}
\end{center}
These two gauges produce exactly the same $\vec{E},\vec{B}$ fields, but do not generate the same internal degrees of freedom, which is why they should be associated. 
Note that the "s-gauge" is by construction {\it the only one} that preserves the property $\partial \varphi/\partial t=\Delta V$ of the "$\varphi$-gauge", when we add-up the two.
In Ref. 
\cite{rovelli}, C. Rovelli compares the gauge parameters that are fixed to "handles" that enable to interact with other fields. Indeed the spin, the parity and the charge
are fundamental (internal) properties that drive interactions. As well, all other (external) properties that define the light field in the small virtual box ($H$, $\vec{P}$ and $\vec{J}, \vec{L}$) can be re-expressed in terms of the scalar fields $\varphi,\varphi\,'$, which can be seen as {\it a generator} for them (see \ref{genflux}).
In this respect, the gauge fixing has been performed on physical grounds, and not as a mathematical {\it ad hoc} simplification \cite{gaugesym}.
Note that $Q$ and $Q\,'$ have {\it nothing to do} with the photon charge (represented by the pseudo-vector $\vec{C}$): they are related to particles (even if virtual), and $\varphi, \varphi\,'$ materialize here the $U(1)$ phase of the common QFT theory. \\

The claim is that the above postulate is {\it generic}, and applies equally well to all situations. Considering the case of guided light in rectangular pipes \cite{waveguide}, the "$s$-gauge" generates an $\vec{S}=0$
solution; light propagating in waveguides must have {\it spin zero}. 
The paradigm shall be extended as well to the case of {\it Gaussian light} \cite{gauss}, which is particularly relevant to quantum technologies; this is outside of the scope of the present work.

\section{The model}
\label{model}

In \ref{gauge} to \ref{propG}, we give a thorough account of the modeling based on a {\it single gauge}, either the "$\varphi$-gauge" or the "$s$-gauge"; this implies the presence of only 2 degrees of freedom (the $x$ and $y$ indexed quadratures).
We present in this Section how to create a new theory where {\it both gauges} apply at the same time, introducing 4 photonic modes. \\

The potentials of the electromagnetic field (vector and scalar) are given as:
\begin{eqnarray}
\vec{A}(\vec{r},t) & =& \vec{A}^\varphi(\vec{r},t)+\vec{A}^s(\vec{r},t) , \\
V(\vec{r},t) & =& V^\varphi(\vec{r},t)+V^s(\vec{r},t),
\end{eqnarray}
Where the $\varphi$ superscript denotes the "$\varphi$-gauge", and the $s$ superscript the "$s$-gauge". These are given by the expressions found in \ref{gauge}, Tabs. \ref{tab_2} and \ref{tab_3} respectively, in which one should substitute the $f_i, \tilde{f}_i$ functions with:
\begin{eqnarray}
f^\varphi_x (z,t) & = & X^\varphi_x \cos(\omega t - \beta z + \Delta \theta )+ Y^\varphi_x \sin(\omega t - \beta z + \Delta \theta ), \label{goodf1} \\
\tilde{f}^\varphi_x (z,t) & = & X^\varphi_x \sin(\omega t - \beta z + \Delta \theta )- Y^\varphi_x \cos(\omega t - \beta z + \Delta \theta ), \\
f^\varphi_y (z,t) & = & X^\varphi_y \cos(\omega t - \beta z + \Delta \theta )+ Y^\varphi_y \sin(\omega t - \beta z + \Delta \theta ), \\
\tilde{f}^\varphi_y (z,t) & = & X^\varphi_y \sin(\omega t - \beta z + \Delta \theta )- Y^\varphi_y \cos(\omega t - \beta z + \Delta \theta ),  \label{goodf4}
\end{eqnarray}
for the former, and:
\begin{eqnarray}
f^s_x (z,t) & = & X^s_x \cos(\omega t - \beta z +  0)+ Y^s_x \sin(\omega t - \beta z + 0), \\
\tilde{f}^s_x (z,t) & = & X^s_x \sin(\omega t - \beta z +  0)- Y^s_x \cos(\omega t - \beta z +  0), \\
f^s_y (z,t) & = & X^s_y \cos(\omega t - \beta z +  0)+ Y^s_y \sin(\omega t - \beta z +  0), \\
\tilde{f}^s_y (z,t) & = & X^s_y \sin(\omega t - \beta z +  0)- Y^s_y \cos(\omega t - \beta z +  0), \label{goodf8}
\end{eqnarray}
for the latter. We take arbitrarily the phase reference (for the propagation origin in $t$ and $\vec{z}$) onto the "$s$-gauge" (the formal "$0$" above); 
but there is no {\it a priori reason} why the "$\varphi$-gauge" should have the same one. We therefore define a phase difference $\Delta \theta$ between the two, Eqs. (\ref{goodf1}-\ref{goodf4}).
Note that the imposed positiveness of the constant $E_m$ introduced in \ref{gauge} is also directly linked to our choice of zero time reference: a $\pi$ shift in all phases accounts for $E_m<0$, which is thus an irrelevant polarity definition, see discussion on 4D-translations below. The signs of the field amplitudes are specifically discussed in \ref{phidual}.

We have therefore introduced four sets of quadratures, which are assumed to be {\it all independent and equivalent}.
As such, they must satisfy commutation rules of the sort:
\begin{eqnarray}
\left[ X^j_i, Y^j_i \right] & = & g_c \left( X^j_i, Y^j_i \right)  \mbox{ \hspace*{1cm} with $i=x,y$ and $j=\varphi,s$ } , \label{X1} \\
\left[ X^j_i, Y^l_k \right] & = & 0 \mbox{  \hspace*{2.9cm} when $i \neq k$ or $j \neq l$} , \\
\left[ X^j_i, X^l_k \right] & = & 0 \mbox{ \hspace*{2.9cm} when $i \neq k$ or $j \neq l$} , \\
\left[ Y^j_i, Y^l_k \right] & = & 0 \mbox{ \hspace*{2.9cm} when $i \neq k$ or $j \neq l$} , \label{X4}
\end{eqnarray}
keeping the problem as generic as it can be.
$g_c$ is therefore (for the time being) a given function, which is characteristic of these modes.
On the model of what is described in \ref{propG}, for each couple of real operators $X^j_i, Y^j_i$, one introduces a {\it complex one} $b^j_i$ such that:
\begin{eqnarray}
X^j_i & = & +\Big( (b^j_i)^\dag + b^j_i \Big), \label{bgenhat} \\
Y^j_i & = & +\mbox{i} \Big( (b^j_i)^\dag - b^j_i \Big) ,\label{bgenhat2}
\end{eqnarray}
which commutators are easily obtained from Eqs. (\ref{X1}-\ref{X4}).
The subtlety here, as compared to the conventional QFT treatment, is that all these modes are of the {\it same nature}: these are all transverse to the propagation (polarized along $\vec{x}$ or $\vec{y}$). There is no specifically longitudinal (along $\vec{z}$), or scalar photon (for the $t$ component). 
The way {\it time} is involved here, is through the relative phase $\Delta \theta$: we combine transverse photons with {\it equivalent ones in another gauge, time-shifted from each other}.
The model leaves us with 8 real gauge coefficients (all the $g^\varphi_i,\tilde{g}^\varphi_i$ and  $g^s_i,\tilde{g}^s_i$ that replace the ones without superscripts in Tabs. \ref{tab_2}, \ref{tab_3}), plus the phase factor $\Delta \theta$, which are yet unknown. \\

These coefficients (and $g_c$) must be obtained, if feasible, from a proper {\it gauge fixing} procedure. 
For this purpose, 
the {\it symmetries} of our problem, the ones of the Poincar\'e group, are essential: our electromagnetic field, described by Eqs. (\ref{Ex}-\ref{Bz}) and Tab. \ref{tab_1}, must comply with them.
This is trivially the case for all 4D-translations, within a phase factor for the $\vec{z},t$ ones which is easily 
removed using a simple change of variables (see \ref{bosontransform}).
From the writing of Eqs. (\ref{goodf1}-\ref{goodf8}), some symmetries have been lost because we arbitrarily chose $\vec{z}$ as direction of propagation: only transverse rotations are eligible, and Lorentz boosts along $\vec{z}$ (at velocity $v$, with $\gamma=1/\sqrt{1-v^2/c^2}$).
These equations are obviously compatible with the latter, performing the replacement:
\begin{eqnarray}  
\omega \rightarrow \omega' &=& \gamma \left( \omega - v \, \beta \right) , \\
\beta \rightarrow \beta' &=& \gamma \left( \beta - \frac{ v \, \omega}{c^2} \right) ,
\end{eqnarray}
which clearly verify $\omega' =c \, |\beta'|$ since  $\omega=c \, |\beta|$ (a property of the $f_i, \tilde{f}_i$ functions, see \ref{gauge}). 
Transverse rotations are presented in  \ref{bosontransform}, building on 
 an essential mathematical tool of the theory: {\it generalized rotations} of the quadratures, which combine a rotation in the transverse plane {\it and} a time-shift (a phase factor). 
 These correspond to the most generic transformation that can be applied to any couple of modes, while preserving the commutation rules {\it and} defining a commutative group.
Finally the discrete symmetries (parity $\cal P$ and time reversal $\cal T$) are addressed in \ref{PandT}.
The conjugation $\cal C$ operation is discussed in \ref{phidual}.

A first consequence of the invariance through an arbitrary rotation around $\vec{z}$ (see \ref{bosontransform} for details), is that the function $g_c$ {\it must be a constant}:
\begin{equation}
g_c \left(X,Y \right) = 2 \,\mbox{i}\, B_c \,\,\, \mbox{with }\,\,\, \Big[b^j_i,(b^j_i)^\dag \Big] = B_c , \,\, \mbox{for all $i$ and $j$}, \label{commute}
\end{equation}
and $B_c$ a complex number which needs to be defined too. We do not impose yet to our modes any specific value for this constant; this must arise from the model. As such, the $b^j_i$ are not yet Dirac operators, and our modes are not yet bosonic. \\

Real and virtual photons are now created by {\it combining the modes} of the two gauges. 
Actually, the generalized rotation involved must reduce to a pure rotation in physical space (meaning with no phase factor), as explained in \ref{bosontransform}. This can be written in matrix form:
\begin{equation}
\left( \begin{array}{c}
X_x \\
X_x^v
\end{array} \right) = \left( R_z \right) \left( \begin{array}{c}
X^s_x \\
X^\varphi_x
\end{array} \right) , \left( \begin{array}{c}
Y_x \\
Y_x^v
\end{array} \right) = \left( R_z \right) \left( \begin{array}{c}
Y^s_x \\
Y^\varphi_x
\end{array} \right) , \label{equaRotxThetaz}
\end{equation}
\begin{equation}
\left( \begin{array}{c}
X_y \\
X_y^v
\end{array} \right) = \left( R_z \right) \left( \begin{array}{c}
X^s_y \\
X^\varphi_y
\end{array} \right) , \left( \begin{array}{c}
Y_y \\
Y_y^v
\end{array} \right) = \left( R_z \right) \left( \begin{array}{c}
Y^s_y \\
Y^\varphi_y
\end{array} \right) , \label{equaRotyThetaz}
\end{equation}
and similarly for the $b^j_i,(b^j_i)^\dag$ operators, with:
\begin{equation}
\left( R_z \right) = \left( \begin{array}{c}
\cos (\theta_z) \\
\sin (\theta_z)
\end{array} \begin{array}{c}
-\sin (\theta_z) \\

\cos (\theta_z)
\end{array}\right) ,
\end{equation}
and $\theta_z$ a (counterclockwise) rotation angle. The same rotation must apply to $x$ and $y$ components, since space is isotropic. $X_x, Y_x$ and $X_y, Y_y$ are thus the true transverse {\it real photon modes}; the quadratures 
with a $v$ superscript correspond to {\it  virtual photons}, keeping the conventional terminology. Note that this shall not be mixed up with the "virtual electrodes" and "virtual charges" wording, which are newly introduced concepts. The virtual photons properties will be addressed below. \\  

Injecting the expressions given above into the definitions of Section \ref{beginning}, one readily computes:
\begin{eqnarray}
H &= & \omega \left( 2 \, \epsilon_0 \,\frac{ E_m^2}{\omega} \, (d w L) \, \Delta_H \right) \left[ +\left(\frac{X_x^2+Y_x^2}{4} + \frac{X_y^2+Y_y^2}{4}\right) \right. \nonumber \\
& &  +\left. \Delta_v \left(\frac{(X^v_x)^2+(Y^v_x)^2}{4} + \frac{(X^v_y)^2+(Y^v_y)^2}{4}\right)\, \right] + \Delta E  , \label{HtoWrite} \\
\vec{P} & =& \beta \, \frac{H}{\omega} \, \vec{z} , \\
\vec{J}  &=& 0 , \\
\vec{L} & = & 0, \\
\vec{K}_L & = & 0.
\end{eqnarray}
As compared to the results obtained with the two gauges taken separately (\ref{propG}), we already notice some peculiarities:
\begin{itemize}
\item The first parenthesis in prefactor of $H$ (which presumably represents $\hbar$, see below) contains   {\it a  renormalization} term $\Delta_H$.
\item There is   {\it an assymmetry} in the energies of real and virtual photons, represented by $\Delta_v$.
\item We have {\it an interaction term} $\Delta E$ between real and virtual photons, which will be clarified in the following.
\end{itemize}
Note also that this Hamiltonian $H$ is {\it completely different} from the one of the usual QFT approach. 

Let us now consider the integral related to the spin operator, the one equal to $\vec{K}_S+\vec{S}$.
For the $\vec{x}$ and $\vec{y}$ components, we split the two pseudo-vectors by assigning to $\vec{K}_S$ the terms corresponding to products of only real quadratures, or only virtual ones. The $\vec{S}$ components are on the other hand identified with all products involving {\it one real}, and {\it one  virtual} quadrature. 
For the $\vec{z}$ component, the spin is obviously the component that involves products of {\it only real} $x$ and $y$ quadratures, while the $\vec{K}_S$ term contains all the others, with at least one virtual quadrature within each product. 
This leads to:
\begin{eqnarray}
&&\!\!\!\!\!\!\!\!\!\!\!\!\!\!\!\!\!\!\!\!\!\!\! \vec{x}. \vec{S}   =    \left( 2 \, \epsilon_0 \,\frac{ E_m^2}{\omega} \, (d w L) \, \Delta_S \right) \left[+\frac{X_y Y_z^{e\!f\!f\,1}- X_z^{e\!f\!f\,1} Y_y }{2} +\frac{X_t^{e\!f\!f\,1} Y_x-X_x Y_t^{e\!f\!f\,1} }{2}\right]\! , \label{spin1} \\
&&\!\!\!\!\!\!\!\!\!\!\!\!\!\!\!\!\!\!\!\!\!\!\! \vec{y}. \vec{S}   =   \left( 2 \, \epsilon_0 \,\frac{ E_m^2}{\omega} \, (d w L) \, \Delta_S \right) \left[+\frac{X_z^{e\!f\!f\,2} Y_x- X_x Y_z^{e\!f\!f\,2}}{2} -\frac{X_y Y_t^{e\!f\!f\,2}-X_t^{e\!f\!f\,2} Y_y }{2} \right] \!,\\
&&\!\!\!\!\!\!\!\!\!\!\!\!\!\!\!\!\!\!\!\!\!\!\! \vec{z}. \vec{S}   =   \left( 2 \, \epsilon_0 \,\frac{ E_m^2}{\omega} \, (d w L) \, \Delta_S \right) \left[+\frac{X_x Y_y- X_y Y_x}{2} \right]\! , \label{spin3}
\end{eqnarray}
with the "effective" quadratures ($e\!f\!f\,1,2$ superscripts, the choice of names will clarify in what follows) defined from sums over the virtual quadratures $X_x^v,Y_x^v, X_y^v, Y_y^v$, weighted by the gauge coefficients $g_i,\tilde{g}_i$ (see below).
As for the energy $H$, a few comments are in order:
\begin{itemize}
\item The prefactor has again {\it a renormalization} term, here $\Delta_S$.
\item We recognize the {\it conventional helicity} operator in $S_z$ (see \ref{propG} and \ref{SpinApp}).
\item The $\vec{x}$ and $\vec{y}$ components involve cross-products that look like spin components; but there appear to be {\it some redundancy}. 
\end{itemize}
To proceed, we should first argue that the symmetry imposed by our original postulate between "$\varphi$-gauge" and "$s$-gauge" implies that {\it the prefactors found in $H$ and $\vec{S}$, leading to $\hbar$, must be equivalent}. Thus $\Delta_H = \Delta_S$ \cite{comment}, which brings: 
\begin{equation}
\tan (\theta_z) = \cos (\Delta \theta) .  \label{thetazDeltatheta}
\end{equation}
This fundamental relationship links the time-delay between the gauges to the transverse rotation angle.
Furthermore, imposing $X_z^{e\!f\!f\,1}=X_z^{e\!f\!f\,2}$ and $Y_z^{e\!f\!f\,1}=Y_z^{e\!f\!f\,2}$ leads to:
\begin{eqnarray}
g_x^\varphi & =& g_x^s + \tilde{g}_x^s \, \tan(\Delta \theta) ,\\
\tilde{g}_x^\varphi & =& \tilde{g}_x^s - g_x^s \, \tan(\Delta \theta) ,\\
g_y^\varphi & =& g_y^s + \tilde{g}_y^s \, \tan(\Delta \theta) ,\\
\tilde{g}_y^\varphi & =& \tilde{g}_y^s - g_y^s \, \tan(\Delta \theta) ,
\end{eqnarray}
which fixes half of the gauge parameters. As a result:
\begin{eqnarray}
&&  \Delta_H =+ 1- \frac{2 \cos(\Delta \theta)^2}{1+\cos(\Delta \theta)^2} , \\
&& \Delta_v =+ \frac{4}{\sin(\Delta \theta )^2} -3 , \\
&&  \mbox{and:} \\
&&  X_z^{e\!f\!f\,1} = -\frac{\sin(\Delta \theta)}{2}  \left[\tilde{g}_x^s \frac{\beta \,w}{\Delta_H} X_x^v -g_x^s\frac{\beta \,w}{\Delta_H} Y_x^v +\tilde{g}_y^s\frac{\beta \,d}{\Delta_H} X_y^v-g_y^s \frac{\beta \,d}{\Delta_H} Y_y^v \right. \nonumber \\
&& \left. -\left(g_x^s\frac{\beta \,w}{\Delta_H} X_x^v +\tilde{g}_x^s\frac{\beta \,w}{\Delta_H} Y_x^v + g_y^s\frac{\beta \,d}{\Delta_H} X_y^v+\tilde{g}_y^s \frac{\beta \,d}{\Delta_H}Y_y^v \right) \tan(\Delta \theta) \right] \! , \label{Xzeff} \\
&&  Y_z^{e\!f\!f\,1} = -\frac{\sin(\Delta \theta)}{2}  \left[ g_x^s \frac{\beta \,w}{\Delta_H}X_x^v +\tilde{g}_x^s \frac{\beta \,w}{\Delta_H}Y_x^v + g_y^s\frac{\beta \,d}{\Delta_H} X_y^v+\tilde{g}_y^s \frac{\beta \,d}{\Delta_H}Y_y^v \right. \nonumber \\
&& \left. +\left(\tilde{g}_x^s\frac{\beta \,w}{\Delta_H} X_x^v -g_x^s\frac{\beta \,w}{\Delta_H} Y_x^v +\tilde{g}_y^s \frac{\beta \,d}{\Delta_H}X_y^v- g_y^s\frac{\beta \,d}{\Delta_H} Y_y^v \right) \tan(\Delta \theta) \right] \!  \label{Yzeff} , \\
&&  X_t^{e\!f\!f\,1} = X_t^{e\!f\!f\,2} =0, \\
&&  Y_t^{e\!f\!f\,1} = Y_t^{e\!f\!f\,2} =0 . 
\end{eqnarray}
At that stage, let us point out that we have $\Delta_H \geq 0$ and $\Delta_v >0$, meaning that the energy stored in the modes is necessarily {\it positive} (which {\it is not} the case in the standard QFT approach, see Introduction), even if it comes at the cost of $\Delta E \neq 0$. 
But for the model to be meaningful, we must also require that $X_z^{e\!f\!f\,1}, Y_z^{e\!f\!f\,1}$ are quadrature operators with {\it the same commutation rules} as the original ones: such that we can indeed identify them with mode operators of the same nature. 
This imposes to the "$s$-gauge" parameters to verify (see \ref{PandT}):
\begin{eqnarray}
\tilde{g}_x^s & = &- \frac{\Delta_H}{\beta w} \,\frac{2 \cos({\cal G})}{\tan(\Delta \theta)} \, \cos (\delta) , \label{gaugecoeff1} \\
g_x^s & = &- \frac{\Delta_H}{\beta w} \,\frac{2 \cos({\cal G})}{\tan(\Delta \theta)} \, \sin (\delta)  ,\\
\tilde{g}_y^s & = & -\frac{\Delta_H}{\beta d} \,\frac{2 \sin({\cal G})}{\tan(\Delta \theta)} \, \cos (\delta)  ,\\
g_y^s & = & -\frac{\Delta_H}{\beta d} \,\frac{2 \sin({\cal G})}{\tan(\Delta \theta)} \, \sin (\delta)  . \label{gaugecoeff4}
\end{eqnarray}
These are essentially the coefficients of a generalized rotation performed on the {\it virtual photons}, characterized by a rotation angle ${\cal G}$ and a phase $\delta$. But this transformation must also comply with the {\it discrete symmetries} of the problem (parity $\cal P$ and time reversal $\cal T$), see \ref{PandT}. Not only this defines the $X_z^{e\!f\!f\,1} = X_z$ and $Y_z^{e\!f\!f\,1} =Y_z$ quadratures, but also the associated $X_t$, $Y_t$ ones.
Note that we use here the historical labels of $z$ and $t$ photons for simplicity; but remember that these have been created in a completely different manner as compared to the standard QFT theory.
Eq. (\ref{HtoWrite}) can be recast with:
\begin{equation}
\frac{(X^v_x)^2+(Y^v_x)^2}{4} + \frac{(X^v_y)^2+(Y^v_y)^2}{4}   =   \frac{X_z^2+Y_z^2}{4} + \frac{X_t^2+Y_t^2}{4} , 
\end{equation}
since this sum is {\it invariant} under a generalized rotation. \\

The properties of the spin operator $\vec{S}$, Eqs. (\ref{spin1}-\ref{spin3}), are discussed in \ref{SpinApp}.
It is actually an essential ingredient of the model, since imposing the canonical spin commutation rules brings:
\begin{eqnarray}
&& 2 \, \epsilon_0 \,\frac{ E_m^2}{\omega} \, (d w L) \, \Delta_H    =   \hbar , \label{hbarlaw} \\
&& B_c   =   1.
\end{eqnarray}
As for the single gauge case (\ref{genflux} and \ref{propG}), the value of the constant $E_m$ is directly determined from $\hbar$; the difference being here the presence of the renormalization factor $\Delta_H$. 
Note the presence of the volume $d w L$ in this expression \cite{waveguide}.
Besides, the commutator constant $B_c$ {\it must be equal to 1}, which qualifies the photon modes as {\it bosons}, and the $b_i^\dag,b_i$ operators as Dirac creation/annihilation operators. 
This is actually an illustration of the Spin-Statistics Theorem \cite{QcohenBook}, in the framework of electromagnetism. \\

We must now explicit the coupling term $\Delta E$, and the $\vec{K}=\vec{K}_S$ pseudo-vector. Reminding the discussion of Section \ref{beginning}, we pose $\vec{K}=\vec{\Pi}+\vec{C}$, with $\vec{\Pi}$ a {\it parity related} pseudo-vector, and $\vec{C}$ a {\it charge related} one. The way we split them resembles how we have split $\vec{K}_S$ from $\vec{S}$ in the above: the terms that contain {\it only} products of virtual quadratures belong to $\vec{\Pi}$, while the ones that have at least one real photon quadrature in the product should be assigned to $\vec{C}$.
This leads to, after some algebra and making use of Eq. (\ref{hbarlaw}):
\begin{eqnarray}
\Delta E & = & \hbar \omega \, \Delta_\pi \, \left[ \sqrt{2} \cos({\cal G}) \cos(\alpha) \left( +\frac{X_y X_t + Y_y Y_t}{2}+\frac{X_x X_z+ Y_x Y_z}{2} \right) \right. \nonumber \\
&+& \left. \sqrt{2} \sin({\cal G}) \cos(\alpha) \left( -\frac{X_x X_t+ Y_x Y_t}{2}+\frac{X_y X_z + Y_y Y_z}{2} \right) \right. \nonumber \\
&+& \left. \sqrt{2} \sin({\cal G}) \sin(\alpha) \left( +\frac{X_t Y_x - X_x Y_t}{2}+\frac{X_y Y_z - X_z Y_y}{2} \right) \right. \nonumber \\
&+& \left.\sqrt{2} \cos({\cal G}) \sin(\alpha) \left( +\frac{X_y Y_t - X_t Y_y}{2}-\frac{X_z Y_x - X_x Y_z}{2} \right)  \right] , \label{deltaE}
\end{eqnarray}
for the energy coupling, and:
\begin{eqnarray}
\vec{x}. \vec{\Pi} & = & \hbar\, \Delta_\pi \, \left[ +2\sqrt{2} \sin({\cal G}) \sin(\alpha) \frac{X_z^2+ Y_z^2 }{4} \right. \nonumber \\
&&\!\!\!\!\!\!\!\!\!\!\!\!\!\!\!  \left. + \sqrt{2} \cos({\cal G}) \sin(\alpha) \frac{X_z X_t+ Y_z Y_t }{2}- \sqrt{2} \cos({\cal G}) \cos(\alpha) \frac{X_z Y_t-X_t Y_z  }{2} \right], \label{Pixcomp} \\
\vec{y}. \vec{\Pi} & = & \hbar\, \Delta_\pi \, \left[ -2\sqrt{2} \cos({\cal G}) \sin(\alpha) \frac{X_z^2+ Y_z^2 }{4} \right. \nonumber \\
&&\!\!\!\!\!\!\!\!\!\!\!\!\!\!\!  \left. + \sqrt{2} \sin({\cal G}) \sin(\alpha) \frac{X_z X_t+ Y_z Y_t }{2}- \sqrt{2} \sin({\cal G}) \cos(\alpha) \frac{X_z Y_t-X_t  Y_z }{2} \right], \\
\vec{z}. \vec{\Pi} & = & \hbar \, \Delta_{\pi z}\, \left( \frac{X_z Y_t - X_t Y_z}{2} \right) ,  \label{Pizcomp}
\end{eqnarray}
for the parity pseudo-vector, and finally:
\begin{eqnarray}
\vec{C} &=& -  \left(\Delta_c \,C_c +\Delta_r \,C_r \right) \vec{z} , \label{chargeQ1} \\
C_c & = & - \hbar\, \frac{\Delta_\pi}{4} \, \left[ \sqrt{2} \sin({\cal G}) \sin(\alpha) \left( +\frac{X_x X_z+ Y_x Y_z }{2}+\frac{X_y X_t+ Y_y Y_t }{2} \right) \right. \nonumber \\
&&\!\!\!\!\!\!\!\!\!\!\!\!\!\!\!  \left. + \sqrt{2} \cos({\cal G}) \sin(\alpha) \left( -\frac{X_y X_z+ Y_y Y_z }{2}+\frac{X_x X_t+ Y_x Y_t }{2} \right)   \right. \nonumber \\
&&\!\!\!\!\!\!\!\!\!\!\!\!\!\!\!  \left. + \sqrt{2} \sin({\cal G}) \cos(\alpha) \left( +\frac{X_z Y_x- X_x Y_z }{2}-\frac{X_y Y_t-  X_t Y_y }{2} \right)  \right. \nonumber \\
&&\!\!\!\!\!\!\!\!\!\!\!\!\!\!\!  \left. + \sqrt{2} \cos({\cal G}) \cos(\alpha) \left( +\frac{X_y Y_z- X_z Y_y }{2}+\frac{X_t Y_x -  X_x Y_t }{2} \right)  \right] ,  \label{chargeQ2} \\
& & \!\!\!\!\! \!\!\!\!\! \!\!\!\!\! \!\!\!\!\! \!\!\!\!\! \!\!\!\!\! C_r    =   - \mbox{sign}(\tan[\Delta \theta] ) \,  \hbar \, \frac{\Delta_\pi}{4} \, \left[ \sqrt{2} \sin({\cal G}) \cos(\alpha) \left( -\frac{X_x X_z+ Y_x Y_z }{2}-\frac{X_y X_t+ Y_y Y_t }{2} \right) \right. \nonumber \\
&&\!\!\!\!\!\!\!\!\!\!\!\!\!\!\!  \left. + \sqrt{2} \cos({\cal G}) \cos(\alpha) \left( +\frac{X_y X_z+ Y_y Y_z }{2}-\frac{X_x X_t+ Y_x Y_t }{2} \right)   \right. \nonumber \\
&&\!\!\!\!\!\!\!\!\!\!\!\!\!\!\!  \left. + \sqrt{2} \sin({\cal G}) \sin(\alpha) \left( +\frac{X_z Y_x- X_x Y_z }{2}-\frac{X_y Y_t-  X_t Y_y }{2} \right)  \right. \nonumber \\
&&\!\!\!\!\!\!\!\!\!\!\!\!\!\!\!  \left. + \sqrt{2} \cos({\cal G}) \sin(\alpha) \left( +\frac{X_y Y_z- X_z Y_y }{2}+\frac{X_t Y_x -  X_x Y_t }{2} \right)  \right] , \label{chargeQ3}
\end{eqnarray}
for the charge pseudo-vector. 
In order to keep the writing as compact as possible, we defined:
\begin{eqnarray}
\Delta_\pi & = & \frac{\sqrt{5+3 \cos(2 \Delta \theta)} }{2 \, |\sin(\Delta \theta) \,  |} ,  \\
\Delta_{\pi z} & = & \frac{2}{\tan(\Delta \theta)^2} ,\\
\Delta_c & = & 2 \,\frac{ 7+5 \cos(2\Delta \theta) }{        5+3 \cos(2\Delta \theta)   } ,   \\
\Delta_r & = &  4\, \frac{ 3+ \cos(2\Delta \theta)  }{      | \tan(\Delta \theta)\, |  \left(5+3 \cos[2\Delta \theta]\right) },
\end{eqnarray}
which are all real positive. Note that $2 \Delta_\pi^2 =\Delta_v$. The angle $\alpha$ that appears in the above is obtained as:
\begin{eqnarray}
\cos(\alpha) &= & - \sqrt{2} \,  | \sin(\Delta \theta)\,| \, \frac{   +\cos( \delta  ) +2 \sin( \delta  )/\tan ( \Delta \theta)   }{\sqrt{5+3 \cos(2 \Delta \theta)}}, \label{cosalpha} \\
\sin(\alpha) &=& + \sqrt{2}\,| \sin(\Delta \theta)\,| \,   \frac{     +2 \cos( \delta  )/\tan ( \Delta \theta) - \sin( \delta  )   }{\sqrt{5+3 \cos(2 \Delta \theta)}} , \label{sinalpha}
\end{eqnarray}
from the original phases $\delta$ and $\Delta \theta$. \\

The properties of the $\vec{\Pi}$ pseudo-vector are discussed in \ref{PiApp}, while those of the $\vec{C}$ pseudo-vector are presented in \ref{CApp}.
We shall include here only the elements relevant to the overall model's presentation.
Note the similarity between $S_z$ (the "helicity charge") given by  Eq. (\ref{spin3}), and Eq. (\ref{Pizcomp}). This leads to the definition of $S_{PT}= \hbar \, (X_z Y_t-X_t Y_z)/2$ in \ref{PiApp}, which will play the role of "parity charge".
From properties given in these two Appendices, plus also \ref{SpinApp}, one obtains:
\begin{eqnarray}
\left[\,  S_z+S_{PT}\,, H \, \right] & = & 0 , \\
\left[\,  C_c+S_{PT}\,, H \, \right] & = & 0 , 
\end{eqnarray}
together with the fact that $S_z$, $S_{PT}$ and $C_c$, $C_r$ are {\it invariants} (under any transform of the Poincar\'e group, especially the rotations around $\vec{z}$).
This implies that $ S_z+S_{PT}$ and $C_c+S_{PT}$ are both {\it constants of motion}. Only one constant of motion is still missing, which should involve $C_r$ in a way or another. 
It turns out that this last relationship is redundant, essentially because $C_c$ and $C_r$ are tight together by a purely (internal) gauge transformation, see \ref{CApp}.  \\

The theory is now formalized, and left with four constants: $\Delta \theta$, which is linked to $\theta_z$ through  Eq. (\ref{thetazDeltatheta}), $\cal G$ and $\delta$ which deserve to be defined. Is it really feasible, is the {\it gauge fixing complete}? This is an intriguing question, since the model is perfectly viable without answering it: all measurable 
properties do {\it not} depend on these gauge parameters, as will become clear in the next 
Section \ref{eigenstates}.

Addressing the final gauge fixing issue requires to introduce the generalized fluxes $\varphi_j, \varphi_j'$ (with $j=x,y, z$ or $t$), and adapt the single gauge formalism of \ref{propG} to the dual gauge case; this is performed in \ref{phidual}.
What is required is that the {\it amplitudes} of these fluxes, which can be directly linked to the virtual charges and currents, 
do comply with the total energy $H$.
This imposes extra constraints through adapted Devoret-like boundary conditions, built on the same $\Delta V, \Delta V'$ and $\Delta A, \Delta A'$ quantities as in the single gauge case.
We obtain remarkably:
\begin{eqnarray}
& & \Delta_H = \frac{1}{3} , \,\,\, \,\,\, \Delta_v = 5 , \label{DeltaHs} \\
& & \Delta_\pi = \sqrt{\frac{5}{2}} ,\,\,  \Delta_{\pi z} = 2 , \label{deltapis} \\
& & \Delta_c = \frac{14}{5} ,\,\,\, \,\,\, \Delta_r = \frac{12}{5} . \label{deltacs}
\end{eqnarray}
Note that the virtual photons are then {\it five times more energetic} than the real ones.
The generalized fluxes (both real and virtual) of the dual gauge approach share the same properties as those of the single gauge cases (\ref{genflux}): they all correspond to scalar fields that propagate at the speed of light $c$, with no rest mass $m=0$ (see \ref{phidual}). 
Besides, ensuring perfect symmetry between the two sets of virtual electrodes leads to:
\begin{equation}
\cos({\cal G})^2 = \sin({\cal G})^2=\frac{1}{2} \,\,\, \mbox{and }\, d=w .
\end{equation}
With this, the writing of the fluxes is {\it equivalent}  on top-bottom and left-right electrode pairs, see \ref{phidual} for details.
Finally, \ref{PiApp} introduces the peculiar commutation rules of the $\vec{\Pi}$ operator. This brings us to propose:
\begin{equation}
\sin(\alpha)^2 =\frac{2}{5} \,\,\,\,\, \mbox{and }\,\cos(\alpha)^2 =\frac{3}{5}. \label{sin2cos2}
\end{equation}
All of these put together in the above expressions does not leave any room for unknowns, {\it except for discrete degeneracies due to the free choice of  signs} for
the different terms appearing in $\Delta E$, $\vec{\Pi}$ and $\vec{C}$, Eqs. (\ref{deltaE}-\ref{chargeQ3}).
This is due to the fact that in the dual gauge formalism, all generalized fluxes involve phase factors which time-shift them from each other: one then cannot define a generic positive amplitude reference, as it is the case in \ref{propG} for a single gauge. As such, $\Delta \theta$, $\theta_z$, $\cal G$ and  $\delta$ are all defined within a $\pm$ sign and a modulo $\pi$, which generates a finite set of possible phase solutions, see \ref{phidual}, Tabs. \ref{tab_4}, \ref{tab_5} and discussion therein.
This is intimately linked to the discrete symmetries of the problem, and especially to time reversal $\cal T$ which requires equivalent solutions to exist (\ref{PandT}).
As well, such redundancies enable to construct the internal conjugation symmetry $\cal C$ (see \ref{phidual}).

\section{Eigenstates of free traveling light}
\label{eigenstates}

We have fixed a new framework for a theory of light, which  must now be solved: this means finding the {\it eigenstates of the Hamiltonian} that also describe in the most practical way all constants of motion.  
We already know that some constraints must emerge, since for instance only the $S_z$ spin component is physical: there must be an equivalent of the {\it  Ward identity} \cite{ward} ensuring that $<S_x>=<S_y>=0$ for accessible states. \\

A conventional transformation enables to diagonalize the helicity operator $S_z$, and define Dirac operators for "left-handed" and "right-handed" photons \cite{loudon}. The similarity of the $S_{PT}$ parity operator with $S_z$  suggests to perform the same transformation onto the virtual bosons. We therefore propose:
\begin{eqnarray}
b_{H,+} & = & + \frac{1}{\sqrt{2}} \, b_x - \frac{\mbox{i}}{\sqrt{2}} \, b_y , \label{BHPlus} \\
b_{H,-} & = & -\frac{\mbox{i}}{\sqrt{2}} \, b_x + \frac{1}{\sqrt{2}} \, b_y , \label{BHMoins} \\
& \mbox{and: } & \\
b_{P,+} & = & + \frac{1}{\sqrt{2}} \, b_z - \frac{\mbox{i}}{\sqrt{2}} \, b_t  , \label{BPPlus} \\
b_{P,-} & = & -\frac{\mbox{i}}{\sqrt{2}} \, b_z + \frac{1}{\sqrt{2}} \, b_t , \label{BPMoins}
\end{eqnarray}
together with their conjugates ($\dag$). These transforms effectively fall in the class of eligible ones, as discussed in \ref{bosontransform}.
The $H$ subscript means "helicity photons" (constructed from the real ones), with a $\pm$ sign mentioning the polarity. Likewise, a $P$ denotes "parity photons" (constructed from the virtual ones), with also a $\pm$ possible choice.  
All the electromagnetic properties can be rephrased in terms of these operators:
\begin{eqnarray}
& &  \!\!\!\!\! \!\!\!\!\! \!\!\!\!\! \!\!\!\!\! \!\!\!\!\! \!\!\!\!\! H  =  \hbar \omega \left[ \left( b_{H,+}^\dag b_{H,+}+\frac{1}{2} + b_{H,-}^\dag b_{H,-}+\frac{1}{2} \right) + \Delta_v \left( b_{P,+}^\dag b_{P,+}+\frac{1}{2} + b_{P,-}^\dag b_{P,-}+\frac{1}{2}\right) \right] \nonumber \\
& & \!\!\!\!\! \!\!\!\!\! \!\!\!\!\! \!\!\!\!\! + \Delta E , \end{eqnarray}
with:
\begin{eqnarray}
&& \!\!\!\!\! \!\!\!\!\! \!\!\!\!\! \!\!\!\!\! \!\!\!\!\! \!\!\!\!\! \Delta E = \hbar \omega \, \sqrt{2} \, \Delta_\pi \left[
+ b_{P,-}^\dag b_{H,-} \, e^{+\mbox{i}(\alpha-{\cal G})}+ b_{P,-} b_{H,-}^\dag \, e^{-\mbox{i}(\alpha-{\cal G})} \right. \nonumber \\
&& \left. + \exp(+2 \mbox{i}  {\cal G}) \, b_{P,+}^\dag b_{H,+} \, e^{+\mbox{i}(\alpha-{\cal G})}+ \exp(-2 \mbox{i}  {\cal G}) \, b_{P,+} b_{H,+}^\dag \, e^{-\mbox{i}(\alpha-{\cal G})} \right] ,
\end{eqnarray}
and $\vec{P} = \beta \, H/\omega \, \vec{z}$; we also remind that $\vec{J}=\vec{L}=0$. The spin operator $\vec{S}$ is obtained as:
\begin{eqnarray}
\vec{x}. \vec{S} & = & \frac{\hbar}{2} \left( b_{P,-}^\dag b_{H,-} +b_{P,-} b_{H,-}^\dag - b_{P,+}^\dag b_{H,+} - b_{P,+} b_{H,+}^\dag \right. \nonumber \\
&& \left. +\mbox{i} \, b_{P,+}^\dag b_{H,-} - \mbox{i} \, b_{P,+} b_{H,-}^\dag +\mbox{i}\, b_{P,-}^\dag b_{H,+} - \mbox{i}\, b_{P,-} b_{H,+}^\dag \right) , \\
\vec{y}. \vec{S} & = & \frac{\hbar}{2} \left( -\mbox{i} \, b_{P,-}^\dag b_{H,-} + \mbox{i} \, b_{P,-} b_{H,-}^\dag - \mbox{i} \, b_{P,+}^\dag b_{H,+} +\mbox{i} \, b_{P,+} b_{H,+}^\dag \right. \nonumber \\
&& \left. + b_{P,+}^\dag b_{H,-} + b_{P,+} b_{H,-}^\dag -  b_{P,-}^\dag b_{H,+} -   b_{P,-} b_{H,+}^\dag \right) , \\
\vec{z}. \vec{S} & = & S_z , \\
S_z & = & \hbar \left( b_{H,+}^\dag b_{H,+} - \, b_{H,-}^\dag b_{H,-} \right) . \label{Szdef}
\end{eqnarray}
The parity pseudo-vector can be recast in:
\begin{eqnarray}
&& \!\!\!\!\!\!\!\!\!\!\!\!\!\!\!\!\!\!\!\!\!\!\!\!\!\!\!\!\!\!\!\!\!\!\!\!\!\!\!\!\!\!\!\!\! \vec{x}. \vec{\Pi}   =   \hbar   \,  \Delta_\pi \left[+\sqrt{2} \sin({\cal G}) \sin(\alpha) \left(+   b_{P,+}^\dag b_{P,+} +  b_{P,-}^\dag b_{P,-}+1 \right)  
-\sqrt{2} \cos({\cal G}) \cos(\alpha) \left( b_{P,+}^\dag b_{P,+} - b_{P,-}^\dag b_{P,-}\right)
 \right. \nonumber \\
&& \!\!\!\!\!\!\!\!\!\!\!\!\!\!\!\!\!\!\!\!\!\!\!\!\!\!\!\!\!\!\!\!\!\!\!\!\!\!\! \left.  +\sqrt{2}   \sin(\alpha) \left( + b_{P,-}^\dag b_{P,+} \,e^{-\mbox{i} {\cal G}}  +    b_{P,-} b_{P,+}^\dag \,e^{+\mbox{i} {\cal G}}  \right)  \right] , \label{xPifunc} \\
&& \!\!\!\!\!\!\!\!\!\!\!\!\!\!\!\!\!\!\!\!\!\!\!\!\!\!\!\!\!\!\!\!\!\!\!\!\!\!\!\!\!\!\!\!\! \vec{y}. \vec{\Pi}   =   \hbar   \,  \Delta_\pi \left[-\sqrt{2} \cos({\cal G}) \sin(\alpha) \left(+   b_{P,+}^\dag b_{P,+} +  b_{P,-}^\dag b_{P,-}+1 \right) 
-\sqrt{2} \sin({\cal G}) \cos(\alpha) \left( b_{P,+}^\dag b_{P,+} - b_{P,-}^\dag b_{P,-}\right)
\right. \nonumber \\
&& \!\!\!\!\!\!\!\!\!\!\!\!\!\!\!\!\!\!\!\!\!\!\!\!\!\!\!\!\!\!\!\!\!\!\!\!\!\!\! \left.  +\sqrt{2}   \sin(\alpha) \left( + \mbox{i} \, b_{P,-}^\dag b_{P,+} \,e^{-\mbox{i} {\cal G}}  - \mbox{i} \,   b_{P,-} b_{P,+}^\dag \,e^{+\mbox{i} {\cal G}}  \right)  \right] , \label{yPifunc} \\
&&\!\!\!\!\!\!\!\!\!\!\!\!\!\!\!\!\!\!\!\!\!\!\!\!\!\!\!\!\!\!\!\!\!\!\!\!\!\!\!\!\!\!\!\!\! \vec{z}. \vec{\Pi}   =  \, \Delta_{\pi z} \, S_{PT} , \\
&&\!\!\!\!\!\!\!\!\!\!\!\!\!\!\!\!\!\!\!\!\!\!\!\!\!\!\!\!\!\!\!\!\!\!\!\!\!\!\!\!\!\!\!\!\! S_{PT}  =  \hbar \left( b_{P,+}^\dag b_{P,+} - \, b_{P,-}^\dag b_{P,-} \right) .
\end{eqnarray}
In the above, $S_z$ the "helicity charge" operator, and $S_{PT}$ the "parity charge" operator are kept explicit.
Finally, the two operators appearing in the $\vec{C}$ pseudo-vector write:
\begin{eqnarray}
C_c & = & \hbar   \,  \frac{\Delta_\pi}{2 \sqrt{2}} \left[ - b_{P,-}^\dag b_{H,-} \, e^{+\mbox{i}(\alpha-{\cal G})}- b_{P,-} b_{H,-}^\dag \, e^{-\mbox{i}(\alpha-{\cal G})} \right. \nonumber \\
&&\!\!\!\!\!\!\!\!\! \left. + \exp(+2 \mbox{i}  {\cal G}) \, b_{P,+}^\dag b_{H,+} \, e^{+\mbox{i}(\alpha-{\cal G})}+ \exp(-2 \mbox{i}  {\cal G}) \, b_{P,+} b_{H,+}^\dag \, e^{-\mbox{i}(\alpha-{\cal G})} \right] , \\
C_r & = &\mbox{sign}(\tan[\Delta \theta] ) \, \hbar   \,  \frac{\Delta_\pi}{2 \sqrt{2}} \left[ + \mbox{i} \, b_{P,-}^\dag b_{H,-} \, e^{+\mbox{i}(\alpha-{\cal G})}-\mbox{i} \, b_{P,-} b_{H,-}^\dag \, e^{-\mbox{i}(\alpha-{\cal G})} \right. \nonumber \\
&&\!\!\!\!\!\!\!\!\! \left. -\mbox{i} \, \exp(+2 \mbox{i}  {\cal G}) \, b_{P,+}^\dag b_{H,+} \, e^{+\mbox{i}(\alpha-{\cal G})} + \mbox{i} \, \exp(-2 \mbox{i}  {\cal G}) \, b_{P,+} b_{H,+}^\dag \, e^{-\mbox{i}(\alpha-{\cal G})} \right] ,
\end{eqnarray}
and $C_c$ will be our "electric charge" operator, $C_r$ being redundant (\ref{CApp}).
Note the similarities across all these expressions; we keep the $\Delta_i$ terms for clarity, even if they are known from Eqs. (\ref{DeltaHs},\ref{deltapis}), as well as the $\cal G$ and $\alpha$ angles which are given within a discrete degeneracy (\ref{phidual}). As such, $\exp(+2 \mbox{i}  {\cal G}) =\pm  \mbox{i}$. \\

Following the conventional terminology, let us define:
\begin{eqnarray}
n_{H,+} & = & < b_{H,+}^\dag b_{H,+} >,  \label{popHP} \\
n_{H,-} & = & < b_{H,-}^\dag b_{H,-} >,  \label{popHM} \\
n_{P,+} & = & < b_{P,+}^\dag b_{P,+} >, \\
n_{P,-} & = & < b_{P,-}^\dag b_{P,-} >, 
\end{eqnarray}
the {\it populations} of the various photon families, with $n_{i,\pm} = 0,1,2, \cdots$ integers. 
The corresponding states form an orthonormal basis for our Hilbert space:
\begin{equation}
\{ |n_{P,+},n_{P,-},n_{H,+}, n_{H,-} > \} \,\,\, \mbox{for all $n_{i,\pm}$} ,
\end{equation}
the vacuum state being obviously $|0,0,0,0 >$.
With today's computing capabilities, it is not too hard to obtain the eigenstates of the Hamiltonian $H$ with exact diagonalization within a subspace $ 0 \leq n_{i,\pm} \leq N_{max}$, with typically $N_{max} \approx 10$. While being quite limited, this is particularly useful for identifying the properties of these states.
And there is actually {\it only one} subspace of states which matches experimental facts, which will be described below. This brings us to postulate:
\begin{center}
\fbox{
    \parbox{0.7\textwidth}{
    {\it Eigenstates of light in free space}: the only relevant eigenstates for the free electromagnetic field are the ones constructed from states with {\it same number of parity and helicity photons} of a given polarity {\it and none} from the other one, namely states such that $n_{P,+}=n_{H,+}$ (with $n_{P,-}=n_{H,-}=0$), or $n_{P,-}=n_{H,-}$ (with  $n_{P,+}=n_{H,+}=0$).
    }
}
\end{center}
By "constructed" we mean that starting from a highly symmetric state with {\it single polarity, and equal parity and helicity}, the effect of the coupling term $\Delta E$ in the Hamiltonian mixes it up with neighboring states differing by $\pm j$ photons.  
The obtained eigenstates write mathematically, 
for $n>0$:
\begin{eqnarray}
|\Psi_{n,+}> & = &  |n,0,n,0> \nonumber \\
&&\!\!\!\!\!\!\!\!\!\!\!\!\!\!\!\!\!\!\!\!\!\!\!\!\!\!\!\!\!\!\!\!\!\!\!\!\!\!\!\!\!\!\!\!\!\!\!\!\!\!\!\!\!\!\!\!\!\!\!\!\!\!\!\!\!\!\!\!\!\!\!\!\!\!\!\!\!\!\!\!\!\!\!\!\!\!\!\!\!\!\!\!\!\!\!\!\!\!\!\!\!     + \! \sum_{j=1}^{n} \! \left( \! +\frac{(+\mbox{i} \, \sqrt{2} \,  \Delta_\pi)^j \, e^{+\mbox{i}\,j \, (\alpha-{\cal G})} }{j! \,(-4)^j } |n+j,0,n-j,0> + \frac{(-\mbox{i} \, \sqrt{2} \,  \Delta_\pi)^j \, e^{-\mbox{i}\,j \, (\alpha-{\cal G})}}{ j! \, 4^j} |n-j,0,n+j,0> \! \right) \! , \label{PsiPlus}
\end{eqnarray}
and similarly:
\begin{eqnarray}
|\Psi_{n,-}> & = &  |0,n,0,n> \nonumber \\
&&\!\!\!\!\!\!\!\!\!\!\!\!\!\!\!\!\!\!\!\!\!\!\!\!\!\!\!\!\!\!\!\!\!\!\!\!\!\!\!\!\!\!\!\!\!\!\!\!\!\!\!\!\!\!\!\!\!\!\!\!\!\!\!\!\!\!\!\!\!\!\!\!\!\!\!\!\!\!\!\!\!\!\!\!\!\!\!\!\!\!\!\!\!\!\!\!\!\!\!\!\!     + \! \sum_{j=1}^{n} \! \left( \! +\frac{( \sqrt{2} \,  \Delta_\pi)^j \, e^{+\mbox{i}\,j \, (\alpha-{\cal G})} }{j! \,(-4)^j } |0,n+j,0,n-j> + \frac{( \sqrt{2} \,  \Delta_\pi)^j \, e^{-\mbox{i}\,j \, (\alpha-{\cal G})}}{ j! \, 4^j} |0,n-j,0,n+j> \! \right) \! , \label{PsiMoins}
\end{eqnarray}
which have norm squared (using Eq. (\ref{deltapis}), $\Delta_\pi = \sqrt{5/2}$): 
\begin{equation}
<\Psi_{n,+} | \Psi_{n,+}> = <\Psi_{n,-} | \Psi_{n,-}> = 1 + 2 \sum_{j=1}^{n} \! \left( \frac{5^j}{16^j \, (j!)^2}  \right).
\end{equation}
This sum converges very rapidly, and the norm can be taken as the $n\rightarrow + \infty$ value:
\begin{equation}
\sqrt{<\Psi_{n,\pm} | \Psi_{n,\pm}>} = \sqrt{ 1+ 2 \left( {\cal I}_0 [ \sqrt{5}/2] -1 \right)} \approx 1.2944\cdots ,
\end{equation}
with ${\cal I}_n(x)$ the modified Bessel function of the first kind. Eqs. (\ref{PsiPlus},\ref{PsiMoins}) are essentially {\it perturbation theory} expansions, taken at order $n$. It turns out that for these states, perturbation theory (if not truncated) leads to the exact result. Note also that these states are all orthogonal to each other. \\

We shall now report on the remarkable properties of the $| \Psi_{n,\pm}>$ eigenstates. At first, they provide:
\begin{equation}
<\Delta E > = < C_c > = < C_r > = < S_x > = < S_y > =0 , \label{mainprop}
\end{equation}
which is precisely what experimental facts require: {\it no charge, and no transverse spin component}.
Besides, we obtain:
\begin{eqnarray}
< \vec{x}.\vec{\Pi} > & = & \hbar   \sqrt{2} \,  \Delta_\pi \left[+ \sin({\cal G}) \sin(\alpha) \mp n \, \cos ({\cal G} \pm \alpha)  \right], \label{xPifin} \\
< \vec{y}.\vec{\Pi} > & = & \hbar   \sqrt{2} \,  \Delta_\pi \left[- \cos({\cal G}) \sin(\alpha) \mp n \, \sin ({\cal G} \pm \alpha)  \right] , \label{yPifin} 
\end{eqnarray}
and finally:
\begin{eqnarray}
< H > & = & \hbar \omega \, \left[ \left(n+1 \right) + \Delta_v \left( n+1 \right) \right] , \label{Hfin}  \\
< S_z > & = & \pm \hbar \, n = \, < S_{PT} > . \label{SPTfin}
\end{eqnarray}
One can show that $\Delta E \, |\Psi_{n,\pm}>$, $C_c \, |\Psi_{n,\pm}>$, $C_r \, |\Psi_{n,\pm}>$, $S_x \, | \Psi_{n,\pm}>$ and $S_y \,  | \Psi_{n,\pm}>$ are all state vectors orthogonal to $|\Psi_{n',\pm,\mp}> \neq |\Psi_{n,\pm}>$ (this is also true for the non-diagonal terms appearing in $\vec{x}.\vec{\Pi}$ and $\vec{y}.\vec{\Pi}$, Eqs. (\ref{xPifunc},\ref{yPifunc}), the second line of each).
This actually means that {\it any state} constructed from the $\{ |\Psi_{n,\pm}> \}$ basis:
\begin{equation}
| \Psi > =\alpha_0 \, |0,0,0,0>+ \sum_{j=1}^{\infty} \alpha_{j,+} \, |\Psi_{j,+}> + \sum_{j=1}^{\infty} \alpha_{j,-} \, |\Psi_{j,-}> ,
\end{equation}
{\it does preserve} the properties Eq. (\ref{mainprop}).
Finally, Eqs. (\ref{xPifin},\ref{yPifin}) and Eqs. (\ref{Hfin},\ref{SPTfin}) must be commented, since they might at first glance look different from what is expected from our common knowledge: zero $<\vec{\Pi}>$ transverse components and no virtual energy $\Delta_v(n+1)$.
The point is that the $b_{H,\pm}, b_{H,\pm}^\dag$ and 
$b_{P,\pm}, b_{P,\pm}^\dag$ operators {\it are not equal}, even though the expectation values $<b_{H,\pm}^\dag b_{H,\pm}>$ and $<b_{P,\pm}^\dag b_{P,\pm}>$ might coincide.
In other words,  restraining the description of the physics at stake to {\it real photon operators only}, then:
\begin{eqnarray}
< \vec{x}.\vec{\Pi} >  \nonumber \\
< \vec{y}.\vec{\Pi} >  \hspace*{3cm} \mbox{should be treated as {\it numbers}, } \nonumber\\
\Delta_v(n+1) \hspace*{2.85cm} \mbox{not operators.} \nonumber\\
<S_{PT} > \nonumber
\end{eqnarray}
This means that if we never couple to $b_{P,\pm}, b_{P,\pm}^\dag$ (virtual) operators in any physical phenomenon, these terms 
are invisible and can simply be ignored.
We then end up with {\it exactly the conventional description of light}. Note that all gauge factors have formally disappeared from this final writing: Hamiltonian and helicity include then only one constant, which is $\hbar$. \\

We must add a final comment on the meaning of this mathematical construction.
Since all symmetries act the same way on $S_z$ and $S_{PT}$, they comply with the chosen eigenstates. Besides, with  
$<C_c>=0$ and $S_{PT}+C_c$ constant of motion, we conclude that $S_z$ becomes a constant of motion in itself. Concomitantly, $S_{PT}$ becomes one as well. 
But it is interesting to point out that these are {\it emergent properties} due to our choice of eigenstates $| \Psi_{n,\pm}>$: this is actually the equivalent of the Ward identity in our framework, which restores {\it observable properties} from a more complex model.
Finally, the fact that virtual photons are not physically accessible is also an important ingredient, leading in the end to the conventional light description.

\section{Conclusion}
\label{conclu}

In the present manuscript, we develop a new model for light traveling in free space.
The starting point is drastically different from the standard QFT approach: here, we remain at a much lower  level, building on basic Quantum Mechanics and Maxwell's relations. The symmetries of the Poincar\'e group are invoked in order to define all constants of motion.

The notion of virtual charges/currents and electrodes are introduced, on the basis of Ref. \cite{waveguide}. These can be seen as responsible for the "confinement" of photons in a small volume of space.
The boundary conditions, that complete Maxwell's relations, are thus a key in the modeling. These are fairly generic since the actual nature of these charges is irrelevant: photons couple to any charge, this is precisely their signature as {\it electromagnetic gauge bosons}. We obtain 
"Devoret-like" expressions defining the generalized fluxes, 
the quantum-conjugate of the virtual charges, which are nothing but the scalar fields from which "real photons" on one hand, and "virtual photons" on the other  originate. 
Similarly, the internal degrees of freedom (which are spin, parity and charge) stem from an angular momentum density, 
defined also on these virtual electrodes.
These scalar fields all propagate at speed of light $c$, and correspond to particles with no rest mass $m=0$.

The way the model is constructed is through a specific {\it gauge fixing} procedure. We call it the {\it dual gauge} paradigm, which assumes that two gauges must apply simultaneously: the "$\varphi$-gauge" responsible for the generalized fluxes, and the "$s$-gauge" that produces the helicity. We combine the two in a very specific manner, that includes a dephasing between the two: this is how {\it time} is involved in the modeling. There is actually no proper $z$ and $t$ photons,  the 4 modes being constructed from two sets of equivalent transverse photons.

The theory succeeds in producing a spin operator $\vec{S}$; but it also leads to the definition of a parity-related one $\vec{\Pi}$, and a charge related one $\vec{C}$. They all verify specific commutation rules, especially with the Hamiltonian, which is how we finally obtain the 3 internal constants of motion: "helicity charge" $S_z$, "parity charge" $S_{PT}$ and "electric charge" $C_c$.
These pseudo-vectors $\vec{S}, \vec{\Pi}$ and $\vec{C}$ also comply with all discrete symmetries in specific manners: 
parity $\cal P$ leaves them unchanged, while both time-reversal $\cal T$ and conjugation $\cal C$ flip their signs. As far as the "external" properties are concerned, 
 $\cal P$ and  $\cal T$ flip the sign of the momentum $\vec{P}$ while  $\cal C$ leaves it unaffected ($H$, $\vec{J}=\vec{L}=0$ being always preserved).
All the gauge parameters we introduce are fixed, within a trivial discrete (sign $\pm$ and modulo $\pi$) degeneracy:
 this solves the (philosophical) debate on {\it surplus variables} from a physical point of view \cite{gaugesym,rovelli}.

The compatibility with observed properties is obtained by selecting the proper eigenstates of the Hamiltonian: these are built from real and virtual photons, the former ones being responsible for helicity and the latter ones for parity. The null photon charge and null transverse spin components appear as an emergent property of these physical states, which verify {\it equality between helicity and parity photon populations}. The construction of these eigenstates is our second postulate, which replaces here the Ward identity of QFT.
As a consequence, photons and anti-photons (linked through the conjugation $\cal C$ operation) are the same particle. \\

Obviously, the presented mathematical construction might explain {\it how} light behaves, but it certainly cannot tell {\it why}. 
Is there a {\it profound reason} that makes the doubling of gauges produce exactly the number of bosons required for a 4D space-time? In the present formalism, this appears as a mere (but necessary) coincidence.
Clearly in order to justify our postulates, a higher level modeling would be required, based on topological arguments and leading to a sound Lagrangian density derivation.
But if one takes it seriously, the implications for the Standard Model are quite remarkable. 
Comparable postulates shall certainly apply to other types of interactions, especially to {\it gravity}. 
The virtual photons are 5 times more energetic than the real ones; this ratio might not be universal, and might depend on the type of gauge bosons considered. And: even though the $z$ and $t$ labeled photons cannot be physically addressed, {\it they must exist} within our framework, precisely because they are created from the same {\it transverse bosons at the origin of the real photons}. In this sense, the name "virtual" is poorly chosen, and "dark" would be better suited.
This is in clear contrast with other QFT approaches where virtual photons appear as a redundancy of the gauge theory \cite{gaugesym}.
Besides, such "dark bosons" might be considered as potential candidates for {\it light dark matter}, without having to introduce a new (and essentially invisible) U(1)-type interaction force, which is what the literature refers to when dealing with "dark photons"
\cite{darkphotons}.

Many questions can be raised from our approach, especially the trivial one: {\it "does it produce any directly experimentally testable properties?"}. The answer might be "yes", if one extends the present work to physical cases like waveguides and Gaussian light. Here, the angular momentum is always $\vec{L}=0$. What do we learn in situations where it is not the case (e.g. Ref. \cite{OAM})? How to handle helicity and parity in a waveguide, for which the spin {\it must be zero}? 
 What does actually the commutation rules of the $\vec{\Pi}$ pseudo-vector represent?
 Similarly, how can we interpret the mathematical expression of $\vec{C}$, knowing that only $<\vec{C}>=0$ is meaningful?
  Are {\it the phases} proper to this model measurable? 
Is it possible to couple to the angular momentum density $\vec{m}_s$, as it is possible to couple to charges? 
And finally, are the virtual photons really impossible to interact with? 
This point has actually already drawn attention (theoretically) in the literature, see Ref. \cite{scalarphi}.
All of these are intriguing open questions for future work.

\section*{Data availability}

No data was used or created for this manuscript. Mathematica\textsuperscript{\textregistered}$\,$ codes are available at the following URL: 
\small{\underline{https://cloud.neel.cnrs.fr/index.php/s/CnnYPKn8XHYZgXa}}.
These correspond to all calculations presented here, and also to the work of Ref. \cite{waveguide}.

\section*{Acknowledgements}

This work has emerged from other ones started in the framework of the European Microkelvin Platform (EMP) collaboration, 
visit: \underline{https://emplatform.eu/}.
It has benefited from stimulating discussions with my student 
 Alexandre Delattre, always keen to dive into disruptive arguments. 
I wish also to thankfully acknowledge enlightening exchanges with many colleagues, both theorists and experimentalists; but it 
does not seem neither realistic nor fair to produce a short list of names, so I prefer to acknowledge a collaborative support from the scientific community as a whole.

\appendix
\normalsize
\section{Gauges of the free field}
\label{gauge}

We follow the same conventions as in Ref. \cite{waveguide}, which modeled microwave light propagation in rectangular waveguides. 
For these, {\it only one degree of freedom} is introduced, represented by the so-called quadratures $X$ and $Y$ (with no units), which are the {\it quantum conjugate observables} that define the light state. 
With the free field, {\it two light polarizations} are possible, which require then two such sets in the conventional modeling.
Their commutators are discussed in \ref{genflux}.
Since we seek solutions traveling in the $\vec{z}$ direction, we introduce the functions:
\begin{eqnarray}
f_x(z,t) & = & X_x \cos(\omega t - \beta z + \theta_0)+ Y_x \sin(\omega t - \beta z + \theta_0)  , \\
\tilde{f}_x(z,t) & = & X_x \sin(\omega t - \beta z + \theta_0)- Y_x \cos(\omega t - \beta z + \theta_0) , \\
& \mbox{and:} &  \nonumber \\
f_y(z,t) & = & X_y \cos(\omega t - \beta z + \theta_0)+ Y_y \sin(\omega t - \beta z + \theta_0)  , \\
\tilde{f}_y(z,t) & = & X_y \sin(\omega t - \beta z + \theta_0)- Y_y \cos(\omega t - \beta z + \theta_0) .
\end{eqnarray} 
The overall phase $\theta_0$ reminds that the choice of the origin in time $t$ and on the $\vec{z}$ axis are arbitrary.
$\beta$ is the wavevector of the light field, positive for a wave traveling in the direction of $\vec{z}$, and 
$\omega>0$ is the angular frequency. The expressions presented in this Appendix shall be adapted in Section \ref{model} of the manuscript, when dealing with {\it four pairs of quadratures}.

The electromagnetic field can be written in a fairly symmetric manner, imposing a perfect factorization between transverse $x,y$ and longitudinal $z$ components \cite{waveguide}:
\begin{eqnarray}
E_x (\vec{r},t) & = & E_m \,\, \Big( g^x_{Ex} (x,y)\, f_x(z,t)+ g^y_{Ex} (x,y)\, f_y(z,t)\Big), \label{Ex} \\
E_y (\vec{r},t) & = & E_m \,\, \Big( g^x_{Ey} (x,y)\, f_x(z,t) +g^y_{Ey} (x,y)\, f_y(z,t)\Big), \\
E_z (\vec{r},t) & = & E_m \,\, \Big( g^x_{Ez} (x,y)\, \tilde{f}_x(z,t)+g^y_{Ez} (x,y) \, \tilde{f}_y(z,t) \Big),
\end{eqnarray}
\vspace*{-7mm}
\begin{eqnarray}
B_x (\vec{r},t) & = & B_m \,\, \Big( g^x_{Bx} (x,y) \, f_x(z,t)+ g^y_{Bx} (x,y) \, f_y(z,t)\Big), \\
B_y (\vec{r},t) & = & B_m \,\, \Big( g^x_{By} (x,y) \, f_x(z,t)+g^y_{By} (x,y) \, f_y(z,t) \Big), \\
B_z (\vec{r},t) & = & B_m \,\, \Big( g^x_{Bz} (x,y) \, \tilde{f}_x(z,t) + g^y_{Bz} (x,y) \, \tilde{f}_y(z,t)\Big). \label{Bz}
\end{eqnarray}
Note that these formulas (compatible with the waveguide treatment) {\it do not} match the Gaussian beam case \cite{gauss}, and would require some adaptation to deal with it.
The units are carried by the (positive) constant $E_m$ in Volt/meter (which will need to be defined, see \ref{genflux}), 
and $B_m = E_m/c$.

We are interested here only in the simplest case: the free-field homogeneous TEM wave. 
Let us decide that the quadratures $X_x,Y_x$ represent    the quantum conjugate observables attached to the $\vec{x}$ electric field polarization, while $X_y,Y_y$ correspond to the $\vec{y}$ one (as depicted in Fig. \ref{fig_1}). Then, the transverse modal functions $g^j_i(x,y)$ are listed in Tab. \ref{tab_1};
the sign choice is such that the transverse field amplitude is taken positive when oriented towards the inside of the small volume $w \times d \times L$.
An equivalent choice could have been made referencing the quadratures to the magnetic field components. 
Maxwell's equations bring trivially $\omega = c \, |\beta|$. 
We take for the box in the direction of propagation {\it periodic boundary conditions}, with  $\beta = 2 \pi \, l/L$, and $l \in \Z^*$ ($\beta \neq 0$, we exclude non-propagating solutions). \\

\begin{table}[h!]
\center
\caption{Modal functions for the free TEM wave (see text). }
\begin{tabular}{@{}llllllll}
\br
Function & & Expression & & Function & & Expression \\
\mr
$g^x_{Ex} (x,y)$ & $=$ & $-1$ & & $g^y_{Ex} (x,y)$ & $=$ & $0$ \\
$g^x_{Ey} (x,y)$ & $=$ & $0$&  & $g^y_{Ey} (x,y)$ & $=$ & $-1$ \\
$g^x_{Ez} (x,y)$ & $=$ & $0$ & & $g^y_{Ez} (x,y)$ & $=$ & $0$ \\
$g^x_{Bx} (x,y)$ & $=$ & $0$  & & $g^y_{Bx} (x,y)$ & $=$ & $+\mbox{sign}(\beta)$ \\
$g^x_{By} (x,y)$ & $=$ & $-\mbox{sign}(\beta)$ &  & $g^y_{By} (x,y)$ & $=$ & $0$ \\
$g^x_{Bz} (x,y)$ & $=$ & $0$ & & $g^y_{Bz} (x,y)$ & $=$ & $0$ \\
\br
\label{tab_1}
\end{tabular}
\end{table}
\normalsize

We adapt here the gauge discussion of Ref. \cite{waveguide}. 
The only relevant expressions for $\Lambda(\vec{r},t)$ are those involving the field quadratures. Any other function leads to a {\it trivial invariance}, with no physical meaning. We therefore write: 
\begin{eqnarray}
\Lambda(\vec{r},t) & = & \Lambda_x(\vec{r},t) + \Lambda_y(\vec{r},t) , \label{gaugeexpr} \\
\Lambda_x(\vec{r},t) &= &\lambda_x(x,y) \, f_x(z,t)+ \tilde{\lambda}_x(x,y) \, \tilde{f}_x(z,t) , \\
\Lambda_y(\vec{r},t) &= &\lambda_y(x,y) \, f_y(z,t)+ \tilde{\lambda}_y(x,y) \, \tilde{f}_y(z,t). 
\end{eqnarray}
The Lorenz gauge condition Eq. (\ref{lorz}) then brings:
\begin{equation}
\frac{\partial^2 \lambda_x(x,y)}{\partial x^2} +  \frac{\partial^2 \lambda_x(x,y)}{\partial y^2} +\left( \frac{\omega^2}{c^2} - \beta^2 \right) \lambda_x(x,y)  =0, \label{gaugeprop}
\end{equation}
and equivalently for $\tilde{\lambda}_x, \lambda_y$ and 
$\tilde{\lambda}_y$ (the $f_i,\tilde{f}_i$ functions being all orthogonal in the sense of linear analysis). 
The parenthesis above is zero, and Eq. (\ref{gaugeprop}) is easily solved, imposing the symmetries of the problem at hand:
\begin{eqnarray}
\lambda_x(x,y) & = & +\gamma_x \, x + g_x , \\
\tilde{\lambda}_x(x,y) & = & +\tilde{\gamma}_x \, x - \tilde{g}_x , \\
\lambda_y(x,y) & = & +\gamma_y \, y + g_y , \\
\tilde{\lambda}_y(x,y) & = & +\tilde{\gamma}_y \, y - \tilde{g}_y .
\end{eqnarray}
Note the sign choices we made for commodity.
In a given gauge, each set of quadratures $X_i, Y_i$ is therefore characterized by only 4 real constants, which will have to be fixed in order to generate both external and internal constants of motion. The gauge fixing will be performed in the first place through the {\it potential differences}:
\begin{eqnarray} 
\Delta V (x,z,t) & =& V(x,d/2,z,t)-V(x,-d/2,z,t) , \\
\Delta A (x,z,t) & = & A_z(x,d/2,z,t)-A_z(x,-d/2,z,t) , \\
& \mbox{and:} & \nonumber \\
\Delta V' (y,z,t) & =& V(w/2,y,z,t)-V(-w/2,y,z,t) , \\
\Delta A' (y,z,t) & = & A_z(w/2,y,z,t)-A_z(-w/2,y,z,t) .
\end{eqnarray}
All non-primed quantities correspond to top, bottom virtual electrodes ($t,b$), while the primed ones stand for left and right ($l,r$). 
These will enable us to re-express the {\it external} constants of motion. 
In the same way, we construct the {\it angular momentum surface density differences}:
\begin{eqnarray}
\Delta \vec{m}_s (x,z,t) & = & \vec{m}_s (x,d/2,z,t) - \vec{m}_s (x,-d/2,z,t) , \\
\Delta \vec{m}_s' (y,z,t) & = & \vec{m}_s (w/2,y,z,t) - \vec{m}_s (-w/2,y,z,t) ,
\end{eqnarray}
for the top-bottom electrodes (no prime), and the left-right ones (primed). 
These will be at the heart of the definition of {\it internal} degrees of freedom. \\

We conclude the Appendix by presenting the two particular gauges which will be invoked in the following. The first one is the "$\varphi$-gauge", see Tab. \ref{tab_2}, while the second one is the "$s$-gauge", see Tab. \ref{tab_3}.

\begin{table}[h!]
\center
\caption{The "$\varphi$-gauge" expressions. }
\begin{tabular}{@{}lll}
\br
Function & & Expression \\
\mr
$A_x (x,y,z,t) $ & $=$ & $0$ \\
$A_y (x,y,z,t) $ & $=$ & $0$ \\
$A_z (x,y,z,t) $ & $=$ & $+\frac{E_m }{\omega }  \beta \Big( x \,f_x (z,t)+ y\, f_y (z,t) \Big) $ \\
& & $+ \frac{E_m }{\omega }  \beta \Big( \!+g_x \,w\,\tilde{f}_x (z,t)+ \tilde{g}_x \,w\,f_x (z,t)+g_y \,d\,\tilde{f}_y (z,t)+ \tilde{g}_y\, d\,f_y (z,t) \, \Big) $ \\
$V (x,y,z,t) $   & $=$ & $+E_m \Big( x \,f_x (z,t)+ y\, f_y (z,t) \Big)$ \\
& & $+ E_m \Big( \!+g_x  \,w\,\tilde{f}_x (z,t)+ \tilde{g}_x \,w\,f_x (z,t)+g_y \,d\,\tilde{f}_y (z,t)+ \tilde{g}_y \,d\,f_y (z,t) \, \Big)$ \\
\br
\label{tab_2}
\end{tabular}
\end{table}
\normalsize

\begin{table}[h!]
\center
\caption{The "$s$-gauge" expressions. }
\begin{tabular}{@{}lll}
\br
Function & & Expression \\
\mr
$A_x (x,y,z,t) $ & $=$ & $+\frac{E_m }{\omega} \tilde{f}_x (z,t) $ \\
$A_y (x,y,z,t) $ & $=$ & $+\frac{E_m }{\omega} \tilde{f}_y (z,t) $ \\
$A_z (x,y,z,t) $ & $=$ & $+ \frac{E_m }{\omega }  \beta \Big( \!+g_x \,w\,\tilde{f}_x (z,t)+ \tilde{g}_x \,w\,f_x (z,t)+g_y \,d\,\tilde{f}_y (z,t)+ \tilde{g}_y\, d\,f_y (z,t) \, \Big)$ \\
$V (x,y,z,t) $   & $=$ & $+ E_m \Big( \!+g_x \,w\,\tilde{f}_x (z,t)+ \tilde{g}_x \,w\,f_x (z,t)+g_y \,d\,\tilde{f}_y (z,t)+ \tilde{g}_y \,d\,f_y (z,t) \, \Big)$ \\
\br
\label{tab_3}
\end{tabular}
\end{table}
\normalsize

For each gauge, the $\gamma_i,\tilde{\gamma}_i$ constants have been fixed, and only the 4 real coefficients $g_i, \tilde{g}_i$ are still unknown (these have been normalized by $g_x \rightarrow g_x \, w E_m/\omega, \tilde{g}_x \rightarrow \tilde{g}_x \, w E_m/\omega$ and $g_y \rightarrow g_y \, d E_m/\omega, \tilde{g}_y \rightarrow \tilde{g}_y \, d E_m/\omega$ in order to have no units).
Consider the simple case where $g_i=\tilde{g}_i=0$. The "$\varphi$-gauge" reduces then to the conventional longitudinal gauge, which we also named Devoret's gauge in Section \ref{paradigm}: this is because it corresponds to the historical treatment performed in quantum electronics, when dealing with guided microwave signals.
On the other hand, the "$s$-gauge" is nothing but the transverse gauge, or Coulomb gauge (used in QED, see Introduction). 
The fundamental 
properties of these two gauges are discussed thoroughly in \ref{propG}.

\section{Decomposition of total angular momentum}
\label{spindecompe}

In this Appendix, we propose to decompose and explain Eq. (\ref{humblet}) using the original Ref. \cite{spin}.
The components of each of the terms appearing therein write:
\begin{eqnarray}
\vec{x}.\left( \epsilon_0 \vec{E} \wedge \vec{A} \right) & = &  \epsilon_0 \Big( E_y(\vec{r},t) A_z(\vec{r},t)- E_z(\vec{r},t) A_y(\vec{r},t) \Big) , \\
\vec{y}.\left( \epsilon_0 \vec{E} \wedge \vec{A} \right) & = &  \epsilon_0 \Big( E_z(\vec{r},t) A_x(\vec{r},t)- E_x(\vec{r},t) A_z(\vec{r},t) \Big) , \\
\vec{z}.\left( \epsilon_0 \vec{E} \wedge \vec{A} \right) & = & \epsilon_0 \Big( E_x(\vec{r},t) A_y(\vec{r},t)- E_y(\vec{r},t) A_x(\vec{r},t) \Big) , 
\end{eqnarray}
for the spin-related term, and:
\begin{eqnarray}
& & \!\!\!\!\! \!\!\!\!\!\!\!\!\!\! \!\!\!\!\! \!\!\!\!\!\!\!\!\!\! \!\!\!\!\! \!\!\!\!\! \!\!\!\!\!  \!\!\!\!\! \!\!\!\!\! \!\!\!\!\! \!\!\!\!\! \vec{x}.\left( \sum_{i}^{x,y,z} \epsilon_0 E_i \,.\, \vec{r} \wedge \vec{\mbox{grad}} A_i \right)   =   \\ 
& & \!\!\!\!\! \!\!\!\!\!\!\!\!\!\! \!\!\!\!\! \!\!\!\!\! \!\!\!\!\! \!\!\!\!\! \!\!\!\!\! \!\!\!\!\!  \!\!\!\!\! \!\!\!\!\! \!\!\!\!\! \!\!\!\!\! \epsilon_0 \Big( E_x(\vec{r},t) \left[ y \frac{\partial A_x(\vec{r},t)}{\partial z} -  z \frac{\partial A_x(\vec{r},t)}{\partial y} \right]+E_y(\vec{r},t) \left[ y \frac{\partial A_y(\vec{r},t)}{\partial z} -  z \frac{\partial A_y(\vec{r},t)}{\partial y} \right] + E_z(\vec{r},t) \left[ y \frac{\partial A_z(\vec{r},t)}{\partial z} -  z \frac{\partial A_z(\vec{r},t)}{\partial y} \right] \Big), \nonumber \\
& & \!\!\!\!\! \!\!\!\!\!\!\!\!\!\! \!\!\!\!\! \!\!\!\!\! \!\!\!\!\! \!\!\!\!\! \!\!\!\!\! \!\!\!\!\! \!\!\!\!\! \!\!\!\!\! \!\!\!\!\! \!\!\!\!\! \vec{y}.\left( \sum_{i}^{x,y,z} \epsilon_0 E_i \,.\, \vec{r} \wedge \vec{\mbox{grad}} A_i \right)  =    \\
&& \!\!\!\!\! \!\!\!\!\!\!\!\!\!\! \!\!\!\!\! \!\!\!\!\! \!\!\!\!\! \!\!\!\!\! \!\!\!\!\! \!\!\!\!\! \!\!\!\!\! \!\!\!\!\! \!\!\!\!\! \!\!\!\!\! \epsilon_0 \Big( E_x(\vec{r},t) \left[ z \frac{\partial A_x(\vec{r},t)}{\partial x} -  x \frac{\partial A_x(\vec{r},t)}{\partial z} \right]+E_y(\vec{r},t) \left[ z \frac{\partial A_y(\vec{r},t)}{\partial x} -  x \frac{\partial A_y(\vec{r},t)}{\partial z} \right] + E_z(\vec{r},t) \left[ z \frac{\partial A_z(\vec{r},t)}{\partial x} -  x \frac{\partial A_z(\vec{r},t)}{\partial z} \right] \Big) , \nonumber\\
& &\!\!\!\!\! \!\!\!\!\!\!\!\!\!\! \!\!\!\!\! \!\!\!\!\!  \!\!\!\!\! \!\!\!\!\! \!\!\!\!\! \!\!\!\!\! \!\!\!\!\! \!\!\!\!\! \!\!\!\!\! \!\!\!\!\!  \vec{z}.\left( \sum_{i}^{x,y,z} \epsilon_0 E_i \,.\, \vec{r} \wedge \vec{\mbox{grad}} A_i \right)  =   \\
& &\!\!\!\!\! \!\!\!\!\! \!\!\!\!\! \!\!\!\!\! \!\!\!\!\! \!\!\!\!\! \!\!\!\!\! \!\!\!\!\! \!\!\!\!\! \!\!\!\!\! \!\!\!\!\! \!\!\!\!\! \!\!\!\!\! \epsilon_0 \Big( E_x(\vec{r},t) \left[ x \frac{\partial A_x(\vec{r},t)}{\partial y} -  y \frac{\partial A_x(\vec{r},t)}{\partial x} \right]+E_y(\vec{r},t) \left[ x \frac{\partial A_y(\vec{r},t)}{\partial y} -  y \frac{\partial A_y(\vec{r},t)}{\partial x} \right] + E_z(\vec{r},t) \left[ x \frac{\partial A_z(\vec{r},t)}{\partial y} -  y \frac{\partial A_z(\vec{r},t)}{\partial x} \right] \Big) \nonumber ,
\end{eqnarray}
for the orbital-related one.
Finally, the angular momentum density $\vec{m}_s$ must be decomposed onto each boundary (of normal $\vec{n}$).
For the top-bottom virtual electrodes we have:
\begin{eqnarray}
\vec{x}. \left( \vec{r} \wedge \vec{A} \, . \, \epsilon_0 \vec{E} \times \vec{n}  \right) & = & \epsilon_0 E_y(\vec{r},t) \Big( y A_z(\vec{r},t)- z A_y(\vec{r},t) \Big) , \\
\vec{y}. \left( \vec{r} \wedge \vec{A} \, . \, \epsilon_0 \vec{E} \times \vec{n}  \right) & = & \epsilon_0 E_y(\vec{r},t) \Big( z A_x(\vec{r},t)- x A_z(\vec{r},t) \Big) , \\
\vec{z}. \left( \vec{r} \wedge \vec{A} \, . \, \epsilon_0 \vec{E} \times \vec{n}  \right) & = & \epsilon_0 E_y(\vec{r},t) \Big( x A_y(\vec{r},t)- y A_x(\vec{r},t) \Big) ,  
\end{eqnarray}
while the left-right ones verify: 
\begin{eqnarray}
\vec{x}. \left( \vec{r} \wedge \vec{A} \, . \, \epsilon_0 \vec{E} \times \vec{n}  \right) & = & \epsilon_0 E_x(\vec{r},t) \Big( y A_z(\vec{r},t)- z A_y(\vec{r},t) \Big) , \\
\vec{y}. \left( \vec{r} \wedge \vec{A} \, . \, \epsilon_0 \vec{E} \times \vec{n}  \right) & = & \epsilon_0 E_x(\vec{r},t) \Big( z A_x(\vec{r},t)- x A_z(\vec{r},t) \Big) , \\
\vec{z}. \left( \vec{r} \wedge \vec{A} \, . \, \epsilon_0 \vec{E} \times \vec{n}  \right) & = & \epsilon_0 E_x(\vec{r},t) \Big( x A_y(\vec{r},t)- y A_x(\vec{r},t) \Big) .  
\end{eqnarray}
The last boundaries of the box correspond to the planes perpendicular to the direction of propagation $\vec{z}$:
\begin{eqnarray}
\vec{x}. \left( \vec{r} \wedge \vec{A} \, . \, \epsilon_0 \vec{E} \times \vec{n}  \right) & = & \epsilon_0 E_z(\vec{r},t) \Big( y A_z(\vec{r},t)- z A_y(\vec{r},t) \Big) , \\
\vec{y}. \left( \vec{r} \wedge \vec{A} \, . \, \epsilon_0 \vec{E} \times \vec{n}  \right) & = & \epsilon_0 E_z(\vec{r},t) \Big( z A_x(\vec{r},t)- x A_z(\vec{r},t) \Big) , \\
\vec{z}. \left( \vec{r} \wedge \vec{A} \, . \, \epsilon_0 \vec{E} \times \vec{n}  \right) & = & \epsilon_0 E_z(\vec{r},t) \Big( x A_y(\vec{r},t)- y A_x(\vec{r},t) \Big) . 
\end{eqnarray}
For the TEM wave, these ones are identically zero. \\

Eq. (\ref{humblet}) holds thanks to the Gauss-like identity:
\begin{equation}
\int \!\!\! \int \! d^2r \, \left( \vec{r} \wedge \vec{A} \, . \, \epsilon_0 \vec{E} \times \vec{n}  \right) =  \int \!\!\! \int\!\!\! \int \! d^3r \, \left( \vec{E} \times \vec{\mbox{grad}}  \right) \left( \vec{r} \wedge \vec{A} \right),
\end{equation}
valid owing to the fact that $\mbox{div}\vec{E}=0$ (no charges present within $w \times d \times L $) \cite{spin}. The volumic vector is explicitly written as:
\begin{eqnarray}
& &\!\!\!\!\!\!\!\!\!\! \!\!\!\!\!\!\!\!\!\! \!\!\!\!\! \!\!\!\!\!  \!\!\!\!\!\vec{x}.  \left( \vec{E} \times \vec{\mbox{grad}}  \right) \left( \vec{r} \wedge \vec{A} \right)  = \\
&& \!\!\!\!\!\!\!\!\!\! \!\!\!\!\!\!\!\!\!\! \!\!\!\!\! \!\!\!\!\!  \!\!\!\!\!E_x(\vec{r},t)\frac{\partial \left[y A_z(\vec{r},t)-z A_y(\vec{r},t)\right]}{\partial x}+E_y(\vec{r},t)\frac{\partial \left[y A_z(\vec{r},t)-z A_y(\vec{r},t)\right]}{\partial y}+E_z(\vec{r},t)\frac{\partial \left[y A_z(\vec{r},t)-z A_y(\vec{r},t)\right]}{\partial z}  \nonumber , \\
& & \!\!\!\!\!\!\!\!\!\! \!\!\!\!\!\!\!\!\!\! \!\!\!\!\! \!\!\!\!\!  \!\!\!\!\!\vec{y}.  \left( \vec{E} \times \vec{\mbox{grad}}  \right) \left( \vec{r} \wedge \vec{A} \right)  = \\
& & \!\!\!\!\!\!\!\!\!\! \!\!\!\!\!\!\!\!\!\! \!\!\!\!\! \!\!\!\!\!  \!\!\!\!\! E_x(\vec{r},t)\frac{\partial \left[z A_x(\vec{r},t)-x A_z(\vec{r},t)\right]}{\partial x}+E_y(\vec{r},t)\frac{\partial \left[z A_x(\vec{r},t)-x A_z(\vec{r},t)\right]}{\partial y}+E_z(\vec{r},t)\frac{\partial \left[z A_x(\vec{r},t)-x A_z(\vec{r},t)\right]}{\partial z} \nonumber, \\
& & \!\!\!\!\!\!\!\!\!\! \!\!\!\!\!\!\!\!\!\! \!\!\!\!\! \!\!\!\!\!  \!\!\!\!\!\vec{z}.  \left( \vec{E} \times \vec{\mbox{grad}}  \right) \left( \vec{r} \wedge \vec{A} \right)  = \\
& & \!\!\!\!\!\!\!\!\!\! \!\!\!\!\!\!\!\!\!\! \!\!\!\!\! \!\!\!\!\!  \!\!\!\!\!  E_x(\vec{r},t)\frac{\partial \left[x A_y(\vec{r},t)-y A_x(\vec{r},t)\right]}{\partial x}+E_y(\vec{r},t)\frac{\partial \left[x A_y(\vec{r},t)-y A_x(\vec{r},t)\right]}{\partial y}+E_z(\vec{r},t)\frac{\partial \left[x A_y(\vec{r},t)-y A_x(\vec{r},t)\right]}{\partial z} \nonumber .
\end{eqnarray}
These expressions, introduced in Section \ref{beginning}, are the starting point for the definition of light's internal degrees of freedom.

\section{Charge, current and generalized flux}
\label{genflux}

Building on the formalism described in Ref. \cite{waveguide}, we introduce in the present Appendix the {\it generalized fluxes} $\varphi (z,t)$ and $\varphi\,' (z,t)$ which are directly related to the virtual surface charges/currents. In the framework of what has been presented in \ref{gauge}, they write:
\begin{eqnarray}
\varphi (z,t) & = & \phi_m \, \tilde{f}_y (z,t), \label{firstPHI} \\
\varphi' (z,t) & = & \phi_m' \, \tilde{f}_x (z,t), \label{firstPHIprime} 
\end{eqnarray}
with the two constants $\phi_m = E_m d/\omega$ and $\phi_m '= E_m w/\omega$, obtained from $E_m$.
The boundary conditions Eqs. (\ref{bound1}-\ref{bound4}) lead to:
\begin{eqnarray}
\sigma_t(x,z,t) & = & + C_d \frac{\partial \varphi (z,t)}{\partial t}, \label{sigmat} \\
\vec{j}_t(x,z,t) & = &-  L_d^{-1} \frac{\partial \varphi (z,t) }{\partial z} \, \vec{z} ,
\end{eqnarray}
for the top virtual electrode; the signs are reversed for the bottom one ($\sigma_b, \vec{j}_b$). As well, the lateral boundary conditions bring:
\begin{eqnarray}
\sigma_l(y,z,t) & = & + C_d\,\!\!' \frac{\partial \varphi\,' (z,t)}{\partial t}, \\
\vec{j}_l(y,z,t) & = &-  L_d'^{\,-1} \frac{\partial \varphi\,' (z,t)}{\partial z} \, \vec{z} , \label{jil}
\end{eqnarray}
with a change of signs for $\sigma_r, \vec{j}_r$.
The above equations contain the {\it capacitances per unit surface}:
\begin{eqnarray}
C_d & = & \frac{\epsilon_0}{d} , \\
C_d\,\!\!' & = & \frac{\epsilon_0}{w} ,
\end{eqnarray}
as well as the {\it inverse inductances per unit surface}:
\begin{eqnarray}
L_d^{-1} & = & \frac{\epsilon_0\, c^2}{  d} , \\
L_d'^{\,-1} & = & \frac{\epsilon_0\, c^2}{  w} .
\end{eqnarray}
Injecting the charges and currents expressions into the conservation law Eq. (\ref{conserve}) leads to the {\it propagation equations}:
\begin{eqnarray}
\frac{\partial^2 \varphi (z,t)}{\partial   z^2} -\frac{1}{c^2} \frac{\partial^2 \varphi (z,t)}{\partial   t^2} & = & 0 , \\
\frac{\partial^2 \varphi\,' (z,t)}{\partial   z^2} -\frac{1}{c^2} \frac{\partial^2 \varphi\,' (z,t)}{\partial   t^2} & = & 0 .
\end{eqnarray}
These are {\it Klein-Gordon} type equations for the scalar fields $\varphi, \varphi\,'$ with mass $m=0$ (the r.h.s. are zero), and propagation velocity the speed of light $c$. This is typical of the TEM wave: for Transverse Electric (TE) and Transverse Magnetic (TM) guided modes, these two properties are {\it not} systematically satisfied \cite{waveguide}. \\

The generalized fluxes can be used to rewrite the constants of motion of {\it external} origin: energy $H$, momentum $\vec{P}$ and (total) angular momentum $\vec{J}$ \cite{waveguide}. They become:
\begin{eqnarray}
H & = & \!\!\! \int_{z=0}^L  \left[ \frac{1}{2}\, C_0^{-1} \left( C_0 \frac{\partial \varphi (z,t)}{\partial t}\right)^{\!\!2} + \frac{1}{2} \, L_0^{-1} \left(  \frac{\partial \varphi (z,t)}{\partial z}\right)^{\!\!2} \right. \nonumber \\
&& \left. +\frac{1}{2}\, C_0\,\!\!'^{-1} \left( C_0\,\!\!' \frac{\partial \varphi\,' (z,t)}{\partial t}\right)^{\!\!2} + \frac{1}{2} \, L_0'^{-1} \left(  \frac{\partial \varphi\,' (z,t)}{\partial z}\right)^{\!\!2}   \right]   dz/L , \label{finalHz} \\
\vec{P} & = &\!\!\!  \int_{z=0}^L     \left[\left( C_0 \frac{\partial \varphi (z,t)}{\partial t} \right)\, \left( -  \frac{\partial \varphi (z,t)}{\partial z} \right) \right. \nonumber \\
& & \left. +\left( C_0\,\!\!' \frac{\partial \varphi\,' (z,t)}{\partial t} \right)\, \left( -  \frac{\partial \varphi\,' (z,t)}{\partial z} \right) \right]  dz/L \, \vec{z} , \label{finalPz} \\
\vec{J} &= & 0 .
\end{eqnarray}
The angular momentum is {\it strictly zero}, and we define  effective capacitances and inductances for the virtual parallel plates:
\begin{eqnarray}
C_0 & = & C_d \, w \, L , \\
L_0^{-1} & = & L_d^{-1} \, w \, L ,
\end{eqnarray}
for the top-bottom pair, and:
\begin{eqnarray}
C_0\,\!\!' & = & C_d\,\!\!' \, d \, L , \\
L_0'^{-1} & = & L_d'^{\,-1} \, d \, L ,
\end{eqnarray}
for left-right. 
The total energy $H$ is actually strictly equal to the {\it surface integral of the charges and currents surface energy density} $H_d$ \cite{waveguide}:
\begin{equation}
H_d = \frac{1}{2} C_d^{-1} \, \sigma_t^2 +\frac{1}{2} L_d \, \vec{j}_t^2 + \frac{1}{2} C_d\,'^{\,-1} \,\sigma_l^2 +\frac{1}{2} L_d' \, \vec{j}_l^2 . \label{Hdensity}
\end{equation}
From the above expressions, we see emerging two new quantities:
\begin{eqnarray}
Q(z,t) & = & C_0 \, \frac{\partial \varphi (z,t)}{\partial t} , \label{Qn} \\
Q\,'(z,t) & = & C_0\,\!\!' \, \frac{\partial \varphi\,' (z,t)}{\partial t}, \label{Qp}
\end{eqnarray}
which verify the commutation rules:
\begin{eqnarray}
\left[\,\varphi (z,t), Q(z,t)\,\right] & = & (2 \,C_0 \, \omega \, \phi_m^2) \, \left[ X_y, Y_y \right]/2 , \\
\left[\,\varphi\,' (z,t), Q\,'(z,t) \, \right] & = & (2 \,C_0\,\!\!' \, \omega \, \phi_m'^{\,2}) \, \left[ X_x, Y_x \right]/2 .   
\end{eqnarray}
Imposing to the quadratures the commutation rule $[X_j,Y_j] = 2\, \mbox{i}$ {\it and } $2 \,C_0 \, \omega \, \phi_m^2 = 2 \,C_0\,\!\!' \, \omega \, \phi_m'^{\,2} = 2 \, \epsilon_0 \, E_m^2 (d w L)/\omega = \hbar$,
which {\it fixes} the constant $E_m$ from Planck's $\hbar$, makes the $\varphi, Q$ variables (and $\varphi\,', Q\,'$ as well) {\it canonical quantum conjugate} ones \cite{waveguide}. 
Note also that assuming the $x$ related quadratures to commute with the $y$ ones, then the primed and non-primed generalized fluxes and charges are also {\it independent variables}.
The quadratures commutation rule is further discussed in \ref{propG} below.\\

Up to this point, all of these characteristics %
 {\it did not directly imply a gauge definition}; this is performed in 
\ref{propG} below, on the basis of the two relevant gauges introduced in \ref{gauge}.
This modeling, which contains only {\it two "real" degrees of freedom}, shall be adapted in Section \ref{model} of the manuscript in order to accommodate the extra {\it two  "virtual" photon} modes.

\section{Properties of two relevant gauges: the $\varphi$-gauge and the $s$-gauge}
\label{propG}

In this Appendix, we present the specific properties of the two gauges described in \ref{gauge}.
But before doing so, we simplify Eqs. (\ref{finalHz},\ref{finalPz}) and rewrite them in terms of the two quadratures:
\begin{eqnarray}
H & = & \hbar \omega \left( \frac{X_x^2+ Y_x^2}{4} + \frac{X_y^2+ Y_y^2}{4} \right) , \label{HXY} \\
\vec{P} & = & \hbar \beta \left( \frac{X_x^2+ Y_x^2}{4} + \frac{X_y^2+ Y_y^2}{4}  \right) \vec{z} , \label{PXY}
\end{eqnarray}
which made use of the property $2 \,C_0 \, \omega \, \phi_m^2 = 2 \,C_0\,\!\!' \, \omega \, \phi_m'^{\,2} = 2 \, \epsilon_0 \, E_m^2 (d w L)/\omega = \hbar$ introduced above in \ref{genflux}; 
in what comes below, each time an $\hbar$ appears we applied the same relations.
These expressions are usually given in terms of the Dirac {\it creation and annihilation operators}:
\begin{eqnarray}
X_x & = & +\left( b_x^\dag + b_x \right), \label{bbhat1} \\
Y_x & = & +\mbox{i} \left( b_x^\dag - b_x \right) ,
\end{eqnarray}
and similarly for the $y$ component \cite{waveguide}. 
In order to match the $[X_i,Y_i]$ commutators mentioned in \ref{genflux}, they must verify the commutation rules:
\begin{equation}
\left[b_x , b_x^\dag \right]= 1, \, \left[b_y , b_y^\dag \right]= 1 ,
\end{equation}
which makes them {\it boson operators}. With $\left[b_x , b_y^\dag \right]=\left[b_x , b_y  \right]=0$, the $x$ and $y$ quadratures are also independent. 
Eqs. (\ref{HXY}, \ref{PXY}) can then be recast in:
\begin{eqnarray}
H & = & \hbar \omega \left( b_x^\dag b_x + \frac{1}{2}+b_y^\dag b_y + \frac{1}{2} \right) ,  \\
| \vec{P} \,| & = & H/c ,
\end{eqnarray}
which are our textbook QED expressions describing  light in free space at low energies \cite{cohen,landau2,loudon}.

\subsection{"$\varphi$-gauge" characteristics}

Let us consider first the "$\varphi$-gauge" which components are given in Tab. \ref{tab_2}.
In addition to $H$, $\vec{P}$ and $\vec{J}=0$ (the {\it external} constants of motion) discussed above, this gauge verifies:
\begin{eqnarray}
\vec{L} & = & 0, \\
& \mbox{and:} \nonumber \\
\vec{K}_L & = & 0 , \\
\vec{S} & = & 0, \\
\vec{z}.\vec{K}_S &=& 0 \label{Kzval}.
\end{eqnarray}
All {\it internal} degrees of freedom are zero, apart from the transverse components of the $\vec{K}_S$ pseudo-vector:
\begin{eqnarray}
& &\!\!\!\!\!\!\!\!\!\!\!\!\!\!\!\!\!\!\!\!\!\!\!\! \!\!\!\!\!\!\!\!\!\!\!\!\!\!\!\!\!\!\!\!\!\!\!\!\!\!\!\!\!\!\!\!\!\!\!\!\!\!\!\!  \vec{x}.\vec{K}_S = -\hbar \left( \!(+\tilde{g}_y)(\beta \, d) \frac{X_y^2+Y_y^2}{4} + (+ g_x)(\beta \, w) \frac{X_x Y_y-X_y Y_x }{4} + (+ \tilde{g}_x)(\beta \, w) \frac{X_x X_y + Y_x Y_y}{4} \right) , \label{Kxval}\\
& &\!\!\!\!\!\!\!\!\!\!\!\!\!\!\!\!\!\!\!\!\!\!\!\! \!\!\!\!\!\!\!\!\!\!\!\!\!\!\!\!\!\!\!\!\!\!\!\!\!\!\!\!\!\!\!\!\!\!\!\!\!\!\!\! \vec{y}.\vec{K}_S = +\hbar \left(\! (+\tilde{g}_x)(\beta \, w) \frac{X_x^2+Y_x^2}{4} + (- g_y)(\beta \, d) \frac{X_x Y_y-X_y Y_x }{4} + (+ \tilde{g}_y)(\beta \, d) \frac{X_x X_y + Y_x Y_y}{4} \right) , \label{Kyval}
\end{eqnarray}
which are directly related to the gauge coefficients $g_i, \tilde{g}_i$. \\

The fundamental property of this gauge is to verify the relations (Devoret's ones):
\begin{eqnarray}
\frac{\partial \varphi(z,t)}{\partial t} & =& \Delta V  , \label{devoret1} \\
\frac{\partial \varphi(z,t)}{\partial z} & =& - \Delta A  , \\
& \mbox{and:} & \nonumber \\
\frac{\partial \varphi\,'(z,t)}{\partial t} & =& \Delta V'  , \\
\frac{\partial \varphi\,'(z,t)}{\partial z} & =& - \Delta A'  ,\label{devoret4}
\end{eqnarray}
which {\it defines} the generalized fluxes from the virtual electrodes potential differences.
This gauge also leads to:
\begin{eqnarray}
\vec{x}. \Delta \vec{m}_s & = & + 2 \sigma_t \left( +\varphi_k - (d/w) \, \frac{x}{2} \, \Delta A' \right) , \\
\vec{y}. \Delta \vec{m}_s & = & + 2 \sigma_t \left( + \frac{x}{2} \, \Delta A \right) , \\
\vec{z}. \Delta \vec{m}_s & = & 0 ,
\end{eqnarray}
and: 
\begin{eqnarray}
\vec{x}. \Delta \vec{m}_s' & = & + 2 \sigma_l \left( - \frac{y}{2} \, \Delta A' \right) , \\
\vec{y}. \Delta \vec{m}_s' & = & + 2 \sigma_l \left( -\varphi_k' + (w/d)\, \frac{y}{2} \, \Delta A \right) , \\
\vec{z}. \Delta \vec{m}_s' & = & 0 .
\end{eqnarray}
Two new gauge-related functions appear, which are defined as:
\begin{eqnarray}
& &\!\!\!\!\!\!\!\!\!\!\!\!\!\! \!\!\!\!\!\!\!\!\!\!\!\!\!\!\!\!\!\!\!\!\!\!\!\! \!\!\!\!\!\!\!\!\!\!\!\!\!\!\!\!\!\!\!\!\!\!\!\!\varphi_k (z,t)   =  +\phi_m \Big( \! -\frac{g_x \, (\beta \,w)}{2}\tilde{f}_x (z,t)+\frac{\tilde{g}_x\, (\beta\,w)}{2}f_x (z,t)-\frac{g_y \,(\beta\,d)}{2}\tilde{f}_y (z,t)+\frac{\tilde{g}_y \, (\beta\,d)}{2}f_y (z,t) \, \Big) , \label{varphis}\\
& &\!\!\!\!\!\!\!\!\!\!\!\!\!\! \!\!\!\!\!\!\!\!\!\!\!\!\!\!\!\!\!\!\!\!\!\!\!\!\!\!\!\!\!\!\!\!\!\!\!\!\!\!\!\!\!\!\!\!\!\!\!\!  \varphi_k' (z,t)   =   +(w/d) \, \varphi_k (z,t). \label{varphisprime}
\end{eqnarray}
As such, $\varphi_k$ can be seen as the {\it generator} of the nonzero $\vec{K}_S$ components. It is {\it fixed} from the gauge, with the choice of $g_i,\tilde{g}_i$ (real) coefficients.

\subsection{"$s$-gauge" characteristics}

The "$s$-gauge" (which components are  given in
Tab. \ref{tab_3}) shares the same $H$, $\vec{P}$ and $\vec{J}=0$ as the previous one, Eqs. (\ref{HXY},\ref{PXY}); 
it also verifies $\vec{L}  =  0$ and $\vec{K}_L  =  0$.
The $\vec{K}_S$ pseudo-vector has the same mathematical writing, with Eqs. (\ref{Kzval},\ref{Kxval},\ref{Kyval}), but is obtained from the specific gauge parameters $g_i,\tilde{g}_i$ of the "$s$-gauge" (presumably {\it different} from those of the "$\varphi$-gauge"). 
And remarkably, 
 its {\it spin internal degree of freedom is different}. One obtains:
\begin{equation}
\vec{S}  =   +\hbar\, \frac{X_x Y_y-X_y Y_x }{2}\, \vec{z} ,
\end{equation}
which is nonzero. We recognize $S_z = \hbar \, (X_x Y_y-X_y Y_x)/2 = +\mbox{i} \hbar \,(b_x b_y^\dag - b_y b_x^\dag) $ 
(having introduced the Dirac operators), which is nothing but the textbook helicity constructed from two linearly polarized (along $x$ and $y$) photon modes \cite{spin,loudon}. \\

As compared to the "$\varphi$-gauge", the peculiarity of the "$s$-gauge" is to verify:
\begin{eqnarray}
 \Delta V & =& 0 , \\
 \Delta A & =& 0 , \\
& \mbox{and:} & \nonumber \\
 \Delta  V'  & =& 0, \\
 \Delta A' & =& 0,
\end{eqnarray}
all potential differences {\it are zero}. 
The transverse $\Delta \vec{m}_s, \Delta \vec{m}_s'$ components of the angular momentum density differences write the same as above in the previous Subsection, but the $\varphi_k, \varphi_k'$ functions are here obtained by means of the "$s$-gauge" $g_i , \tilde{g}_i$ coefficients.
Furthermore, the $\vec{z}$ components are now {\it nonzero}:
\begin{eqnarray}
\vec{z}. \Delta \vec{m}_s & = & +2 \sigma_t \left( + \frac{\varphi_s}{2} \right) , \\
\vec{z}. \Delta \vec{m}_s' & = &+2 \sigma_l \left(-\frac{\varphi_s'}{2}  \right), \\
& \mbox{with:} & \nonumber \\
\varphi_s  (z,t) & = & +\phi_m \, \tilde{f}_x (z,t)  , \label{varphisbis} \\
\varphi_s' (z,t) & = & +\phi_m' \, \tilde{f}_y (z,t) ,\label{varphisbisprime}
\end{eqnarray}
which is {\it the origin} of the helicity appearing in $\vec{S}$.
Note the resemblances between the $\varphi_s, \varphi_s'$ functions above 
and the $\varphi,\varphi'$ definitions, Eqs. (\ref{firstPHI},\ref{firstPHIprime}).
The $\varphi_s, \varphi_s'$ and $\varphi_k, \varphi_k'$ also share the same propagation equation as the generalized fluxes: they propagate at the speed of light $c$. 

\section{Generalized rotations}
\label{bosontransform}

Let us consider two {\it independent} mode operators $b_i, b_j$, corresponding to quadratures $X_i, Y_i$ and $X_j, Y_j$.
We want to analyze here how (and why) one can transform these two into {\it equivalent} primed quantities, by means of the most generic linear operation.

We must first define the meaning of independent, and equivalent $b_i, b_j$ operators: independent means commuting with each other, and equivalent means $b_i', b_j'$ having the same commutation properties as the original ones. 
From this statement, it becomes immediately clear that {\it if such transformations are allowed}, the only way to propagate the commutators of a pair of operators is if they are {\it equal} to the same constant:
\begin{equation}
\left[ b_i , b_i^\dag \right] = \left[ b_j , b_j^\dag \right] = B_c  \, .
\end{equation}
The operation that fulfills our requirements exists, and writes in matrix form:
\begin{equation}
\left( \begin{array}{c}
b_i' \\
b_j'
\end{array} \right) = \left( 
\begin{array}{c}
e^{\mbox{i} \,\delta_1} \cos{\Theta} \\
-e^{\mbox{i} \,\delta_2} \sin{\Theta}
\end{array} 
\begin{array}{c}
e^{\mbox{i} \,\delta_1} \sin{\Theta} \,  e^{-\mbox{i} \,\phi}  \\
e^{\mbox{i} \,\delta_2} \cos{\Theta} \,  e^{-\mbox{i} \,\phi}
\end{array} \right) \left( \begin{array}{c}
b_i \\
b_j
\end{array} \right)  .
\end{equation}
Falling into this class of transformations, we have the well-known one that enables to create left and right polarized photons from $x$, $y$ polarized ones \cite{loudon}.
A similar approach is used in Section \ref{eigenstates} of the manuscript. \\

But this generic operation {\it does not} lead to the definition of a commutative group.
This is however mandatory for our theory, which requires to combine symmetry operations represented by such matrices: it must be possible to apply one regardless of the others applied before, therefore they must commute. 
This imposes to restrain the transformations to a subclass of the type: 
\begin{equation}
\left( \begin{array}{c}
b_i' \\
b_j'
\end{array} \right) =e^{\mbox{i} \,\delta} \left( 
\begin{array}{c}
 \cos{\Theta} \\
-  \sin{\Theta}
\end{array} 
\begin{array}{c}
  \sin{\Theta}   \\
  \cos{\Theta}  
\end{array} \right) \left( \begin{array}{c}
b_i \\
 e^{-\mbox{i} \,\phi} \, b_j
\end{array} \right) , \label{genrot}
\end{equation}
which does generate a commutative group. 
We call these {\it generalized rotations}, since they involve a rotation angle $\Theta$ {\it and} a phase factor $\delta$. 
Note that we kept in Eq. (\ref{genrot}) the phase $\phi$ (representing a time shift between the two original modes),  but it appears {\it directly within} the definition of the mode operator $b_j$.
This is precisely what has been done with the "$\varphi$-gauge" definition in Eqs. (\ref{goodf1}-\ref{goodf4}): the delay between modes is in-built in the gauge representation (through $\Delta \theta$), since by no means can such a phase factor appear within a symmetry-related operation.
 In the following, we will simply consider $
 e^{-\mbox{i} \,\phi} \, b_j \rightarrow b_j  $.
 
The simplest such symmetry operation corresponds to $\Theta = 0$: the generalized rotation reduces to $e^{\mbox{i} \,\delta} \, \mbox{Id}$, 
and using Eqs. (\ref{bgenhat},\ref{bgenhat2}), one easily computes that:
\begin{eqnarray}
X_i' & = & \cos(\delta) X_i -\sin (\delta) Y_i, \\
Y_i' & = & \cos(\delta) Y_i +\sin (\delta) X_i, \\
X_j' & = & \cos(\delta) X_j -\sin (\delta) Y_j, \\
Y_j' & = & \cos(\delta) Y_j +\sin (\delta) X_j,
\end{eqnarray} 
which is nothing but an identical (counterclockwise) {\it rotation of each set of quadratures themselves}. 
If one applies the same phase factor $\delta$ transformation to all 4 modes introduced in Section \ref{model}, one easily shows from  
 Eqs. (\ref{goodf1}-\ref{goodf8}) that this global phase 
  corresponds to a change in the $t=0, z=0$ reference. 
  This change of origin vanishes from the equations if we use the primed operators instead of the non-primed, and is therefore irrelevant. \\
  
Consider now a true transverse rotation of $X_x, Y_x$ and $X_y, Y_y$ quadratures (applying equally well to the $s$ and $\varphi$ gauges; we omit the superscripts here). It is expressed as:
\begin{equation}
\left( \begin{array}{c}
X_x' \\
X_y'
\end{array} \right) = \left( 
\begin{array}{c}
 \cos{\Theta} \\
-  \sin{\Theta}
\end{array} 
\begin{array}{c}
  \sin{\Theta}   \\
  \cos{\Theta}  
\end{array} \right) \left( \begin{array}{c}
X_x \\
  X_y
\end{array} \right) ,  \label{realrot}
\end{equation}
for a clockwise $\Theta$, and likewise for the $Y_x, Y_y$ quadratures. This is a simple consequence of a transform of type Eq. (\ref{genrot}), with $\delta=0$ and $b_x, b_y$ operators.
These operations {\it are allowed} by construction, since they represent one of the symmetries of the Poincar\'e group. In itself, it answers the questioning appearing at the very beginning of this Appendix (on the relevance of mode transformations), which lead to the conclusion that all $b_i$ operators must share the same constant commutator. This fact is therefore guaranteed. \\

We shall now address the gauge transformations, the one involving the angle $\theta_z$, and then the one involving ${\cal G}, \delta$.
The first of them given by Eqs. (\ref{equaRotxThetaz}, \ref{equaRotyThetaz}), which applies equally well to $x$ and $y$ components because of space isotropy, does not contain any (global) phase factor $\delta$ (which would be irrelevant, as pointed out above). It thus corresponds to a simple spatial rotation, {\it but between quadratures of the two different gauges}. 
On the other hand, the gauge coefficients $g_i^s, \tilde{g}_i^s$ of 
Eqs. (\ref{gaugecoeff1}-\ref{gaugecoeff4}) are built from a generalized rotation which transforms the two sets of $X^v_i, Y^v_i$ quadratures into $X_z, Y_z$ and $X_t, Y_t$ (see \ref{PandT}). This one {\it does contain a relevant phase} $\delta$, because it corresponds to an extra time delay appearing between the real and virtual components (the former ones are not affected by this phase, which is thus not global). 

Finally, an important property of these transformations is that {\it gauge operations and spatial rotations do commute}, even though they do not combine the same sets of modes. As a result, when applying a rotation  to the {\it real space}, which therefore applies equally well to $s$ and $\varphi$ gauge components, 
it propagates to our final modes such that:
\begin{equation}
\left( \begin{array}{c}
X_z' \\
X_t'
\end{array} \right) = \left( 
\begin{array}{c}
 \cos{\Theta} \\
-  \sin{\Theta}
\end{array} 
\begin{array}{c}
  \sin{\Theta}   \\
  \cos{\Theta}  
\end{array} \right) \left( \begin{array}{c}
X_z \\
  X_t
\end{array} \right) ,  \label{realZrot} 
\end{equation}
and equivalently for $Y_z,Y_t$, which just means that {\it virtual modes turn the same way as real ones}, Eq. (\ref{realrot}).
This might seem counter intuitive, but it actually simply reflects the way virtual modes are constructed in the present theory: $z$ and $t$ modes are {\it not} of a different nature than the $x$ and $y$ ones, as opposed to the conventional QFT theories.
 This result will be used in \ref{SpinApp},  \ref{PiApp} and \ref{CApp} when discussing the properties of $\vec{S}$ , $\vec{\Pi}$ and $\vec{C}$ operators respectively.

\section{Gauge coefficients and discrete symmetries: parity $\cal P$ and time reversal $\cal T$}
\label{PandT}

The "$s$-gauge" coefficients $g_i^s, \tilde{g}_i^s$ must be chosen, {\it if possible}, such that from Eqs. (\ref{Xzeff},\ref{Yzeff}) proper $X_z, Y_z$ quadratures are produced. Introducing $b_z, b_z^\dag$ operators, this  
requires that:
\begin{equation}
\!\!\!\!\!\!\!\!\!\!\!\!\!\!\!\!\!\!\!\!\!\!\!\!\!\!\!\!\!\!\!\!\! b_z = b_x^v \left( -\frac{1}{2} \frac{\beta \, w}{\Delta_H} ( \tilde{g}_x^s + \mbox{i}\, g_x^s) \tan(\Delta \theta) e^{\mbox{i} \,\Delta \theta} \right) + b_y^v \left( -\frac{1}{2} \frac{\beta \, d}{\Delta_H}( \tilde{g}_y^s + \mbox{i} \,g_y^s) \tan(\Delta \theta) e^{\mbox{i} \,\Delta \theta} \right)\! ,  
\end{equation}
or equivalently with the complex conjugate expression (the gauge $g_i^s, \tilde{g}_i^s$ being all {\it reals}).
This is effectively possible only if the quantities in parenthesis correspond to a generalized rotation (see \ref{bosontransform} above) of the sort (in matrixform):
\begin{equation}
e^{\mbox{i} \,(\delta+\Delta \theta)} \left( 
\begin{array}{c}
 \cos{{\cal G}} \\
-  \sin{{\cal G}}
\end{array} 
\begin{array}{c}
  \sin{{\cal G}}   \\
  \cos{{\cal G}}  
\end{array} \right) , 
\end{equation}
which leads to the gauge coefficients given in Eqs. (\ref{gaugecoeff1}-\ref{gaugecoeff4}). This immediately implies for the associated $t$ operator:
\begin{equation}
b_t = - b_x^v \, \left( e^{\mbox{i} \,(\delta+\Delta \theta)} \sin{{\cal G}} \right)  +  b_y^v \, \left( e^{\mbox{i} \,(\delta+\Delta \theta)} \cos{{\cal G}} \right) ,
\end{equation}
together with its complex conjugate. ${\cal G}$ is the "$s$-gauge" rotation angle, and $\delta$ its phase factor. \\

But the gauge coefficients are not only intimately linked to a generalized rotation: they are also impacted by the four space-time {\it discrete symmetries}, which effect on the physical properties (namely $H, \vec{P}, \vec{J}, \vec{L}$ for the external and $\vec{\Pi}, \vec{C}, \vec{S}$ for internal ones)  must be considered.
The transverse mirror symmetries ($x \rightarrow -x$ and $y \rightarrow -y$) can be shown to leave all these parameters unaltered: these are true symmetries of the problem at hand. We are left with parity $\cal P$ and time reversal $\cal T$ which must be properly analyzed.  \\

Parity $\cal P$ consists in the symmetry $z \rightarrow - z$, which is equivalent to performing:
\begin{equation}
\beta \rightarrow - \beta ,
\end{equation}
in all Eqs. (\ref{goodf1}-\ref{goodf8}). As a result, only one parameter is modified such that:
\begin{equation}
\vec{P} \rightarrow - \vec{P} ,
\end{equation}
all others being unaffected, which is quite expected. 
However, it is worth mentioning that the sign of the gauge coefficients $g_i^s, \tilde{g}_i^s$ are {\it also reversed}, since they depend on $1/\beta$. All gauge angles ($\theta_z$, $\cal G$, $\delta$ and obviously $\Delta \theta$) must {\it not} be modified.

Consider now time reversal $\cal T$: $t \rightarrow - t$.
In order to know its impact on the real space $\vec{E}, \vec{B}$ fields,  one should analyze how this operation affects the $f^j_i$ functions of Eqs. (\ref{goodf1}-\ref{goodf8}). 
It is then  straightforward to realize that time reversal is equivalent to operating {\it simultaneously}:
\begin{eqnarray}
\beta & \rightarrow & - \beta , \\
\Delta \theta & \rightarrow & - \Delta \theta , \\
Y_i^j & \rightarrow & - Y_i^j \,\,\,\,\,\,\,\,\,\, \mbox{with $i=x,y$ and $j=\varphi,s$}\, .
\end{eqnarray}
The first line leads to $\vec{P} \rightarrow - \vec{P}$, as in the previous paragraph. 
The polarity flip of $\Delta \theta$ is actually compatible with $\theta_z$ unchanged {\it or} $\theta_z \rightarrow \theta_z + \pi$, see \ref{phidual}. 
The last line actually implies that {\it all} $Y_i$ quadratures must flip sign (since this change propagates through all gauge transforms).
As for parity, the gauge must certainly be affected in a way or another. 
In order to define how, we should analyze how $\Delta E$, $\vec{S}$, $\vec{\Pi}$ and $\vec{C}$ are modified by this symmetry. Our common requirements are that  
$\Delta E$ must be immune to the $t$ sign change, while the others must reverse sign, see \ref{SpinApp}, \ref{PiApp} and \ref{CApp} respectively.
This is consistent with $\alpha \rightarrow - \alpha$, which from Eqs. (\ref{cosalpha},\ref{sinalpha}) arises from:
\begin{equation}
\delta \rightarrow - \delta   ,
\end{equation}
while $\cal G$ must remain unchanged. This transformation of $\delta$ is compatible with the set of solutions found in \ref{phidual}, which makes it lawful. 
As a result $g_x^s, g_y^s$ flip signs, while $\tilde{g}_x^s , \tilde{g}_y^s$ are unaltered.
Note that while $E_m>0$ is always preserved by construction, symmetries {\it can flip signs} of  the actual generalized fluxes amplitudes describing our 4 bosons, see discussion in \ref{phidual}. The last, purely internal, discrete symmetry (conjugation $\cal C$) is also discussed therein.

\section{Properties of the $\vec{S}$ operator}
\label{SpinApp}

Let us define the following operators, which introduce Planck's constant $\hbar$ explicitly:
\begin{eqnarray}
& & S_x =\hbar \, \frac{X_y Y_z - X_z Y_y}{2} , \,\,\,\,\,\,\,\,\,\,\,\,  S^t_x =\hbar \, \frac{X_y Y_t - X_t Y_y}{2} , \label{spinx} \\
& & S_y =\hbar \, \frac{X_z Y_x - X_x Y_z}{2} , \,\,\,\,\,\,\,\, \,\,\,\, S^t_y =\hbar \, \frac{X_t Y_x - X_x Y_t}{2} , \\
& & S_z = S^t_z = \hbar \, \frac{X_x Y_y - X_y Y_x}{2} .\label{spinz}
\end{eqnarray}
Remember that $z$ and $t$ modes are created from the same gauge transform, and are thus of same nature (as opposed to conventional QFT approaches, from which we only borrow the names). The above operators with and without the superscript represent therefore equivalent formulations. These are the components of two pseudo-vectors. We also define:
\begin{eqnarray}
S^2 & = & S_x^2+S_y^2 + S_z^2 , \\
(S^t)^2 & = & (S^t_x)^2+(S^t_y)^2 + (S^t_z)^2 , 
\end{eqnarray}
which are nothing but their norms squared. It turns out that these verify:
\begin{eqnarray}
\left[ S_x, S_y\right] &=& \mbox{i}\hbar\, S_z , \\
\left[ S_y, S_z\right] &=& \mbox{i}\hbar\, S_x , \\
\left[ S_z, S_x\right] &=& \mbox{i}\hbar\, S_y , \\
\left[ S_i, S^2\right] &=& 0 \,\, \mbox{for $i=x,y,z$} \,,
\end{eqnarray}
and similarly with the $t$ superscript, if and only if:
\begin{equation}
\left[ X_j, Y_j \right] =  2\,\mbox{i}  \,\, \mbox{for $j=x,y,z,t$} ,
\end{equation}
which from Eq. (\ref{commute}) leads to $B_c=1$. These properties are those of a {\it spin}, and are fulfilled if the corresponding $b_i,b_i^\dag$ operators are those of {\it bosons} (they are Dirac annihilation/creation operators).
Note that the $t$ and non-$t$ spin components do not all commute with each other, for instance $[S_x,S_x^t]=\mbox{i}\hbar\, S_{PT}$ with $S_{PT}$ defined in the following \ref{PiApp} as $\hbar (X_z Y_t - X_t Y_z)/2$. \\

Consider now the components without superscript, and let us operate a transverse rotation (around $\vec{z}$) of arbitrary clockwise angle $\Theta$. From \ref{bosontransform}, the same rotation applies to $x,y$ and $z,t$ quadratures, Eqs. (\ref{realrot},\ref{realZrot}), and leads to the new pseudo-vector $\vec{S}^{\,\Theta}$:
\begin{eqnarray}
&& \!\!\!\!\!\!\!\!\!\!\!\!\!\!\!\!\!\!\!\!\!\!\!\!\!\!\!\!\!\!\!\! \vec{x}.\vec{S}^{\,\Theta} = S_x^{\,\Theta}   =   +S_x \cos(\Theta)^2 - S_y \cos(\Theta)\sin(\Theta)- S^t_x \cos(\Theta)\sin(\Theta) + S^t_y \sin(\Theta)^2,\\
&& \!\!\!\!\!\!\!\!\!\!\!\!\!\!\!\!\!\!\!\!\!\!\!\!\!\!\!\!\!\!\!\! \vec{y}.\vec{S}^{\,\Theta} = S_y^{\,\Theta}   =   +S_x \cos(\Theta) \sin(\Theta)  + S_y \cos(\Theta)^2 - S^t_x \sin(\Theta)^2 - S^t_y \cos(\Theta) \sin(\Theta)  ,\\
&& \!\!\!\!\!\!\!\!\!\!\!\!\!\!\!\!\!\!\!\!\!\!\!\!\!\!\!\!\!\!\!\! \vec{z}.\vec{S}^{\,\Theta} = S_z^{\,\Theta}    =   S_z , 
\end{eqnarray}
with norm $S^{\,\Theta}$ such that:
\begin{equation}
(S^{\,\Theta} )^2= (S_x^{\,\Theta})^2 + (S_y^{\,\Theta})^2 + (S_z^{\,\Theta})^2 .
\end{equation}
$\vec{S}^{\,\Theta}$ fulfills the spin rules, as the original components do. Furthermore, we have:
\begin{eqnarray}
S_x^{\,\Theta\,=0}  & =  & S_x^{\,\Theta\,=\pi} = S_x , \\
S_y^{\,\Theta\,=0}  & =  & S_y^{\,\Theta\,=\pi} = S_y ,
\end{eqnarray}
and: 
\begin{eqnarray}
S_x^{\,\Theta\,=\pm \pi/2}  & =  & + S^t_y , \\
S_y^{\,\Theta\,=\pm \pi/2}  & =  & - S^t_x ,
\end{eqnarray}
which are the specific rotation symmetry properties of our generic spin operator.
From Eqs. (\ref{spin1}-\ref{spin3}), we see that imposing:
\begin{eqnarray}
\vec{x}. \vec{S} & = & S_x , \\
\vec{y}. \vec{S} & = & S_y , \\
\vec{z}. \vec{S} & = & S_z , 
\end{eqnarray}
requires 
$2 \, \epsilon_0 \,\frac{ E_m^2}{\omega} \, (d w L) \, \Delta_H    =   \hbar$, Eq. (\ref{hbarlaw}), which fixes $E_m$ from $\hbar$. The gauge choice performed in Section \ref{model} is such that the $t$ component of the spin {\it does not appear}, that is $\Theta=0,\pi$ is taken by construction. A similar property will emerge in \ref{PiApp} for the $\vec{\Pi}$ operator.  

Let us analyze as well the effect of discrete symmetries on the spin. From the definitions of \ref{PandT}, parity $\cal P$ leaves all spin components unaffected. On the other hand, time reversal $\cal T$ flips the sign of all $Y_i$ components, which leads to:
\begin{eqnarray}
\vec{S} \rightarrow - \vec{S} ,
\end{eqnarray}
the law applying equally well to all spin operators defined in this Appendix (regardless of the superscript). Finally, the conjugation $\cal C$ also flips the sign of $\vec{S}$, see \ref{phidual}. \\

The $z$ component of the spin $S_z$ is not affected by rotations: it is an invariant of the Poincar\'e group. Computing the commutator with the Hamiltonian, we obtain:
\begin{eqnarray}
&&\left[ S_z, H \right]   =   \left[ S_z, \Delta E \right] = \nonumber \\
&  &\!\!\!\!\!\!\!\!\!\!\!\!\!\!\!  +\mbox{i} \, \hbar^2 \omega \, \Delta_\pi \, \left[ -\sqrt{2} \sin({\cal G}) \cos(\alpha) \left( +\frac{X_x X_z+ Y_x Y_z }{2}+\frac{X_y X_t+ Y_y Y_t }{2} \right) \right. \nonumber \\
&&\!\!\!\!\!\!\!\!\!\!\!\!\!\!\!  \left. - \sqrt{2} \cos({\cal G}) \cos(\alpha) \left( -\frac{X_y X_z+ Y_y Y_z }{2}+\frac{X_x X_t+ Y_x Y_t }{2} \right)   \right. \nonumber \\
&&\!\!\!\!\!\!\!\!\!\!\!\!\!\!\!  \left. + \sqrt{2} \sin({\cal G}) \sin(\alpha) \left( +\frac{X_z Y_x- X_x Y_z }{2}-\frac{X_y Y_t-  X_t Y_y }{2} \right)  \right. \nonumber \\
&&\!\!\!\!\!\!\!\!\!\!\!\!\!\!\!  \left. + \sqrt{2} \cos({\cal G}) \sin(\alpha) \left( +\frac{X_y Y_z- X_z Y_y }{2}+\frac{X_t Y_x -  X_x Y_t }{2} \right) \right] . \label{SzHcom} 
\end{eqnarray}
As such, $S_z$ {\it is not} a constant of motion in itself (the r.h.s. above is not zero). But combining Eq. (\ref{SzHcom}) with other similar expressions (linked to $\vec{\Pi}$ and $\vec{C}$, see following Appendices),  proper {\it internal} constants of motion are presented in Section \ref{model}.\\

We finally comment on the equivalent writing of the spin components, Eqs. (\ref{spinx}-\ref{spinz}), which involves the $b_i,b_i^\dag$ operators:
\begin{eqnarray}
S_x & =& \mbox{i} \hbar \left(b_y b_z^\dag - b_z b_y^\dag \right) ,\\
S_y & =& \mbox{i} \hbar \left(b_z b_x^\dag - b_x b_z^\dag \right) ,\\
S_z & =& \mbox{i} \hbar \left(b_x b_y^\dag - b_y b_x^\dag \right) , 
\end{eqnarray}
and likewise with the $t$ spin superscript and $b_t,b_t^\dag$ instead of $b_z,b_z^\dag$. 
While the pseudo-vector $\vec{S}$ is a spin (meaning it fulfills the commutation rules presented above), this construction from 3 bosons leads to specific properties that we want to point out here. We focus on the "conventional" spin operator (the one without $t$ superscript), and thus on the $x$, $y$ and $z$ photons only.
 
The Dirac operators are directly used in Section \ref{eigenstates}, when dealing with the eigenstates of the light field. These are constructed from {\it hybrid states} obtained in particular from the $x,y$ components, using Eqs. (\ref{BHPlus}-\ref{BHMoins}).  The $b_{H,+}, b_{H,+}^\dag$ and $b_{H,-}, b_{H,-}^\dag$ cration/annihilation operators enable to define eigenstates of the $S_z$ operator, see Eq. (\ref{Szdef}). 
Eqs. (\ref{popHP},\ref{popHM}) introduce the  {\it populations} $n_{H,+}=< b_{H,+}^\dag b_{H,+} >$ and $n_{H,-}=< b_{H,-}^\dag b_{H,-} >$ corresponding to them. Similarly, we define here:
\begin{equation}
n_z = < b_{z}^\dag b_{z}> ,
\end{equation}
the population of the $z$ photon. The norm squared $S^2$ can be recast in:
\begin{eqnarray}
S^2 & = & \left( b_{H,+}^\dag b_{H,+} - b_{H,-}^\dag b_{H,-}\right)^2  + \left( b_{H,+}^\dag b_{H,+} \right) + \left( b_{H,-}^\dag b_{H,-} \right) + 2 \, \left( b_{z}^\dag b_{z} \right)\nonumber \\
& &  + 2 \, \left( b_{H,+}^\dag b_{H,+} \right) \, \left( b_{z}^\dag b_{z} \right) + 2 \, \left( b_{H,-}^\dag b_{H,-} \right) \, \left( b_{z}^\dag b_{z} \right) \nonumber \\
&& + 4 \mbox {i} \, \left( b_{H,+}^\dag b_{H,-}^\dag b_z^2 -  b_{H,+}  b_{H,-}  (b_z^\dag)^2 \right) .
\end{eqnarray}
We consider in the following the states of the form:
\begin{equation}
| n_{H,+} , n_{H,-} , n_z >,
\end{equation}
with $|0,0,0>$ the vacuum state. Especially, the {\it single photon} states read:
\begin{eqnarray}
|\,+> & = & | n_{H,+}=1 , n_{H,-}=0 , n_z=0 >, \\
|\,-> & = & | n_{H,+}=0 , n_{H,-}=1 , n_z=0 >, \\
|\,\,0> & = & | n_{H,+}=0 , n_{H,-}=0 , n_z=1 > .
\end{eqnarray} 
These verify:
\begin{eqnarray}
S_x |\,\pm > & = & |0,0,0> , \\
S_y |\,\pm > & = & |0,0,0> , \\
S_x |\,\,0 > & = & |0,0,0> , \\
S_y |\,\,0 > & = & |0,0,0> ,
\end{eqnarray}
and:
\begin{eqnarray}
S_z |\,\pm > & = & \pm \hbar \, |\,\pm > , \\
S_z |\,\,0 > & = & |0,0,0> , \\
S^2 |\,\pm > & = & 2 \hbar^2 \,|\,\pm > , \\
S^2 |\,\,0 > & = & 2 \hbar^2 \,|\,\,0 > .
\end{eqnarray}
We recover {\it precisely all properties of a quantum spin 1}, having identified its 3 possible eigenstates; as one would have expected.
We can also look at other relevant states for experiments:
\begin{eqnarray}
| \, n+ > & = & | n_{H,+} \gg ( n_{H,-}, n_{z}, 1 )\, , n_{H,-}  , n_z  >, \\
| \, n- > & = & | n_{H,+} \, , n_{H,-} \gg (n_{H,+}, n_{z}, 1 ) \, , n_z  >,
\end{eqnarray}
which typically represent "left" and "right" circularly polarized waves in our common language. We do not discuss the $n_z \gg 1$ case which is unphysical (since dealing with a virtual photon). These states verify:
\begin{eqnarray}
S_x |\,n\pm > & = & |0,0,0> , \\
S_y |\,n\pm > & = & |0,0,0> , 
\end{eqnarray}
and:
\begin{eqnarray}
S_z |\,n+ > & \approx & + \hbar \, n_{H,+} \, |\,n + > , \\
S_z |\,n- > & \approx & - \hbar \, n_{H,-} \, |\,n - > , \\
S^2 |\,n\pm > & \approx & \hbar^2 \, n_{H,\pm}^2 \, |\,n\pm > .
\end{eqnarray}
In these cases, the spin  looks like a {\it classical 
vector} aligned along $\vec{z}$, of norm $\hbar \, n_{H,\pm}$.

\section{Properties of the $\vec{\Pi}$ operator}
\label{PiApp}

We will follow in this Appendix a procedure similar to the one adopted for the spin in \ref{SpinApp}. We introduce:
\begin{eqnarray}
\Pi_x & = & \hbar\, \sqrt{\frac{5}{2}} \, \left[ +2\sqrt{2} \sin({\cal G}) \sin(\alpha) \frac{X_z^2+ Y_z^2 }{4} \right. \nonumber \\
&&\!\!\!\!\!\!\!\!\!\!\!\!\!\!\!  \left. + \sqrt{2} \cos({\cal G}) \sin(\alpha) \frac{X_z X_t+ Y_z Y_t }{2}- \sqrt{2} \cos({\cal G}) \cos(\alpha) \frac{X_z Y_t-X_t Y_z  }{2} \right] \!, \\
 \Pi_y & = & \hbar\, \sqrt{\frac{5}{2}}  \, \left[ -2\sqrt{2} \cos({\cal G}) \sin(\alpha) \frac{X_z^2+ Y_z^2 }{4} \right. \nonumber \\
&&\!\!\!\!\!\!\!\!\!\!\!\!\!\!\!  \left. + \sqrt{2} \sin({\cal G}) \sin(\alpha) \frac{X_z X_t+ Y_z Y_t }{2}- \sqrt{2} \sin({\cal G}) \cos(\alpha) \frac{X_z Y_t-X_t  Y_z }{2} \right] \!, \\
\Pi_z & = & S_{PT} ,  
\end{eqnarray}
as well as:
\begin{eqnarray}
 \Pi^t_x & = & \hbar\, \sqrt{\frac{5}{2}} \, \left[ -2\sqrt{2} \sin({\cal G}) \sin(\alpha) \frac{X_t^2+ Y_t^2 }{4} \right. \nonumber \\
&&\!\!\!\!\!\!\!\!\!\!\!\!\!\!\!  \left. + \sqrt{2} \cos({\cal G}) \sin(\alpha) \frac{X_z X_t+ Y_z Y_t }{2}+ \sqrt{2} \cos({\cal G}) \cos(\alpha) \frac{X_z Y_t-X_t Y_z  }{2} \right] \!, \\
 \Pi^t_y & = & \hbar\, \sqrt{\frac{5}{2}}  \, \left[ +2\sqrt{2} \cos({\cal G}) \sin(\alpha) \frac{X_t^2+ Y_t^2 }{4} \right. \nonumber \\
&&\!\!\!\!\!\!\!\!\!\!\!\!\!\!\!  \left. + \sqrt{2} \sin({\cal G}) \sin(\alpha) \frac{X_z X_t+ Y_z Y_t }{2}+ \sqrt{2} \sin({\cal G}) \cos(\alpha) \frac{X_z Y_t-X_t  Y_z }{2} \right] \!, \\
\Pi^t_z & = & S_{PT} .  
\end{eqnarray}
The two $z$ components are identical, and we specifically define:
\begin{equation}
S_{PT} = \hbar \, \frac{X_z Y_t - X_t Y_z}{2} ,
\end{equation}
which plays an important role in Section \ref{eigenstates} of the manuscript (justifying the name in subscript "PT": it refers to parity and time reversal).
The commutation rules of these operators are quite peculiar. They verify:
\begin{eqnarray}
\left[ \Pi_x+\Pi_x^t, \Pi_y+\Pi_y^t \right]&=& 4 \mbox{i} \hbar \left( \frac{5}{2} \sin(\alpha)^2 \right) \left(\Pi_z+\Pi_z^t  \right) , \label{Pilaw1} \\
\left[ \Pi_y+\Pi_y^t, \Pi_z+\Pi_z^t \right]&=& 4 \mbox{i} \hbar  \left(\Pi_x+\Pi_x^t  \right) , \\
\left[ \Pi_z+\Pi_z^t, \Pi_x+\Pi_x^t \right]&=& 4 \mbox{i} \hbar  \left(\Pi_y+\Pi_y^t  \right) . \label{Pilaw3}
\end{eqnarray}
These resemble, but {\it are not} the commutation rules of a spin. In order to have equivalent laws for each axis, we must impose:
\begin{equation}
\sin(\alpha)^2   = \frac{2}{5}  ,
\end{equation}
which preserves then isotropy of space. The norm squared of these  pseudo-vectors' sum   writes:
\begin{equation}
 (\Pi+\Pi^t)^2   =   (\Pi_x+\Pi^t_x)^2 + (\Pi_y+\Pi^t_y)^2 + (\Pi_z+\Pi^t_z)^2 , 
\end{equation}
and we have:
\begin{equation}
\left[ \Pi_i+\Pi_i^t,(\Pi+\Pi^t)^2  \right]   =  0  \,\, \mbox{for $i=x,y,z$} \, ,
\end{equation}
which completes Eqs. (\ref{Pilaw1}-\ref{Pilaw3}). \\

We now analyze how the pseudo-vector without superscript transforms under a rotation around $\vec{z}$, of (clockwise) angle $\Theta$. As for the spin, using the quadratures's rotation properties of \ref{bosontransform} we obtain $\vec{\Pi}^{\,\Theta}$:
\begin{eqnarray}
&& \!\!\!\!\!\!\!\!\!\!\!\!\!\!\!\!\!\!\!\!\!\!\!\!\!\! \Pi^{\,\Theta}_x   =   +\Pi_x \cos(\Theta)^2 - \Pi_y \cos(\Theta)\sin(\Theta)-\Pi^t_x \sin(\Theta)^2 -\Pi^t_y \cos(\Theta)\sin(\Theta) , \\
&& \!\!\!\!\!\!\!\!\!\!\!\!\!\!\!\!\!\!\!\!\!\!\!\!\!\! \Pi^{\,\Theta}_y   = +\Pi_x \cos(\Theta)\sin(\Theta) + \Pi_y \cos(\Theta)^2 +\Pi^t_x \cos(\Theta)\sin(\Theta)  -\Pi^t_y  \sin(\Theta)^2 , \\
&& \!\!\!\!\!\!\!\!\!\!\!\!\!\!\!\!\!\!\!\!\!\!\!\!\!\! \Pi^{\,\Theta}_z   =   \Pi_z  .
\end{eqnarray}
The rotation symmetries of this pseudo-vector are:
\begin{eqnarray}
\Pi_x^{\,\Theta\,=0}  & =  & \Pi_x^{\,\Theta\,=\pi} = \Pi_x , \\
\Pi_y^{\,\Theta\,=0}  & =  & \Pi_y^{\,\Theta\,=\pi} = \Pi_y ,
\end{eqnarray}
as for the spin, and: 
\begin{eqnarray}
\Pi_x^{\,\Theta\,=\pm \pi/2}  & =  & - \Pi^t_x , \\
\Pi_y^{\,\Theta\,=\pm \pi/2}  & =  & - \Pi^t_y ,
\end{eqnarray}
which are specific to the parity-related operator. 
Since Eq. (\ref{hbarlaw}) is guaranteed by the spin properties (\ref{SpinApp}), and Eq. (\ref{deltapis}) by the generalized fluxes (\ref{phidual}), the $\vec{\Pi}$ components of Eqs. (\ref{Pixcomp}-\ref{Pizcomp}) write:
\begin{eqnarray}
\vec{x}. \vec{\Pi} & = & \Pi_x , \\
\vec{y}. \vec{\Pi} & = & \Pi_y , \\
\vec{z}. \vec{\Pi} & = & \Pi_z + \Pi^t_z .  
\end{eqnarray}
These relationships are fixed by the gauge choice, but note the difference with the spin: the $z$ component involves actually both $t$-superscripted and non-superscripted terms, leading to $\vec{z}. \vec{\Pi} = 2 S_{PT}$.  

The discrete symmetries act on the $\vec{\Pi}$ pseudo-vector as they do on the spin $\vec{S}$, for the same reasons. Parity $\cal P$ leaves all components unchanged, while time reversal $\cal T$ (which produces a $Y_i \rightarrow - Y_i$ flip and $\alpha \rightarrow - \alpha$, see \ref{PandT}) leads to: 
\begin{eqnarray}
\vec{\Pi} \rightarrow - \vec{\Pi} ,
\end{eqnarray}
the sign of all components is reversed (the law applies equally well for all superscripts). The conjugation $\cal C$ operation has the same effect $\vec{\Pi} \rightarrow - \vec{\Pi} $, see \ref{phidual}.\\ 

Like $S_z$, the operator $S_{PT}$ is unaffected by the symmetries of the Poincar\'e group: it is an invariant. Its commutator with the Hamiltonian $H$ is calculated as:
\begin{eqnarray}
&&\left[ S_{PT}, H \right]   =   \left[ S_{PT}, \Delta E \right] = \nonumber \\
&  &\!\!\!\!\!\!\!\!\!\!\!\!\!\!\!  -\mbox{i} \, \hbar^2 \omega \, \Delta_\pi \, \left[ -\sqrt{2} \sin({\cal G}) \cos(\alpha) \left( +\frac{X_x X_z+ Y_x Y_z }{2}+\frac{X_y X_t+ Y_y Y_t }{2} \right) \right. \nonumber \\
&&\!\!\!\!\!\!\!\!\!\!\!\!\!\!\!  \left. - \sqrt{2} \cos({\cal G}) \cos(\alpha) \left( -\frac{X_y X_z+ Y_y Y_z }{2}+\frac{X_x X_t+ Y_x Y_t }{2} \right)   \right. \nonumber \\
&&\!\!\!\!\!\!\!\!\!\!\!\!\!\!\!  \left. + \sqrt{2} \sin({\cal G}) \sin(\alpha) \left( +\frac{X_z Y_x- X_x Y_z }{2}-\frac{X_y Y_t-  X_t Y_y }{2} \right)  \right. \nonumber \\
&&\!\!\!\!\!\!\!\!\!\!\!\!\!\!\!  \left. + \sqrt{2} \cos({\cal G}) \sin(\alpha) \left( +\frac{X_y Y_z- X_z Y_y }{2}+\frac{X_t Y_x -  X_x Y_t }{2} \right) \right] , \label{SPTHcom} 
\end{eqnarray}
which is {\it not } zero ($S_{PT}$ is not a constant of motion), but is actually {\it exactly opposite} to the one of  $S_z$. This leads to the definition of our first internal constant of motion in Section \ref{model}: $S_z+S_{PT}$, the {\it sum} of the helicity and parity operators, the two properties at the core of Section \ref{eigenstates}.

\section{Properties of the $\vec{C}$ operator}
\label{CApp}

The charge pseudo-vector has been defined in Eqs. (\ref{chargeQ1}-\ref{chargeQ3}), introducing two constants $\Delta_c, \Delta_r$ arising from the modeling: $\vec{C}=-(\Delta_c C_c+\Delta_r C_r)\,\vec{z}$.
 It is thus split in two components, $C_c$ and $C_r$ which are both carried by the $\vec{z}$ direction. 
More profoundly, one can verify that both are invariants of the Poincar\'e group (especially they are not affected by transverse rotations).

The discrete symmetries described in \ref{PandT} act on these two as they do on all components of the $\vec{\Pi}$ and $\vec{S}$ pseudo-vectors. $C_c$ and $C_r$ are unaffected by parity $\cal P$, while time reversal $\cal T$ leads to:
\begin{eqnarray}
C_c & \rightarrow & - C_c , \\
C_r & \rightarrow & - C_r ,
\end{eqnarray}
since it flips signs for $Y_i$, $\alpha$ and $\Delta \theta$. The conjugation $\cal C$ flips signs as well (\ref{phidual}).
Furthermore, it is easy to show that the transformation $\delta \rightarrow \delta+ \mbox{sign}(\tan[\Delta \theta])\, \pi/2$ links them together:
\begin{equation}
C_r \rightarrow C_c ,
\end{equation}
and reversely with $\delta \rightarrow \delta - \mbox{sign}(\tan[\Delta \theta])\, \pi/2$. This {\it is not} an allowed gauge transformation, since it does not connect to each other sets of gauge parameters that represent a photon state  (see \ref{phidual}). What this transformation means is simply that, if the value of $<C_c>$ is given, automatically $<C_r>$ is fixed: these {\it are not} two independent variables, and only one of them is necessary to represent the charge. We chose $C_c$, for reasons that become clear below. \\

Computing the commutators with $H$ of these two operators, we obtain:
\begin{equation}
\left[ C_{c}, H \right]   =  + \left[ S_{z},H \right] = - \left[ S_{PT},H \right] , \label{CHcom1}
\end{equation}
for the first one, and:
\begin{eqnarray}
&&\!\!\!\!\!\!\!\!\!\!\!\!\!\!\!\!\!\!\!\!\!\!\!\!\!\!\!\!\!\! \left[ C_{r}, H \right]   =      -\mbox{sign}(\tan[\Delta \theta])\, \mbox{i} \, \hbar^2 \omega \, \Delta_\pi \, \left[ \sqrt{2} \sin({\cal G}) \sin(\alpha) \left( +\frac{X_x X_z+ Y_x Y_z }{2}+\frac{X_y X_t+ Y_y Y_t }{2} \right) \right. \nonumber \\
&&\!\!\!\!\!\!\!\!\!\!\!\!\!\!\!  \left. +\sqrt{2} \cos({\cal G}) \sin(\alpha) \left( -\frac{X_y X_z+ Y_y Y_z }{2}+\frac{X_x X_t+ Y_x Y_t }{2} \right)   \right. \nonumber \\
&&\!\!\!\!\!\!\!\!\!\!\!\!\!\!\!  \left. + \sqrt{2} \sin({\cal G}) \cos(\alpha) \left( +\frac{X_z Y_x- X_x Y_z }{2}-\frac{X_y Y_t-  X_t Y_y }{2} \right)  \right. \nonumber \\
&&\!\!\!\!\!\!\!\!\!\!\!\!\!\!\!  \left. + \sqrt{2} \cos({\cal G}) \cos(\alpha) \left( +\frac{X_y Y_z- X_z Y_y }{2}+\frac{X_t Y_x -  X_x Y_t }{2} \right) \right. \nonumber \\
&&\!\!\!\!\!\!\!\!\!\!\!\!\!\!\!  \left. +  \sqrt{\frac{5}{2}} \left( S_{PT}-S_z \right) \right] , \label{CHcom2} 
\end{eqnarray}
for the second.
Both {\it are not} constants of motion, since the r.h.s. of Eqs. (\ref{CHcom1},\ref{CHcom2}) are nonzero. 
However, Eq. (\ref{CHcom1}) demonstrates that $C_c+S_{PT}$ is one of them, related to internal symmetries.
The simplicity of this result justifies why we take $C_c$ as "electric charge". 
In the end, among the 10 constants of motions imposed by the Poincar\'e group, only one is missing, which should involve $C_r$ to some extent. 
We actually do not need to explicit this last relationship, since it is redundant; in Section \ref{eigenstates}, only the expressions composed with $S_z$, $S_{PT}$ and $C_c$ will be necessary to construct the eigenstates of light.

\section{Virtual charges, generalized fluxes, and angular momentum density in the dual gauge picture: polarities and conjugation symmetry $\cal C$. }
\label{phidual}

It is the peculiar properties of the "$\varphi$-gauge" and "$s$-gauge" that underlie the {\it gauge fixing} realized within the dual gauge postulate. We must thus explicit the new rules that replace here the ones presented in
\ref{propG}. To this end, we define the generalized fluxes corresponding to each photon family:
\begin{eqnarray}
& &\!\!\!\!\!\!\!\!\!\!\!\!\!\!\!\!\!\!\!\!\!\!\!\!\!\!\!\!\!\! \varphi_x' (z,t)    =   \left(\frac{E_m\,w}{\omega}\right) \Delta_{H\,ampl} \, \left[X_x \sin(\omega t - \beta z + \Delta \theta )- Y_x \cos(\omega t - \beta z + \Delta \theta ) \right] , \label{fluxdef1} \\
&&\!\!\!\!\!\!\!\!\!\!\!\!\!\!\!\!\!\!\!\!\!\!\!\!\!\!\!\!\!\! \varphi_y (z,t)    =  \left(\frac{E_m\,d}{\omega}\right) \Delta_{H\,ampl} \, \left[X_y \sin(\omega t - \beta z + \Delta \theta )- Y_y \cos(\omega t - \beta z + \Delta \theta ) \right] , \\
&&\!\!\!\!\!\!\!\!\!\!\!\!\!\!\!\!\!\!\!\!\!\!\!\!\!\!\!\!\!\! \varphi_z (z,t)    =  \left(\frac{E_m\,d}{\omega}\right) \Delta_{v\,ampl\,s} \, \left[X_z \cos(\omega t - \beta z + \delta  )+ Y_z \sin(\omega t - \beta z + \delta  ) \right] , \\
&&\!\!\!\!\!\!\!\!\!\!\!\!\!\!\!\!\!\!\!\!\!\!\!\!\!\!\!\!\!\! \varphi_t (z,t)    =  \left(\frac{E_m\,d}{\omega}\right) \Delta_{v\,ampl\,c} \, \left[X_t \cos(\omega t - \beta z + \delta  )+ Y_t \sin(\omega t - \beta z + \delta  ) \right] , \\
&&\!\!\!\!\!\!\!\!\!\!\!\!\!\!\!\!\!\!\!\!\!\!\!\!\!\!\!\!\!\! \varphi_z' (z,t)    =  \left(\frac{E_m\,w}{\omega}\right) \Delta_{v\,ampl\,c} \, \left[X_z \cos(\omega t - \beta z + \delta  )+ Y_z \sin(\omega t - \beta z + \delta  ) \right] , \\
&&\!\!\!\!\!\!\!\!\!\!\!\!\!\!\!\!\!\!\!\!\!\!\!\!\!\!\!\!\!\! \varphi_t' (z,t)    =  \left(\frac{E_m\,w}{\omega}\right) \Delta_{v\,ampl\,s} \, \left[X_t \cos(\omega t - \beta z + \delta  )+ Y_t \sin(\omega t - \beta z + \delta  ) \right] . \label{fluxdef6}
\end{eqnarray}
As for the single gauge case, we use a prime to denote parameters corresponding to lateral electrodes. 
Note a few peculiarities: the virtual photon fluxes are present {\it on both electrodes}, their phase is different from the one of the real photons, and all amplitudes are {\it renormalized}, 
by parameters that can be {\it positive or negative} (see discussion below).
In particular:
\begin{equation}
\Delta_{H\,ampl} = - \sin(\theta_z) \,.
\end{equation}
But {\it the virtual charges and currents}  living on the electrodes must proceed from these fluxes through adapted expressions of the type of Eqs. (\ref{sigmat}-\ref{jil}), \ref{genflux}, see final Subsection of this Appendix.   
As such, their energy density when integrated over the electrode surfaces should give us the total energy $H$.
This implies:
\begin{equation}
\Delta_{H}= \Delta_{H\,ampl}^2 \, , \label{deltaHval}
\end{equation}
which fixes the possible values for $\theta_z$, and through Eq. (\ref{thetazDeltatheta}) of $\Delta \theta$. All are summarized in Tab. \ref{tab_4}, with the corresponding {\it signs} of the fluxes amplitudes present in Eqs. (\ref{fluxdef1}-\ref{fluxdef6}). 
Remarkably, these choices all lead to Eqs. (\ref{DeltaHs}-\ref{deltacs}).
The virtual amplitude renormalizations become then:
\begin{eqnarray}
\Delta_{v\,ampl\,s} & = & - \mbox{sign}(\sin[\Delta \theta])\, \Delta_{H\,ampl}\,\sqrt{2} \, \sin({\cal G}), \label{calG1} \\
\Delta_{v\,ampl\,c} & = & - \mbox{sign}(\sin[\Delta \theta]) \,\Delta_{H\,ampl}\, \sqrt{2} \, \cos({\cal G}) .\label{calG2}
\end{eqnarray}
The virtual electrodes are part of a {\it gedankenexperiment}, and as such there is no reason to introduce an asymmetry between top-bottom and left-right pairs. A completely symmetric treatment of the fluxes amplitudes is obtained if one considers a square section box containing the electromagnetic field ($d=w$), and choosing ${\cal G} = \pm \pi/4$ modulo $\pi$. Again only {\it the signs} of the fluxes amplitudes as defined in Eqs. (\ref{fluxdef1}-\ref{fluxdef6}) can be different, and all possibilities are presented in Tab. \ref{tab_4}.
Since the phases appearing in these expressions have a profound meaning ($\Delta \theta$ and $\delta$ are gauge angles), we deliberately kept the sign degeneracy on the amplitude factors only (instead of incorporating it in each specific phase).

\begin{table}[h!]
\center
\caption{Gauge angles $\theta_z, \Delta \theta$ and $\cal G$ obtained by the final gauge fixing, and corresponding polarities of the fluxes amplitudes $\Delta_{H\,ampl},\Delta_{v\,ampl\,c},\Delta_{v\,ampl\,s}$. The signs of the tilded functions components $\tilde{\varphi}_i$ and of $\varphi_k, \varphi_s$ (with and without prime) are straightforwardly deduced  (see text). 
}
\begin{tabular}{@{}llllllll}
\br
  &  Polarities  & & &   & Angles  &   \\
$\varphi_x', \varphi_y$ & $\varphi_z, \varphi_t'$ & $\varphi_t, \varphi_z'$ & & $\theta_z$ & $\Delta \theta$ & $\cal G$ \\
\mr
$-$  & $+$  & $+$ & & $\theta_{z\,0}$  &  $\Delta \theta_0$ & ${\cal G}_0$  \\
$-$  & $-$  & $-$ & & $\theta_{z\,0}$  &  $-\Delta \theta_0$ & ${\cal G}_0$  \\
$+$  & $+$  & $+$ & & $-\theta_{z\,0}$  &  $\Delta \theta_0+\pi$ & ${\cal G}_0$  \\
$+$  & $-$  & $-$ & & $-\theta_{z\,0}$  &  $-\Delta \theta_0+\pi$ & ${\cal G}_0$  \\
$+$  & $-$  & $-$ & & $\theta_{z\,0}+\pi$  &  $\Delta \theta_0$ & ${\cal G}_0$  \\
$+$  & $+$  & $+$ & & $\theta_{z\,0}+\pi$  &  $-\Delta \theta_0$ & ${\cal G}_0$  \\
$-$  & $-$  & $-$ & & $-\theta_{z\,0}+\pi$  &  $\Delta \theta_0+\pi$ & ${\cal G}_0$  \\
$-$  & $+$  & $+$ & & $-\theta_{z\,0}+\pi$  &  $-\Delta \theta_0+\pi$ & ${\cal G}_0$  \\
$-$  & $-$  & $+$ & & $\theta_{z\,0}$  &  $\Delta \theta_0$ & $-{\cal G}_0$  \\
$-$  & $+$  & $-$ & & $\theta_{z\,0}$  &  $-\Delta \theta_0$ & $-{\cal G}_0$  \\
$+$  & $-$  & $+$ & & $-\theta_{z\,0}$  &  $\Delta \theta_0+\pi$ & $-{\cal G}_0$  \\
$+$  & $+$  & $-$ & & $-\theta_{z\,0}$  &  $-\Delta \theta_0+\pi$ & $-{\cal G}_0$  \\
$+$  & $+$  & $-$ & & $\theta_{z\,0}+\pi$  &  $\Delta \theta_0$ & $-{\cal G}_0$  \\
$+$  & $-$  & $+$ & & $\theta_{z\,0}+\pi$  &  $-\Delta \theta_0$ & $-{\cal G}_0$  \\
$-$  & $+$  & $-$ & & $-\theta_{z\,0}+\pi$  &  $\Delta \theta_0+\pi$ & $-{\cal G}_0$  \\
$-$  & $-$  & $+$ & & $-\theta_{z\,0}+\pi$  &  $-\Delta \theta_0+\pi$ & $-{\cal G}_0$  \\
$-$  & $-$  & $-$ & & $\theta_{z\,0}$  &  $\Delta \theta_0$ & ${\cal G}_0+\pi$  \\
$-$  & $+$  & $+$ & & $\theta_{z\,0}$  &  $-\Delta \theta_0$ & ${\cal G}_0+\pi$  \\
$+$  & $-$  & $-$ & & $-\theta_{z\,0}$  &  $\Delta \theta_0+\pi$ & ${\cal G}_0+\pi$  \\
$+$  & $+$  & $+$ & & $-\theta_{z\,0}$  &  $-\Delta \theta_0+\pi$ & ${\cal G}_0+\pi$  \\
$+$  & $+$  & $+$ & & $\theta_{z\,0}+\pi$  &  $\Delta \theta_0$ & ${\cal G}_0+\pi$  \\
$+$  & $-$  & $-$ & & $\theta_{z\,0}+\pi$  &  $-\Delta \theta_0$ & ${\cal G}_0+\pi$  \\
$-$  & $+$  & $+$ & & $-\theta_{z\,0}+\pi$  &  $\Delta \theta_0+\pi$ & ${\cal G}_0+\pi$  \\
$-$  & $-$  & $-$ & & $-\theta_{z\,0}+\pi$  &  $-\Delta \theta_0+\pi$ & ${\cal G}_0+\pi$  \\
$-$  & $+$  & $-$ & & $\theta_{z\,0}$  &  $\Delta \theta_0$ & $-{\cal G}_0+\pi$  \\
$-$  & $-$  & $+$ & & $\theta_{z\,0}$  &  $-\Delta \theta_0$ & $-{\cal G}_0+\pi$  \\
$+$  & $+$  & $-$ & & $-\theta_{z\,0}$  &  $\Delta \theta_0+\pi$ & $-{\cal G}_0+\pi$  \\
$+$  & $-$  & $+$ & & $-\theta_{z\,0}$  &  $-\Delta \theta_0+\pi$ & $-{\cal G}_0+\pi$  \\
$+$  & $-$  & $+$ & & $\theta_{z\,0}+\pi$  &  $\Delta \theta_0$ & $-{\cal G}_0+\pi$  \\
$+$  & $+$  & $-$ & & $\theta_{z\,0}+\pi$  &  $-\Delta \theta_0$ & $-{\cal G}_0+\pi$  \\
$-$  & $-$  & $+$ & & $-\theta_{z\,0}+\pi$  &  $\Delta \theta_0+\pi$ & $-{\cal G}_0+\pi$  \\
$-$  & $+$  & $-$ & & $-\theta_{z\,0}+\pi$  &  $-\Delta \theta_0+\pi$ & $-{\cal G}_0+\pi$  \\
\br
\label{tab_4}
\end{tabular}
\end{table}
\normalsize

The last angle that must be found is $\delta$. This one is obtained from \ref{PiApp}, considering the commutation properties of the $\vec{\Pi}$ pseudo-vector. 
This fixes the possible values of $\alpha$ with Eq. (\ref{sin2cos2}), knowing that $\alpha \rightarrow - \alpha$ is part of the time reversal symmetry properties (\ref{PandT}). With the help of Eqs. (\ref{cosalpha},\ref{sinalpha}), all possible $\delta$ are then defined. These do not change the signs of the above-mentioned amplitudes themselves, but rather the ones of the different terms appearing in $\Delta E$, $\vec{\Pi}$ and $\vec{C}$ through the polarities of $\cos(\alpha)$ and $\sin(\alpha)$, see Tab. \ref{tab_5}.

Any gauge parameter $\theta_z, \Delta \theta, {\cal G}$ and $\delta$ appears then within a sign $\pm$ degeneracy and a modulo $\pi$. This defines the {\it legitimate transformations} that link equivalent photon states. $\cal P$, $\cal T$ and $\cal C$ must consist of such symmetries only. On the other hand, the $\delta \rightarrow \delta \pm \pi/2$ transformation that flips $C_c \leftrightarrow C_r$ {\it is not} part of these operations: $C_c$ and $C_r$ are different quantities, but they are strictly linked to each other. \\

The "Devoret-like" gauge expressions become in the dual gauge case:
\begin{eqnarray}
&&  +\frac{ \partial \varphi_y}{\partial t} + \frac{ \partial \varphi_z}{\partial t} +\frac{ \partial  \varphi_t}{\partial t} =  \Delta V   , \\
&& + \frac{ \partial \varphi_y}{\partial z} + \frac{ \partial \varphi_z}{\partial z} +\frac{ \partial  \varphi_t}{\partial z} = -\Delta A  , \\
&\mbox{and:} &  \nonumber \\
&& +\frac{ \partial \varphi_x'}{\partial t} + \frac{ \partial \varphi_z'}{\partial t} -\frac{ \partial  \varphi_t'}{\partial t} =\Delta V'  , \\
&& +\frac{ \partial \varphi_x'}{\partial z} + \frac{ \partial \varphi_z'}{\partial z} -\frac{ \partial  \varphi_t'}{\partial z} = - \Delta A'   .
\end{eqnarray}
Note the sign change for $\varphi_t'$. All photons are represented in these formulas, but if one considers only the  generalized fluxes related to {\it real bosons}, we recover exactly Eqs. (\ref{devoret1}-\ref{devoret4}), which are characteristic of the "$\varphi$-gauge".

The equations fulfilled by the angular momentum density
are obtained as:
\begin{eqnarray}
\vec{x}. \Delta \vec{m}_s & = & + 2 \sigma_t \left( +\varphi_k - (d/w) \, \frac{x}{2} \, \Delta A' \right) , \\
\vec{y}. \Delta \vec{m}_s & = & + 2 \sigma_t \left( + \frac{x}{2} \, \Delta A \right) , \\
\vec{z}. \Delta \vec{m}_s & = & +2 \sigma_t \left( + \frac{\varphi_s}{2} \right) ,
\end{eqnarray}
and: 
\begin{eqnarray}
\vec{x}. \Delta \vec{m}_s' & = & + 2 \sigma_l \left( - \frac{y}{2} \, \Delta A' \right) , \\
\vec{y}. \Delta \vec{m}_s' & = & + 2 \sigma_l \left( -\varphi_k' + (w/d)\, \frac{y}{2} \, \Delta A \right) , \\
\vec{z}. \Delta \vec{m}_s' & = &+2 \sigma_l \left(-\frac{\varphi_s'}{2}  \right), 
\end{eqnarray}
which actually {\it look exactly like }  the relations obtained in \ref{propG} for the "$s$-gauge" (we keep here  $d$ and $w$ for the sake of a clear comparison with this previous Appendix).
But the flux-like terms appearing in the above expressions write:
\begin{eqnarray}
&& \!\!\!\!\!\!\!\!\!\!\!\!\!\!\!\!\!\!\!\!\!\!\!\!\!\!\!\!\!\!\!\!\!\!\!\!\!\!\varphi_k(z,t)  =  \left(\frac{E_m\,d}{\omega}\right) \Delta_{H\,ampl} \, \left[X_z \cos(\omega t - \beta z + \delta_k  )+ Y_z \sin(\omega t - \beta z + \delta_k  ) \right], \\
&& \!\!\!\!\!\!\!\!\!\!\!\!\!\!\!\!\!\!\!\!\!\!\!\!\!\!\!\!\!\!\!\!\!\!\!\!\!\!\varphi_k'(z,t)  =  \left(\frac{E_m\,w}{\omega}\right) \Delta_{H\,ampl} \, \left[X_z \cos(\omega t - \beta z + \delta_k  )+ Y_z \sin(\omega t - \beta z + \delta_k  ) \right], \\
&& \!\!\!\!\!\!\!\!\!\!\!\!\!\!\!\!\!\!\!\!\!\!\!\!\!\!\!\!\!\!\!\!\!\!\!\!\!\!\varphi_s(z,t)   =  \sqrt{2}  \left(\frac{E_m\,d}{\omega}\right) \Delta_{H\,ampl} \, \left[X_x \sin(\omega t - \beta z + \Delta \theta_s  )- Y_x \cos(\omega t - \beta z + \Delta \theta_s  ) \right] \nonumber \\
&& \!\!\!\!\!\!\!\!\!\!\!\!\!\!\!\!\!\!\!\!\!\!\!\!\!\!\!\!\!\!\!\!\!\!\!\!\!\! + \frac{1}{ \sqrt{2}} \left(\frac{E_m\,d}{\omega}\right) \Delta_{v\,ampl\,c} \, \left[X_z \cos(\omega t - \beta z + \delta_s  )+ Y_z \sin(\omega t - \beta z + \delta_s  ) \right] \nonumber \\
&& \!\!\!\!\!\!\!\!\!\!\!\!\!\!\!\!\!\!\!\!\!\!\!\!\!\!\!\!\!\!\!\!\!\!\!\!\!\! -\frac{1}{ \sqrt{2}} \left(\frac{E_m\,d}{\omega}\right) \Delta_{v\,ampl\,s} \, \left[X_t \cos(\omega t - \beta z + \delta_s  )+ Y_t \sin(\omega t - \beta z + \delta_s  ) \right] , \label{varphis2}  \\
&& \!\!\!\!\!\!\!\!\!\!\!\!\!\!\!\!\!\!\!\!\!\!\!\!\!\!\!\!\!\!\!\!\!\!\!\!\!\!\varphi_s'(z,t)   =   \sqrt{2}  \left(\frac{E_m\,w}{\omega}\right) \Delta_{H\,ampl} \, \left[X_y \sin(\omega t - \beta z + \Delta \theta_s  )- Y_y \cos(\omega t - \beta z + \Delta \theta_s  ) \right] \nonumber \\
&& \!\!\!\!\!\!\!\!\!\!\!\!\!\!\!\!\!\!\!\!\!\!\!\!\!\!\!\!\!\!\!\!\!\!\!\!\!\! + \frac{1}{ \sqrt{2}} \left(\frac{E_m\,w}{\omega}\right) \Delta_{v\,ampl\,s} \, \left[X_z \cos(\omega t - \beta z + \delta_s  )+ Y_z \sin(\omega t - \beta z + \delta_s  ) \right] \nonumber \\
&& \!\!\!\!\!\!\!\!\!\!\!\!\!\!\!\!\!\!\!\!\!\!\!\!\!\!\!\!\!\!\!\!\!\!\!\!\!\! +\frac{1}{ \sqrt{2}} \left(\frac{E_m\,w}{\omega}\right) \Delta_{v\,ampl\,c} \, \left[X_t \cos(\omega t - \beta z + \delta_s  )+ Y_t \sin(\omega t - \beta z + \delta_s  ) \right] . \label{varphisprime2}
\end{eqnarray}
These replace Eqs. (\ref{varphis},\ref{varphisprime}) and Eqs. (\ref{varphisbis},\ref{varphisbisprime}) obtained in the single gauge case. 
Note the specific phases appearing here ($k$ and $s$ subscripts), 
given in Tab. \ref{tab_6}, 
which are deduced (with uniqueness) from the original $\Delta \theta, \delta$. 

We finally see that the dual gauge approach {\it combines} the properties of the "$\varphi$-gauge" and "$s$-gauge" together: it leads to both the definition of generalized fluxes $\varphi_i, \varphi_i'$ (for all bosons, $i=x,y,z,t$), which enable to express external properties ($H$ and $\vec{P}$), and of angular momentum related functions $\varphi_j, \varphi_j'$ (with $j=k,s$) that enable to derive the internal properties ($\vec{S}, \vec{\Pi}$ and $\vec{C}$). 
Interestingly, $\varphi_k, \varphi_k'$ involve only the $z$ component, while $\varphi_s, \varphi_s'$ contain all of them.
The $\sqrt{2}$ factors appearing in Eqs. (\ref{varphis2},\ref{varphisprime2}) are a consequence of the {\it phase differences} between these expressions and the surface charges $\sigma_t,\sigma_l$, see following Subsection.
In order to achieve this mathematical construction, the boundary conditions (namely the virtual electrodes) appear to be a key ingredient. 

\begin{table}[h!]
\center
\caption{Eligible values of $\delta$ and corresponding $\cos(\alpha)$ and $\sin(\alpha)$ polarities (see text). The only entry here is the $\Delta \theta$ parameter. }
\begin{tabular}{@{}llllllll}
\br
$\Delta \theta$  &  & sign of $\cos(\alpha)$ & sign of $\sin(\alpha)$ &  & $\delta$ &  \\
\mr
$\Delta \theta_0$  &   & $+$ & $+$  &   & $-\delta_1$   &   \\
$\Delta \theta_0$  &   & $-$ & $-$  &   & $\pi-\delta_1$   &   \\
$\Delta \theta_0$  &   & $+$ & $-$  &   & $\pi+\delta_0$   &   \\
$\Delta \theta_0$  &   & $-$ & $+$  &   & $\delta_0$   &   \\
$-\Delta \theta_0$  &   & $+$ & $+$  &   & $\pi-\delta_0$   &   \\
$-\Delta \theta_0$  &   & $-$ & $-$  &   & $ -\delta_0$   &   \\
$-\Delta \theta_0$  &   & $+$ & $-$  &   & $\delta_1$   &   \\
$-\Delta \theta_0$  &   & $-$ & $+$  &   & $\pi+\delta_1$   &   \\
$\Delta \theta_0+\pi$  &   & $+$ & $+$  &   & $-\delta_1$   &   \\
$\Delta \theta_0+\pi$  &   & $-$ & $-$  &   & $\pi-\delta_1$   &   \\
$\Delta \theta_0+\pi$  &   & $+$ & $-$  &   & $\pi+\delta_0$   &   \\
$\Delta \theta_0+\pi$  &   & $-$ & $+$  &   & $\delta_0$   &   \\
$-\Delta \theta_0+\pi$  &   & $+$ & $+$  &   & $\pi-\delta_0$   &   \\
$-\Delta \theta_0+\pi$  &   & $-$ & $-$  &   & $ -\delta_0$   &   \\
$-\Delta \theta_0+\pi$  &   & $+$ & $-$  &   & $\delta_1$   &   \\
$-\Delta \theta_0+\pi$  &   & $-$ & $+$  &   & $\pi+\delta_1$   &   \\
\br
\label{tab_5}
\end{tabular}
\end{table}
\normalsize

\subsection{Virtual charges and currents}

Virtual charges and currents are defined from the boundary conditions Eqs. (\ref{bound1}-\ref{bound4}).
After some manipulations, one obtains:
\begin{eqnarray}
\sigma_t(x,z,t) & = & + C_d \left[ \frac{\partial \tilde{\varphi}_y (z,t)}{\partial t} + \Delta_\pi \left( \frac{\partial \tilde{\varphi}_z (z,t)}{\partial t}+ \frac{\partial \tilde{\varphi}_t (z,t)}{\partial t}\right) \right], \label{sigma2G1}  \\
\vec{j}_t(x,z,t) & = &-  L_d^{-1} \left[ \frac{\partial \tilde{\varphi}_y (z,t) }{\partial z} + \Delta_\pi \left( \frac{\partial \tilde{\varphi}_z (z,t)}{\partial z}+ \frac{\partial \tilde{\varphi}_t (z,t)}{\partial z}\right)\right] \, \vec{z} ,
\end{eqnarray}
for the top virtual electrode, and:
\begin{eqnarray}
\sigma_l(x,z,t) & = & + C_d' \left[ \frac{\partial \tilde{\varphi}_x' (z,t)}{\partial t} + \Delta_\pi \left( \frac{\partial \tilde{\varphi}_z' (z,t)}{\partial t}- \frac{\partial \tilde{\varphi}_t' (z,t)}{\partial t}\right) \right],   \\
\vec{j}_l(x,z,t) & = &-  L_d'^{\,-1} \left[ \frac{\partial \tilde{\varphi}_x' (z,t) }{\partial z} + \Delta_\pi \left( \frac{\partial \tilde{\varphi}_z' (z,t)}{\partial z}- \frac{\partial \tilde{\varphi}_t' (z,t)}{\partial z}\right)\right] \, \vec{z} , \label{sigma2G2}
\end{eqnarray}
for the left one; signs are reversed for bottom and right electrodes, as in the single gauge case. The capacitances $C_d, C_d'$ and inverse inductances $L_d^{-1}, L_d'^{\,-1}$ per unit surface have been given in \ref{genflux}. Note the sign change in front of $\tilde{\varphi}_t'$.
One can easily show that  Eq. (\ref{conserve}) leads to the  propagation equations for the real fields:
\begin{eqnarray}
\frac{\partial^2 \tilde{\varphi}_y (z,t)}{\partial   z^2} -\frac{1}{c^2} \frac{\partial^2 \tilde{\varphi}_y (z,t)}{\partial   t^2} & = & 0 , \\
\frac{\partial^2 \tilde{\varphi}_x' (z,t)}{\partial   z^2} -\frac{1}{c^2} \frac{\partial^2 \tilde{\varphi}_x' (z,t)}{\partial   t^2} & = & 0 ,
\end{eqnarray}
and similar ones for the virtual fields $z,t$ (with and without primes equivalently).  
As in \ref{genflux}, this qualifies {\it all generalized fluxes as massless scalar fields}, propagating at the speed of light $c$, see Ref. \cite{waveguide}. One can also define virtual charges $\tilde{Q}_j, \tilde{Q}_j'$, using equations similar to Eqs. (\ref{Qn},\ref{Qp}): each generalized flux and virtual charge pair verifies the canonical commutation rules, as in the single gauge case.

The expressions Eqs. (\ref{sigma2G1}-\ref{sigma2G2}) resemble very much the single gauge ones, 
 Eqs. (\ref{sigmat}-\ref{jil}), but the virtual fluxes appear with a $\Delta_\pi$ prefactor. Besides, we 
have put a {\it tilde} on top of these generalized fluxes functions, because they actually differ from the original ones Eqs. (\ref{fluxdef1}-\ref{fluxdef6}) by {\it their phases}:
\begin{eqnarray}
& &\!\!\!\!\!\!\!\!\!\!\!\!\!\!\!\!\!\!\!\!\!\!\!\!\!\!\!\!\!\! \tilde{\varphi}_x' (z,t)    =   \left(\frac{E_m\,w}{\omega}\right) \Delta_{H\,ampl} \, \left[X_x \sin(\omega t - \beta z + \tilde{\Delta \theta} )- Y_x \cos(\omega t - \beta z + \tilde{\Delta \theta} ) \right] ,   \\
&&\!\!\!\!\!\!\!\!\!\!\!\!\!\!\!\!\!\!\!\!\!\!\!\!\!\!\!\!\!\! \tilde{\varphi}_y (z,t)    =  \left(\frac{E_m\,d}{\omega}\right) \Delta_{H\,ampl} \, \left[X_y \sin(\omega t - \beta z + \tilde{\Delta \theta} )- Y_y \cos(\omega t - \beta z + \tilde{\Delta \theta} ) \right] , \\
&&\!\!\!\!\!\!\!\!\!\!\!\!\!\!\!\!\!\!\!\!\!\!\!\!\!\!\!\!\!\! \tilde{\varphi}_z (z,t)    =  \left(\frac{E_m\,d}{\omega}\right) \Delta_{v\,ampl\,s} \, \left[X_z \cos(\omega t - \beta z + \tilde{\delta}  )+ Y_z \sin(\omega t - \beta z + \tilde{\delta}  ) \right] , \\
&&\!\!\!\!\!\!\!\!\!\!\!\!\!\!\!\!\!\!\!\!\!\!\!\!\!\!\!\!\!\! \tilde{\varphi}_t (z,t)    =  \left(\frac{E_m\,d}{\omega}\right) \Delta_{v\,ampl\,c} \, \left[X_t \cos(\omega t - \beta z + \tilde{\delta}  )+ Y_t \sin(\omega t - \beta z + \tilde{\delta} ) \right] , \\
&&\!\!\!\!\!\!\!\!\!\!\!\!\!\!\!\!\!\!\!\!\!\!\!\!\!\!\!\!\!\! \tilde{\varphi}_z' (z,t)    =  \left(\frac{E_m\,w}{\omega}\right) \Delta_{v\,ampl\,c} \, \left[X_z \cos(\omega t - \beta z + \tilde{\delta}  )+ Y_z \sin(\omega t - \beta z + \tilde{\delta} ) \right] , \\
&&\!\!\!\!\!\!\!\!\!\!\!\!\!\!\!\!\!\!\!\!\!\!\!\!\!\!\!\!\!\! \tilde{\varphi}_t' (z,t)    =  \left(\frac{E_m\,w}{\omega}\right) \Delta_{v\,ampl\,s} \, \left[X_t \cos(\omega t - \beta z + \tilde{\delta}  )+ Y_t \sin(\omega t - \beta z + \tilde{\delta}  ) \right] . 
\end{eqnarray}
These $\tilde{\Delta \theta}, \tilde{\delta}$
are obtained from the original ones $ \Delta \theta, \delta$ with uniqueness,
 and are listed in Tab. \ref{tab_6}.
This is a peculiarity which arises from the dual gauge approach: the phases that appear in the Devoret expressions are actually {\it different}  from the ones involved in the charge/current formulas.
If this leads to a measurable phenomenon remains an intriguing open question.
Note that all non-tilded $\varphi_j,\varphi_j'$ functions (including the ones with $k$ and $s$ indexes) do conform to the same propagation equations as the tilded generalized fluxes: all fields propagate at the speed of light $c$. \\

Following Ref. \cite{waveguide}, as for the single gauge situation discussed in \ref{genflux} we can define from virtual charges and currents a {\it surface energy density} Eq. (\ref{Hdensity}), which is thus expressed in terms of the $\tilde{\varphi}_i, \tilde{\varphi}_i'$ functions.
The surface integral of $H_d$ over the virtual electrodes leads to our Hamiltonian $H$ if one imposes 
Eq. (\ref{deltaHval}). 
This has been used in the previous Section when fixing all phases, see Tabs. \ref{tab_4} and \ref{tab_5} for a summary of all possible configurations (which differ only by some polarities).
One also obtains a simple relationship between $ \alpha$ and $\tilde{\delta},   \tilde{\Delta \theta}$, see Tab. \ref{tab_6}.

\begin{table}[h!]
\center
\caption{Extra angles obtained from the original gauge ones, when describing angular momentum density and charges/currents (see text). The only entry here is the $\Delta \theta$ parameter. }
\hspace*{-1.5cm}
\begin{tabular}{@{}llllllllllll}
\br
 Gauge   phase &  &  & $\vec{m}_s$ &  \hspace*{-1cm} related  & & & Charge related  & &  &\\
$\Delta \theta$ &   &  & $\delta_k$ & $\Delta \theta_s$ & $\delta_s$ & & $\tilde{\Delta \theta}$ & $\cos(\tilde{\delta})$ & $\sin(\tilde{\delta})$  & $\alpha$ \\
\mr
$\Delta \theta_0$   & & & $+\Delta \theta_0+ \pi$ &  $\,\,\,\pi$  &  $\delta-\frac{\pi}{4} $ & & $+\Delta \theta_0 +\pi $  & $\frac{\cos(\delta)-3\sin(\delta)}{\sqrt{10}}$ & $\frac{3\cos(\delta)+\sin(\delta)}{\sqrt{10}}$ & $\tilde{\delta}-\tilde{\Delta \theta} -\frac{\pi}{2}$ \\
$-\Delta \theta_0$  & & & $-\Delta \theta_0 $  & $\,\,\,\pi$ & $\delta+\frac{\pi}{4} $ &    & $-\Delta \theta_0  $ & $\frac{-\cos(\delta)-3\sin(\delta)}{\sqrt{10}}$ & $\frac{3\cos(\delta)-\sin(\delta)}{\sqrt{10}}$ & $\tilde{\delta}-\tilde{\Delta \theta}  +\frac{\pi}{2}$\\
$\Delta \theta_0+\pi$ & &  & $+\Delta \theta_0 $ &  $\,\,\,0$ & $\delta-\frac{\pi}{4} $  & & $+\Delta \theta_0 $ & $\frac{\cos(\delta)-3\sin(\delta)}{\sqrt{10}}$ & $\frac{3\cos(\delta)+\sin(\delta)}{\sqrt{10}}$& $\tilde{\delta}-\tilde{\Delta \theta} +\frac{\pi}{2}$ \\
$-\Delta \theta_0+\pi$& &  & $-\Delta \theta_0+ \pi$ & $\,\,\,0$  &  $\delta+\frac{\pi}{4} $ & & $-\Delta \theta_0 +\pi $  & $\frac{-\cos(\delta)-3\sin(\delta)}{\sqrt{10}}$ & $\frac{3\cos(\delta)-\sin(\delta)}{\sqrt{10}}$ & $\tilde{\delta}-\tilde{\Delta \theta} -\frac{\pi}{2}$ \\
\br
\label{tab_6}
\end{tabular}
\end{table}
\normalsize

The phases appearing in Tabs. \ref{tab_4}, \ref{tab_5} and \ref{tab_6} are obtained from the following ones:
\begin{eqnarray}
\theta_{z\,0} & = & \arctan \left( \frac{1}{\sqrt{2} } \right) , \\
\Delta \theta_0 & = & \frac{\pi}{4} , \\
\delta_0 & = & \arctan \left( \frac{-\sqrt{2}+2\sqrt{3}}{2\sqrt{2}+\sqrt{3} } \right) , \\
\delta_1 & = & \arctan \left( \frac{+\sqrt{2}+2\sqrt{3}}{2\sqrt{2}-\sqrt{3} } \right) , \\
{\cal G}_0 & = & \frac{\pi}{4} , 
\end{eqnarray}
all defined within $[0,\pi/2]$. 
Looking carefully at Tabs. \ref{tab_4} and \ref{tab_5}, we see that we have $32=2^5$ different fluxes  {\it and} $\{\cos(\alpha),\sin(\alpha)\}$ polarity configurations.
This is a trivial degeneracy, which comes from the free $\pm$ choice of polarities that we have: all the time-shifts are incommensurate and one cannot define a proper reference which could be taken for all (as was the case in the single gauge situation, with a {\it unique} $E_m>0$ imposed).

However, more interestingly each state 
 is {\it four times degenerate}. 
Consider first configurations with identical $\cal G$. 
This is a first twofold degeneracy that enables also to connect states $\delta \rightarrow - \delta$, which is required for the time reversal symmetry (\ref{PandT}). 
The second twofold degeneracy connects states with ${\cal G} \rightarrow {\cal G}+\pi$. This symmetry corresponds to a swap of top-bottom electrodes on one hand, and left-right on the other. Since these electrodes are virtual, one could wonder in the first place if this symmetry contains any physical information or not. 
To answer the point, consider the combined transformation:
\begin{eqnarray}
{\cal G} &\rightarrow & {\cal G}+\pi , \\
\Delta \theta &\rightarrow & - \Delta \theta, \\
Y_i^j & \rightarrow & - Y_i^j \,\,\,\,\,\,\,\,\,\, \mbox{with $i=x,y$ and $j=\varphi,s$}\, , \\
& \mbox{and:} \nonumber \\
\delta & \rightarrow & - \delta +\pi ,
\end{eqnarray}
which resembles somehow time reversal $\cal T$. Like this one, it is a legitimate operation that links equivalent states, see Tab. \ref{tab_5}. What it actually does is that it {\it flips the sign of all internal properties}:
\begin{eqnarray}
\vec{S} & \rightarrow & - \vec{S} , \\
\vec{\Pi} & \rightarrow & - \vec{\Pi} , \\
\vec{C} & \rightarrow & - \vec{C} , 
\end{eqnarray}
while preserving all external ones ($H, \vec{P}$ and $\vec{J}=\vec{L}=0$). In this sense, this is a "full conjugation" of the photon state, which combines {\it charge conjugation} with an $\vec{S}$ and $\vec{\Pi}$ polarity flip.
As well, its effect on the signs of the gauge coefficients is actually complementary to the time reversal $\cal T$ symmetry: $\tilde{g}_x^s , \tilde{g}_y^s$ flip signs, while $g_x^s  , g_y^s$   are unaltered.
 Interestingly:
\begin{equation}
{\cal C.P.T } = \mbox{Id},
\end{equation}
namely applying the 3 discrete symmetries together leaves the problem unchanged.
Note that restricting the Hilbert space to the physical states verifying $<\vec{C}>=0$ (see Section \ref{eigenstates}), the conjugation $\cal C$ amounts to a simple flip of helicity (and parity): if charge conjugation defines anti-photons from photons, we see here that these two share the same nature. \\

For the sake of completeness, we finally give the extra phases $\Delta \theta_s, \tilde{\Delta \theta}$ and $\delta_k, \delta_s, \tilde{\delta}$ constructed form the original gauge parameters in Tab. \ref{tab_6}. 
These values are uniquely defined from a given set of gauge phases; they do not bring any extra degeneracy. 
Note the equality $\delta_k = \tilde{\Delta \theta}$.


\begin{thebibliography}{100}
\bibitem{comment} One could argue that $\Delta_H = - \Delta_S$ is also an acceptable solution, generating a photon spin with negative signature (a $-\hbar$ appearing in all expressions). It turns out that this does not work mathematically.
\end{thebibliography}
\end{document}